\pdfoutput=1
\documentclass[12pt,reqno]{article}
\usepackage[markup=underlined]{changes}
\RequirePackage{etex}

\usepackage{kpfonts}
\usepackage{inconsolata}

\usepackage[greek,english]{babel}
\usepackage{etex}
\usepackage{graphicx}
\usepackage{setspace}
\usepackage{amsmath}
\usepackage{amsfonts}
\usepackage{titlesec}
\usepackage{pictex}
\usepackage{comment}
\usepackage{amsthm}
\usepackage{graphics,epsfig,verbatim,bm,latexsym,url,amsbsy}
\usepackage{rotating}
\usepackage[authoryear,round]{natbib}
\usepackage{float}
\usepackage[position=bottom]{subfig}
\usepackage{mathrsfs}
\usepackage{multirow}
\usepackage{array}
\usepackage{bigints}
\usepackage{bbm}
\usepackage{geometry}
\usepackage{bbm}
\usepackage[ruled,vlined]{algorithm2e}
\usepackage{soul}
\usepackage{mathtools}
\usepackage{amssymb}
\usepackage{enumitem}
\usepackage{microtype}

\DeclarePairedDelimiterX{\inp}[2]{\langle}{\rangle}{#1, #2}

\usepackage{hyperref}
\hypersetup{
	colorlinks=true,
	linkcolor=red!50!black,
	citecolor=blue!50!black,
	filecolor=magenta,      
	urlcolor=blue!80!black
}
\usepackage[nameinlink,noabbrev,sort,capitalise]{cleveref}
\crefname{appendix}{Appendix}{Appendices}

\newcommand{\el}{\ell}

\newcommand{\Ex}{\mathbb{E}}

\renewcommand{\Pr}{\mathbb{P}}

\DeclareMathOperator*{\argmax}{argmax}

\theoremstyle{definition} \newtheorem{example}{Example}
\theoremstyle{definition} 
\theoremstyle{plain} \newtheorem{corollary}{Corollary}
\theoremstyle{definition} 
\theoremstyle{definition} \newtheorem{definition}{Definition}
\theoremstyle{definition} 
\theoremstyle{plain} \newtheorem{lemma}{Lemma}
\theoremstyle{plain} \newtheorem{theorem}{Theorem}
\theoremstyle{definition} \newtheorem{assumption}{Assumption}
\theoremstyle{definition} 
\theoremstyle{definition} 
\theoremstyle{plain}\newtheorem{proposition}{Proposition}
\theoremstyle{definition} 
\theoremstyle{definition} 
\theoremstyle{definition} 
\theoremstyle{definition} 
\theoremstyle{definition} 
\theoremstyle{definition} 

\theoremstyle{definition} 
\usepackage{mathtools}
\titlespacing*{\paragraph}{0pt}{1.25ex plus 1ex minus .2ex}{0.5em}

\makeatletter
\renewcommand\@makefnmark{\hbox{\@textsuperscript{\scriptsize\@thefnmark}}}
\makeatother

\titleformat{\section}
		{\bfseries\center \large \MakeUppercase}
         {\thesection}
        {0.5em}
        {}
        []

\titleformat{\subsection}[runin]
        {\normalfont\bfseries}
        {\thesubsection}
        {0.5em}
        {\addperiod}
        []
\newcommand{\addperiod}[1]{#1.}
\allowdisplaybreaks

\titlespacing*{\subsection}{0pt}{0.5ex plus 0.2ex minus 0.1ex}{0.25em}

\title{
\textls[-30]{\textbf{\MakeUppercase{Robust Technology Regulation}}}\thanks{Koh: MIT Department of Economics; \protect\texttt{\url{ajkoh@mit.edu}}.  Sanguanmoo: MIT Department of Economics; \protect\texttt{\url{sanguanm@mit.edu}}. \\~\\
We are grateful to Stephen Morris, Drew Fudenberg, and Daron Acemoglu for guidance, support, and many conversations. We also thank Ian Ball, Simon Board, Roberto Corrao, Théo Durandard, Piotr Dworczak, Matt Elliott, Andrea Galeotti, Joshua Gans, Alexis Ghersengorin, Gillian Hadfield, Anton Korinek, Jiangtao Li, Zihao Li, Jonathan Libgober, Daniel Luo, Parag Pathak, Jean Tirole, Mark Whitmeyer, Alex Wolitzky, Frank Yang, Weijie Zhong, as well as audiences at the 2025 SITE session on `Dynamic Games', the 2025 AEA Meeting session on `Policy Implications of Transformative AI', Johns Hopkins, and MIT Theory and Public Finance lunches for helpful comments. Andrew Koh acknowledges support from the Gordon B. Pye Dissertation Fellowship and the Centre for the Governance of AI (GovAI) where part of this work was done. First posted version: August 2024. }
}

\date{
\today 
\\ 
\small{\emph{latest version 
\href{https://www.dropbox.com/scl/fo/4hg5dl9lnj7m16cy8uexa/AB6xhts95af9HTDX8ZDL50w?rlkey=hzkpf8zqyen07nmgyudjrid4e&st=3jb5gsng&dl=0}{\underline{here}}}}
}

\author{\makebox[.3\linewidth]{Andrew Koh}\\{MIT} \\ \footnotesize{Job Market Paper} 
\and \makebox[.3\linewidth]{Sivakorn Sanguanmoo}\\{MIT}}

\begin{document}

\maketitle 

\setstretch{1.2}

\begin{abstract}
We analyze how uncertain technologies should be robustly regulated  and how regulation should evolve with new information. An \emph{adaptive sandbox} comprising a zero marginal tax up to an evolving quantity limit is (i) \emph{robust}: it delivers optimal payoff guarantees when the agent's learning process and/or preferences are chosen adversarially; (ii) \emph{dominant}: it outperforms other robust and regular mechanisms across all agent learning processes and preferences; (iii) \emph{time-consistent}: it is the only robust mechanism that can be implemented without commitment. Robustness is \emph{important}: absent robust regulation, worst-case payoffs can be arbitrarily poor and are induced by weak but growing optimism that encourages excessive risk-taking. Our results offer optimality foundations for existing policy and speak directly to current debates around managing emerging technologies.
\end{abstract}

\thispagestyle{empty}

\clearpage 
\setcounter{page}{1}

\section{Introduction} 

The first robot from Greco-Roman antiquity was Talos, a giant bronze automaton who hurled boulders at passing ships. 
The Greeks could not build Talos, but they did try
---Archimedes built catapults that rained `immense masses of stones... with incredible din and speed' (Plutarch, \emph{Marcellus} 14-19). Such technologies were seen, in antiquity, as something to be unleashed.\footnote{The description of Talos is given by Apollonius Rhodius in the Argonautica (4.1638-1688). The word `automaton' (\textgreek{αὐτόματον}) 
was first used in the Iliad (5.749) to describe the automatic wheeled tripos of Hephaestus. Aeschylus glorifies Prometheus for giving humans `all greatness' and `all arts' (\textgreek{τέχνη}/techne).}
Restraint, by contrast, is a more modern preoccupation that accompanied the industrial revolution and the rise of the regulatory state. Recent advancements across robotics, biology, and artificial intelligence mean that we could soon build our own Talos. What are we to make of these new and perhaps transformative technologies? How, if at all, should we regulate them? 

Motivated by these questions, we build a simple theory of technology regulation that takes three `stylized facts' seriously:
\begin{enumerate}
    \item \textbf{Uncertainty} about whether a new technology is beneficial or harmful;\footnote{For instance, there is substantial amount of disagreement among experts and forecasters on the tail risks posed by artificial intelligence \citep{karger2023forecasting} where the 25th and 75th percentile disagree by a factor of about $100$.} 
    \item \textbf{Misalignment} between the firm developing the technology and wider society;
    \item \textbf{Learning} as the technology is developed. 
\end{enumerate}

In our model, a real-valued {technology level} reflects the extent of irreversible R\&D or deployment, more of which is beneficial if the technology is safe, but harmful if it is dangerous. The {firm} developing the technology is less risk-averse than the {regulator} whose preferences reflect those of wider society. This differential risk aversion might arise if the firm does not internalize potential harms, stands to gain disproportionately from the potential benefits, or is caught up in an `R\&D arms race'. Both parties learn as the technology is developed, but the firm learns more than the regulator and might do so in quite varied ways e.g., extra independent or correlated information, or simply observe the same information more quickly. 

The firm's equilibrium technology choices are jointly shaped by \emph{policy}---the regulator's chosen paths of taxes or subsidies that might evolve with new information, \emph{preferences}---the firm's willingness to develop the technology given its beliefs, and \emph{learning}---what the firm has learnt thus far, what it might yet learn as it develops the technology further, and how it expects future regulation to unfold. We develop broad principles for regulating new technologies, as well as guidance for how regulation should evolve with new information.

Economically, we argue for an \emph{adaptive sandbox} that imposes a zero marginal tax on the technology up to an evolving quantity limit. This quantity limit adapts with the regulator's interim information, loosening and tightening as the regulator grows more or less optimistic. The adaptive sandbox is (i) \emph{robust}: it delivers the best payoff guarantee across all agent learning processes and/or preferences; (ii) \emph{undominated} and \emph{dominates} all robust mechanisms that satisfy a basic regularity property; and (iii) \emph{time-consistent}: it is the only robust mechanism that can be implemented without commitment. We further argue that robustness is \emph{important}: sans robust regulation, the worst-case learning process is natural and progressively induces the firm to take on excessive risk---this can magnify even small wedges in preferences to yield unboundedly poor regulator payoffs. Taken together, our results offer a principled defense of quantity over price instruments for technology regulation, as well as sharp refinements among different kinds of quantity instruments.\footnote{As \cite{weitzman1974prices} notes, ``the average economist in the Western marginalist tradition has at least a vague preference toward indirect control by prices, just as the typical non-economist leans toward the direct regulation of quantities.''} 

Conceptually, our framework offers a new perspective on how mechanisms should simultaneously \emph{evolve} with new information (adaptivity) and \emph{safeguard} against the worst-case (robustness). Our robustness notion is new, and requires optimal mechanisms to learn under ambiguity---but it also reduces to more familiar robustness notions when the regulator does not learn. Our notion of adaptive mechanisms lives in the rich middle ground between ex ante and ex post regulation \citep{laffont1993theory}---we use it to shed light on how policy might avoid pitfalls associated with ex ante regulation which fails to exploit interim information, and {ex post} regulation (e.g., liability regimes) which relies excessively on verification and/or tort law.\footnote{Our results also lend precision and guidance to a recent trend in environmental and AI policy advocating for `anticipatory' or `adaptive' regulation \citep{bennear2019adaptive,reuel2024generative,schrepel2025adaptive}.} 

Practically, our results speak directly to existing debates around how new and potentially transformative technologies should be regulated. The adaptive sandbox in our model qualitatively resembles real-world policies (`regulatory sandboxes') that have recently been introduced to regulate new financial technologies,\footnote{By 2020, regulators in over 50 jurisdictions have adopted some form of sandbox to regulate new financial technologies \citep{appaya2020global}. By 2023, approximately 100 sandbox initiatives have been implemented \citep{OECD2023sandbox}.} and are increasingly deployed to regulate artificial intelligence.\footnote{Article 57 of the EU AI Act \citep{EU2024AI} stipulates that member states must establish AI regulatory sandboxes by August 2026.} Despite their growing prevalence, surprisingly little is known---either in theory or practice---about these policies. We bridge this gap by making precise when and why these kinds of adaptive quantity limits work well and, in so doing, offer foundations for the oft-articulated goal of sandboxes as providing a ``dynamic, evidence-based approach to regulation'' \citep{OECD2023sandbox}.

\textbf{Outline of results.}  

\underline{Robustness.} We are interested in mechanisms that perform well even if we are prepared to assume little about what agents might learn, or what preferences they hold. \emph{Learning-robust} mechanisms deliver optimal worst-case guarantees across all learning processes the agent might face. \emph{Dually-robust} mechanisms deliver optimal worst-case guarantees across all learning processes and agent preferences. We show that an adaptive sandbox mechanism consisting of a zero marginal tax on the technology up to an evolving hard limit---beyond which further R\&D or deployment is disallowed---is both learning- and dually-robust. This is driven by two key properties:
\begin{itemize}[itemsep=1pt, topsep=1pt, leftmargin = 1.5em]
    \item \emph{Aligned sensitivity to information}: before the sandbox limit, the zero marginal tax minimally distorts the shape of relative incentives between firm and regulator. To show this, we develop new comparative static results relating risk-aversion to the quantity of experimentation under arbitrary learning processes. This implies that under any learning process, the sandbox positively correlates stopping incentives: if the firm wishes to stop pushing the technology, then so would the regulator \emph{if} they had access to the firm's private information.  
    \item \emph{Minimal sensitivity to information}: at the sandbox limit, firms are forced to ignore their interim information and stop pushing the technology. This safeguards the principal against learning processes we call \emph{weak optimism} that anti-correlate continuation incentives: the agent grows progressively more optimistic and thus wishes to continue development or deployment, but not quickly enough---if the regulator held the same belief they would prefer to stop. Hard limits safeguard against such learning processes. 
\end{itemize}

\underline{Undominated and dominant.} Why might we prefer sandboxes over other robust mechanisms (of which there are infinitely many)? We offer a simple but, we think, compelling reason: the adaptive sandbox is undominated, and dominates all other dually-robust mechanisms that satisfy a basic regularity property that rules out vacillating between taxing and rewarding pushing the technology further. That is, for any learning process and any agent preference---even those away from the worst case---the adaptive sandbox delivers higher regulator payoffs.\footnote{Weakly higher for every learning process and agent preference, and strictly higher for some learning process and agent preference.} 
Dominance is driven by a third key property: 

\begin{itemize}[itemsep=1pt, topsep=1pt, leftmargin = 1.5em]
    \item \emph{Maximal sensitivity to information}: across all mechanisms that induce aligned sensitivity (a necessary condition for robustness), our adaptive sandbox keeps the firm maximally sensitive to interim adverse information: under any learning process and any firm preference, sandboxes do better at positively correlating the events that the firm and regulator both wish to halt.
\end{itemize} 

\begin{figure}[h!]  
\centering
\captionsetup{width=1.0\linewidth}
    \caption{Outline of results} 
    \subfloat[Desiderata for mechanisms]{\includegraphics[width=0.5\textwidth]{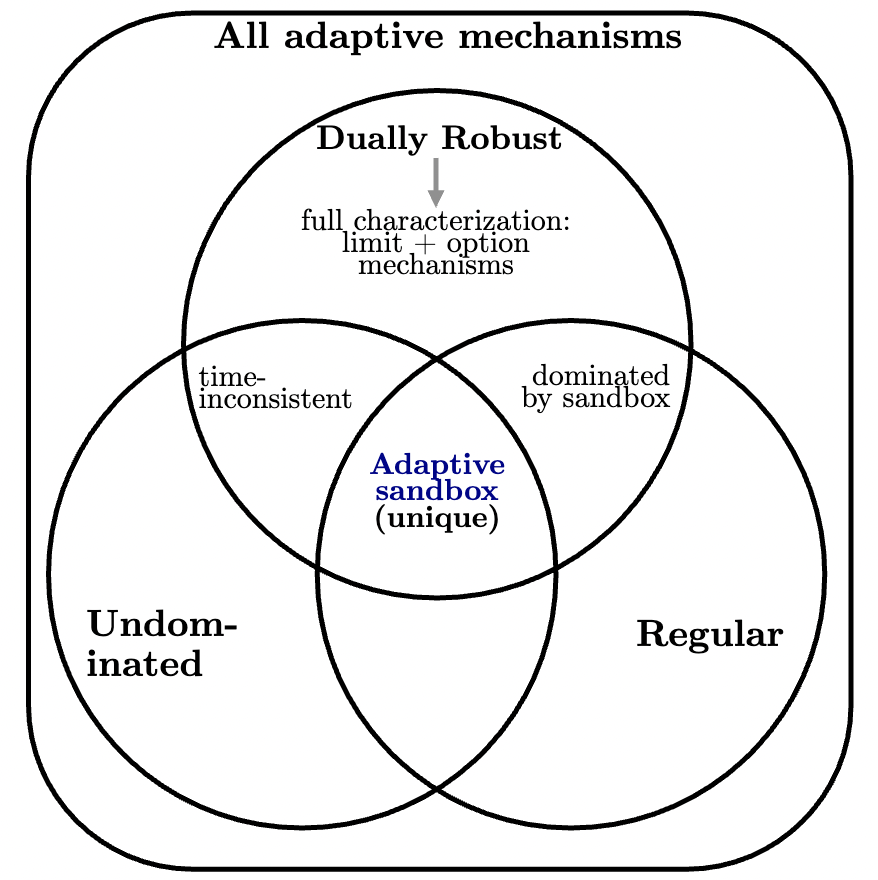}}
    \hspace{0.8em}
    \subfloat[Worst-case sans robustness]
    {\includegraphics[width=0.465\textwidth]{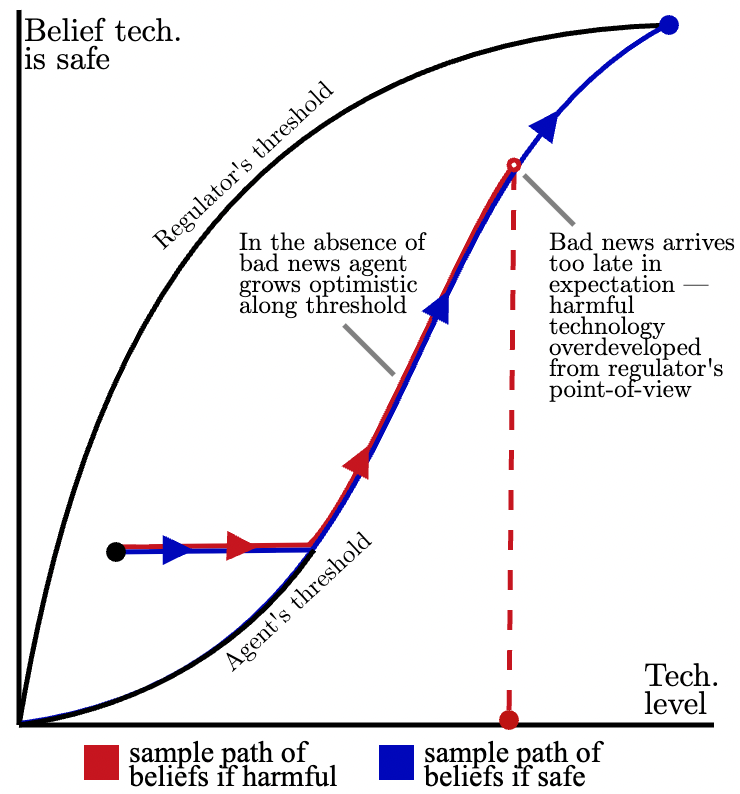}}
    \label{fig:outline}
\end{figure}

Taken together, robustness, dominance, and regularity are illustrated in \cref{fig:outline} (a) where each circle represents a distinct desideratum. The \textcolor{blue!70!black}{optimal adaptive sandbox} is the unique adaptive mechanism that fulfills all three properties. 

\underline{\smash{Time-consistency.}} Beyond evaluating mechanisms according to their ex ante performance (as does robustness and dominance), we might also be interested in whether our regulator has incentives to follow-through ex interim, at every possible history both on and off the equilibrium path of play.\footnote{Indeed, such problems of commitment loom large in monopoly or banking regulation \citep{laffont1993theory}.} The adaptive sandbox is time-consistent: whenever the agent halts or continues pushing the technology, whatever the regulator has learnt thus far, she has no incentive to revise the mechanism. Time-consistency is driven by a kind of positive selection induced by the sandbox's aligned sensitivity: if an agent stops, the regulator must infer that her own optimal technology level has already passed. This, in turn, prevents a Coase conjecture-like logic---in which the agent strategically halts to induce more lenient regulation---from taking hold. What is more, \emph{all} other dually-robust mechanisms are time-inconsistent---each of them presents the regulator's future selves with the temptation to deviate to a different mechanism such as to screen the agent's interim belief.  

\underline{\smash{The importance of robustness.}} We do not think robustness is always---or even often!---a good guide to policy. But we argue that it is especially \emph{important} in the context of regulating risky technologies: sans robust regulation, the worst-case learning process is both natural and delivers poor regulator payoffs. To make this precise, we fully characterize the form and value of the worst-case learning process under non-robust mechanisms. They takes the form of an inhomogeneous `bad news' Poisson process so that, in the absence of news that the technology is harmful, the agent becomes gradually more confident in its safety---but only at a rate that keeps her indifferent between continuing and halting the technology. We call such learning processes \emph{weak optimism}, as illustrated in \cref{fig:outline} (b). Weak optimism hurts welfare by maximizing the distribution of false positives i.e., the technology is overdeveloped or deployed when it is harmful (\textcolor{red!80!black}{red path}), while inducing no false negatives i.e., the technology is always maximized when it is beneficial (\textcolor{blue}{blue  path}). In the absence of robust regulation, even small wedges in risk aversion can be substantially magnified by weak optimism to induce unboundedly poor payoffs for the regulator.  Hard limits safeguard against exactly this. 

Weak optimism arises naturally whenever the harms of new technologies are lumpy and extreme (e.g., pandemic induced by biotechnology; market crash induced by new financial instruments; AI catastrophe from misalignment) since the absence of disaster induces growing optimism that the technology is safe. Silicon Valley assures us that the risks from new technologies are under control because \emph{they} learn as the technology is developed. Thus writes Sam Altman, CEO of OpenAI: ``the best way to make an AI system safe is by iteratively and gradually releasing it... learning from experience'' \citep{altman2025}.\footnote{Likewise argues Dario Amodei: ``We always monitor the models... so that we have a continuous process where we don't get taken by surprise.'' \citep{amodei2025CFR}} 
But our results imply that in the absence of stringent regulation, the possibility of learning per se should not assure \emph{us}---in fact, a little learning is dangerous, and can deliver worse outcomes than no learning at all. Indeed, a number of AI scientists pushing the technological frontier have raised concerns that powerful AI systems can be dangerous because of misalignment \citep{greenblatt2024alignment}, loss of human control  \citep{kulveit2025gradual}, or wider socioeconomic disruptions, 
but nonetheless persist with R\&D or deployment of these technologies---we think this is consistent with weak optimism. That the worst-case learning process is both qualitatively and quantitatively important suggest that our notions of robustness should be a first-order concern for policymakers.

\paragraph{Related literature.}  
Our work relates directly to the literature on how uncertain technologies should be regulated when the private sector is misaligned\footnote{See \cite{jones2016life,jones2024ai,jones2025much,aschenbrenner2024existential} for analysis of the welfare-maximizing solution without misaligned incentives. \cite{acemoglu2021harms} surveys potential socioeconomic harms posed by AI systems used in prediction; \cite{hadfield2025agents} survey risks distinctive to AI agents that can plan and execute complex tasks over long horizons; \cite{feyen2021fintech} surveys the promises and perils of new financial technologies.} and thus relate to \cite{acemoglu2024regulating,gans2025regulating,guerreiro2025regulating} who study regulation under a public and known learning process in which the agent does not have better information. Our work differs in several substantial ways. First, our agent has superior information that can be leveraged. Second, we seek optimal performance guarantees across all learning processes and/or preferences. Third, we analyze adaptive policies that evolve with the regulator's information. Our optimal policies thus differ substantially. Nonetheless, our characterization of the worst-case learning processes speaks directly to the policies proposed by these papers, and makes precise when and how price-instruments like Pigouvian taxes or coarse quantity-instruments like quotas perform poorly.

Our first desideratum is robustness and thus relates to the literature on robust mechanism design that seeks performance guarantees over information structures \citep{bergemann2005robust} or preferences \citep{armstrong1995delegating}. 
Work here has primarily focused on static mechanisms.\footnote{In the context of pricing and auctions, see e.g. \citep*{bergemann2015limits,bergemann2017first,du2018robust,brooks2023structure}. See \cite{frankel2014aligned} for a more recent analysis of robustness to preferences. Also related is \cite{guo2019robust} who study robust monopoly regulation and \cite{carroll2017robustness} who studies robust multidimensional screening.} But R\&D is a fundamentally dynamic process intertwined with learning---our regulator thus seeks payoff guarantees over all adaptive information structures more informative than the principal's own. In the special case with no principal learning, this is related to \cite{libgober2021informational} who analyze informationally robust dynamic pricing.\footnote{See also \cite*{li2022sequentially} who study dynamic pricing under robustness concerns without commitment; in our setting, the regulatory sandbox mechanism is time-consistent. \cite{hannan1957approximation,vovk1990aggregating} are early analyses on robustness in dynamic environments; \cite{chassang2013calibrated,penta2015robust} are important early work on dynamically robust mechanisms.} 
Even here, an important difference is their monopolist only seeks to maximize profits; by contrast, our principal is concerned with shaping the \emph{joint} distribution over equilibrium technology levels and the state. Thus, our optimal mechanisms and their underlying economic forces differ substantially. With principal learning, our robustness notion is conceptually new because optimal mechanisms must `learn under ambiguity' \citep{epstein2007learning}, but retains the spirit of informational robustness. 

Our second desideratum is dominance which is in the tradition of \cite{wald1950,savage1972foundations}. We show the optimal adaptive sandbox is undominated in the sense of \cite*{borgers2024undominated}. Undominated does not imply robust, nor does robust imply undominated \citep{borgers2017no}. We show, however, that dominance and robustness can be combined to obtain sharp prescriptions: our adaptive sandbox is the unique mechanism that is robust, undominated, and regular. In particular, it dominates all other robust and regular mechanisms and, in this regard, is in similar spirit to work by \cite{guo2023regret,dworczak2022preparing} who rank robust policies away from the worst-case in context of static project choice and persuasion. Our dominance notion is demanding, and requires us to evaluate the performance of mechanisms under \emph{any} learning process and agent preference. To do so, we establish new comparative static results relating risk-aversion to optimal experimentation that hold for general payoffs and learning processes---including those that are non-continuous and non-Markov---thereby nesting, unifying, and generalizing past work \citep{chancelier2009risk,keller2019note}.

We also relate to the literature on instrument choice pioneered by \cite{weitzman1974prices} though our (i) policy is fully flexible and can adapt to the regulator's evolving information; and (ii) the agent's experimentation problem is dynamic. When the regulator does not learn, our sandbox has a fixed boundary and thus resembles the static `interval delegation' solutions that often arise in delegation problems \citep*{holmstrom1978incentives,amador2006commitment} though the forces driving our results are different. In our setting, the zero marginal tax ensures that any deviation by nature to a different learning process must improve the principal's payoff, while the hard limit is necessitated by robustness concerns. Our results additionally offer a rationale for `delegation-like' instruments to emerge endogenously under concerns for \emph{both} robustness and dominance.\footnote{If the principal wanted a robust \emph{or} undominated policy, not all of them are `delegation-like'.}

Finally, our results relate to work on optimal experimentation. Much of this literature has assumed a particular exogeneous learning processes for tractability e.g., conclusive good news \citep*{keller2005strategic} or bad news \citep*{keller2015breakdowns}; see, e.g., \cite{guo2016dynamic}. By contrast, our agent's learning process is private and chosen adversarially and flexibly. Normatively, this allows us to encode non-Bayesian uncertainty about the learning process. Economically, this allows us to identify the form of worst-case learning processes that robust mechanisms safeguard against. To do so, we recast nature's problem as one of dynamic information design and have found it helpful to draw on techniques from some of our earlier work which analyzed these problems \citep*{koh2022attention,koh2024persuasion}.\footnote{The literature on dynamic information design in stopping problems has focused on a designer with preferences over stopping levels \citep*{ely2020moving,orlov2020persuading,koh2022attention,saeedi2024getting}, or over both times and actions \citep*{koh2024persuasion}.}  However, the present setting differs in meaningful ways because our fictitious designer (nature) has state-dependent preferences which is novel to the literature. Our outer problem of optimizing dynamic mechanisms is related to the elegant work of \cite{kruse2015optimal,kruse2019inverse} who study the `inverse stopping problem' of designing a boundary to implement a target stopping level under a known diffusion (Markov with continuous sample paths). A key difference is that our regulator evaluates policies according to the worst-case over all learning processes, including those that are non-Markov and non-continuous.

\section{Model}\label{sec:model}

\subsection{Technology and uncertainty}
The \emph{technology state} takes values in $\Theta = \{0,1\}$  where $\theta = 1$ corresponds to the technology being beneficial and $\theta = 0$ corresponds to being harmful.\footnote{All results in this paper extends to any set of ordered states with the exception of \cref{thrm:worstcase} on the form of the worst-case learning process that needs to be modified.} The \emph{technology space} is a compact set $\mathcal{L} \subseteq \mathbb{R}$ that is either discrete or continuous and we write $L := \max \mathcal{L}$. Depending on the specific technology in question, the level $l \in \mathcal{L}$ might correspond to the degree of R\&D, extent of deployment, and so on. Importantly, we assume that technology is irreversible.\footnote{See, e.g., \cite{weitzman1998recombinant} for an influential model of growth building on this idea that once we have learnt how to build something e.g., a bomb, synthetic pathogen, etc. we cannot unlearn it. Technology might also be locked-in because of networked adoption, behavioral biases, decentralization (as with the release of AI model weights), or even because the technology itself resists displacement \citep{russell2019human}.}  

\subsection{Learning} There is a \emph{principal} representing a regulator, and an \emph{agent} representing a firm developing the technology. They share a common prior $\mu_{0-} \in \Delta(\Theta) := [0,1]$, where we use the notation `$0-$' to denote the ex ante level such as to accommodate changes in beliefs exactly at level $0$. Both principal and agent learn about $\theta$ as the technology is developed. 

The principal's learning process is represented by a right-continuous filtration $\mathcal{G} := (\mathcal{G}_l)_{l \in \mathcal{L}}$ generated by some right-continuous signal process $(X_l)_{l \in \mathcal{L}}$ with measure $\mathbb{P}^{\theta}$.\footnote{That is, $\mathbb{P}^{\theta}$ is the probability measure of the signal process over the Skorokhod space $\mathbb{D}(\mathcal{L};\mathbb{R}^d)$ when the technology state is $\theta$. There is an appropriately filtered probability space in the background that we suppress in the main text to focus on the economics; details are in \cref{app:proofs}.} 
We assume the following technical conditions that will simplify our exposition:
\begin{itemize}
    \item[(i)] $\Pr^{0}$ and $\Pr^1$ are mutually absolutely continuous; and
    \item[(ii)] if $\mathbb{P}^1 \neq \mathbb{P}^0$ then for any $M,\epsilon > 0$,  $\mathbb{P}\left(\log \frac{d\mathbb{P}^1}{d\mathbb{P}^0}\big|_{\mathcal{G}_{l+\epsilon}} < -M \, \big| \,\mathcal{G}_l \right) > 0$.
\end{itemize}
These conditions are fulfilled whenever $(X_l)_{l \in \mathcal{L}}$ is a jump diffusion (e.g., a L\`evy process with mutually absolutely continuous diffusion coefficients and jump measures) though the principal's learning might be more generally path-dependent, containing information both about state $\theta$ as well as about the informativeness of the future signal process. The agent learns (weakly) more: her learning process is represented by the filtration $(\mathcal{F}_l)_l$ that is finer than the principal's: 
\[
\mathcal{G}_l \subseteq \mathcal{F}_l \text{ for all $l \in \mathcal{L}$.
} \tag{R}
\label{eqn:refinement}
\]
The set of possible learning processes for the agent is given by
\[
\mathbb{F} := \Big\{(\mathcal{F}_l)_l: \text{$\mathcal{F}$ is a right-continuous filtration fulfilling \eqref{eqn:refinement}}
\Big\}
\]
and we use $\mathsf{\mu}_l := \Ex[\theta \mid \mathcal{F}_l]$ to denote the agent's beliefs about the state. 
The agent's beliefs  $(\mu_l)_l$ evolve (i) as a martingale by Bayes' rule; and (ii) right-continuously since $(\mathcal{F}_l)_l$ is right continuous. 

We emphasize that the set of agent learning processes $\mathbb{F}$ is rich, capturing not simply how the agent's information about the state evolves, but also about the informativeness of the agent's and principal's future learning process. This allows us to capture various ways in which the private sector might be more informed, as the following examples illustrate: 
\begin{enumerate}
    \item \textbf{No principal learning.} $\mathcal{G}_l = \mathcal{G}_{0-}$ a.s. We have already noted that the agent's beliefs $\mu_l = \Ex[\theta \mid \mathcal{F}_l]$ evolves as a c\`adl\`ag martingale. Moreover, any c\`adl\`ag martingale---potentially non-Markov or non-continuous---can be induced by some learning process $\mathcal{F} \in \mathbb{F}$. 
    \item \textbf{Independent extra information.} The agent observes the principal's signal process $(X_l)_l$ and, in addition, observes an additional conditionally independent signal process $(Y_l)_l$. 
    
    \item \textbf{Correlated extra information.} The agent observes the principal's signal process $(X_l)_l$ and, in addition, observes an additional signal process $(Y_l)_l$ with law adapted to the natural filtration $\mathcal{G}$ induced by $(X_l)_l$. 
    \item \textbf{Learning with a lag.} The agent observes the signal process $(X_l)_l$ and the principal observes the process $(Y_l)_l$ which is simply the agent's information but with a lag of $\gamma \geq 0$: $Y_{l + \gamma \wedge L} = X_{l}$.
\end{enumerate}

\subsection{Preferences} If the agent develops the technology up to level $l$ at state $\theta$, the agent's utility is $u(\theta,l)$ and the principal's utility is $v(\theta,l)$.\footnote{We extend the second argument of $v$  from $\mathcal{G}$ to the positive reals, that is, $v: \Theta \times \mathbb{R}_{\geq 0} \to \mathbb{R}$ which is without loss since $\mathcal{L} \subseteq \mathbb{R}_{\geq 0}$.} It will be helpful to work with indirect utilities by defining 
\[
U(\mu,l) := \Ex_{\theta \sim \mu} \big[u(\theta,l)\big] \quad \text{and} \quad V(\mu,l) := \Ex_{\theta \sim \mu} \big[v(\theta,l)\big].  
\]
Increasing the technology level is beneficial if and only if the state is good i.e., $v(1,\cdot)$ is strictly increasing while $v(0,\cdot)$ is strictly decreasing. We assume $v(1,\cdot)$ and $v(0,\cdot)$ are continuously differentiable and normalize payoffs from zero technology to zero so $V(\mu,0) = 0$. 

\begin{figure}[H]  
\centering
\captionsetup{width=0.9\linewidth}
    \caption{Illustration of assumptions on $V$ and $U$} 
    \subfloat[Increasing in optimism]{\includegraphics[width=0.48\textwidth]{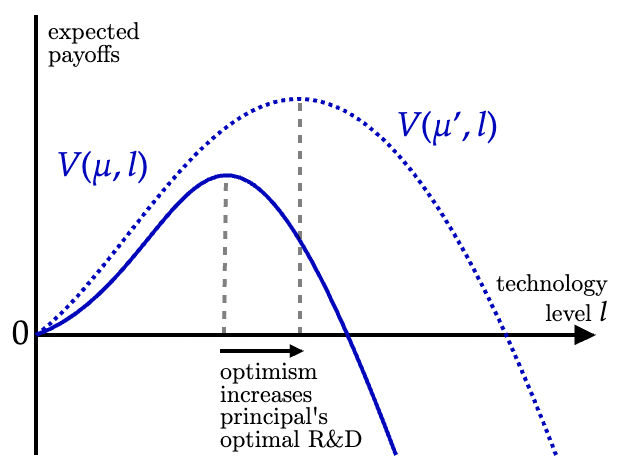}}
    \subfloat[Principal risk aversion]
    {\includegraphics[width=0.48\textwidth]{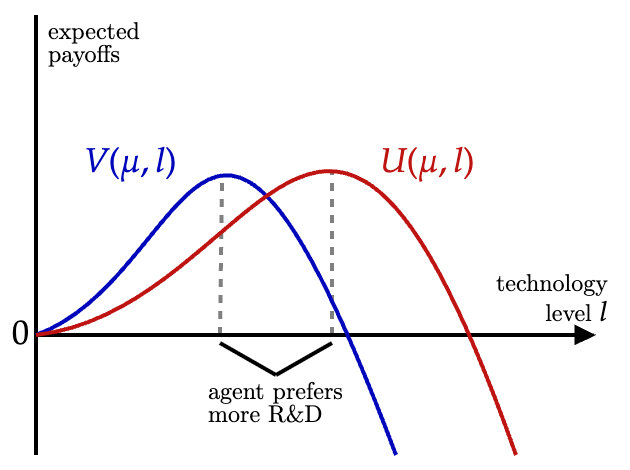}}
    \label{fig:assumption_illust}
\end{figure}

These assumptions imply that the principal prefers a higher technology level when she is more optimistic\footnote{See \cite{milgrom1994monotonicity}.} as illustrated in \cref{fig:assumption_illust} (a) where the principal's indirect utility at belief $\mu \in (0,1)$ is given by the solid blue line, and the same for a more optimistic belief $\mu' > \mu$ is given by the dotted blue line.\footnote{In \cref{fig:assumption_illust} (a) $V(\mu,\cdot)$ is depicted as quasiconcave but this is not assumed.}

We maintain the following assumption on the agent's preference $U$: 
\begin{assumption}[Principal is more risk averse]  \label{assumption:reg} There exists an increasing convex function $g: \mathbb{R} \to \mathbb{R}$ such that $u(\theta,l) = g\circ v(\theta,l)$.
\end{assumption}

Let $\mathbb{U}$ denote the set of agent preferences which fulfill \cref{assumption:reg}. An implication of \cref{assumption:reg} is that for each belief $\mu$, the principal's static choice of technology level (sans further information) is weakly lower than that of the agent's. This is illustrated in \cref{fig:assumption_illust} (b) where, at the belief $\mu$, the location of the peak of the agent's indirect utility is further out than that of the principal's. 

\cref{assumption:reg} might capture intrinsic differences in risk preferences, but also arises naturally in the presence of externalities e.g., if the costs of developing or deploying the uncertain technology are only partially borne, if the benefits from the technology are disproportionately enjoyed by the firm, or if multiple firms are competing in an `R\&D arms race' so that the ensuing equilibrium is as if a representative firm is more risk-seeking. We illustrate these via a range of concrete examples in Online Appendix \ref{app:examples}.\footnote{We develop an example of an economy in which consumption depends on the technology state $\theta \in \Theta$ and level $l \in \mathcal{L}$ and, in so doing, relate this to \cite{jones2024ai}.}

\subsection{Adaptive mechanisms}  Let $\overline{\mathbb{R}} := \mathbb{R} \cup \{-\infty,+\infty\}$ denote the extended reals. 
 An adaptive mechanism is an $\overline{\mathbb{R}}$-valued stochastic process $(\phi_l)_{l \in \mathcal{L}}$ adapted to the principal's filtration $\mathcal{G}$ such that, if the agent stops at technology level $l$, the regulator imposes a tax of $\phi_l$. The set of adaptive mechanisms is thus

\[
\Phi := \left\{
(\phi_l)_{l \in \mathcal{L}}
: \, \begin{aligned}
&\text{(i) $(\mathcal{G}_l)_l$-adapted;} \\[-0.5ex]
&\text{(ii) pathwise lower semi-continuous;} \\[-0.5ex] 
&\text{(iii) class (D) on $\big\{|\phi_l| < +\infty\big\}$.\footnotemark}
\end{aligned} \right\}.
\]
\footnotetext{Recall that a stochastic process $(X_l)_{l}$ is class (D) if the set $\{X_{\ell}: \text{$\ell$ is a $\mathcal{G}$-adapted stopping level}\}$ is uniformly integrable. Class (D) processes form the smallest class that ensures an optimal stopping level for the agent facing any mechanism $\phi \in \Phi$ exists.}

Condition (i) states that the mechanism cannot `look into the future'---it can only condition on present information; conditions (ii) and (iii) are mild technical conditions that guarantee the existence of solutions to the agent's stopping problem. Stopping the technology at level $l$ under belief $\mu$ yields the expected payoffs: 
\[
    \underbrace{U^{\phi}(\mu,l) := U(\mu,l) - \phi_l}_{\substack{\text{Agent's} \\ \text{payoff}}} \quad \text{and} \quad \underbrace{V(\mu,l).}_{\substack{\text{Principal's} \\\text{payoff}}} 
\]
Note that $\phi_l$ does not feature in the principal's payoff because we are interested in efficiency: if $\phi$ is a tax, this might be paired with a lump sum rebate; alternatively, $\phi$ might reflect non-monetary instruments (e.g., legal, administrative, or regulatory barriers).\footnote{Our results can be modified to accommodate revenue motives for the regulator (see \cref{prop:revenue} in Online Appendix \ref{app:ext}). Neither additional randomization nor elicitation of type reports can improve upon the principal's payoff gaurantee (see \cref{prop:novalue} in Online Appendix \ref{app:ext}).} By letting adaptive mechanisms take values in the extended reals, we capture both `price instruments' $(\phi \in \mathbb{R})$, `quantity' instruments' ($\phi \in \{-\infty, +\infty\}$), and mixtures thereof.\footnote{Such `mixed' instruments are in the spirit of the closing remarks of \cite{weitzman1974prices}.} 
Say mechanisms $\phi, \phi'$ are \emph{equivalent} if $\phi = \phi' + \alpha$ for some $\mathcal{G}_{0-}$-measurable random variable $\alpha$. Equivalent mechanisms induce identical incentives for the agent (conditional on participation). 

Our framework offers a unified umbrella for analyzing regulatory timing \citep{laffont1993theory}. Adaptive mechanisms live in the rich middle ground between ex ante regulation that fails to make use of interim information, and ex post regulation that is often infeasible.\footnote{ex post regulation achieves efficiency by conditioning on the true state e.g., via liability regimes \citep{shavell1984liability} which has been proposed for regulating AI \citep{gans2025regulating}. Among the standard objections to ex post regulation e.g., the weakness of tort law, limited liability, etc., we highlight one that is distinctive to technology regulation: state-contingent regulation is feasible only if it induces sufficient technology levels for the regulator to learn in every state. But this is self-defeating since the point of regulation is to shape the joint distribution over technology levels and states.} \cref{table:reg_timing} illustrates.
 \begin{table}[h]
 \centering
 \hspace{2em}
\begin{tabular}{|c|c|c|}
\multicolumn{1}{c}{ex ante} & \multicolumn{1}{c}{ex interim (`adaptive')} & \multicolumn{1}{c}{ex post} \\
\hline
\parbox{3cm}{\centering \vspace{0.2em} Policy is\\$\mathcal{G}_{0-}$-adapted} & \parbox{3cm}{\centering \vspace{0.2em} Policy is\\$(\mathcal{G}_l)_l$-adapted} & \parbox{3cm}{\centering \vspace{0.2em} Policy is\\$\mathcal{G}_{\infty}$-adapted} \\
\hline
Feasible & Feasible & Infeasible \\
\hline
Suboptimal & Second-best & First-best\\
\hline
\end{tabular}
\caption{Comparison of regulatory timing} \label{table:reg_timing} 
\end{table}

\subsection{Agent's problem} 
\begin{figure}[H]  
\centering
\vspace{-0.5em}
\captionsetup{width=1.0\linewidth}
    \caption{How policy, learning, and preferences shape outcomes} 
    {\includegraphics[width=0.68\textwidth]{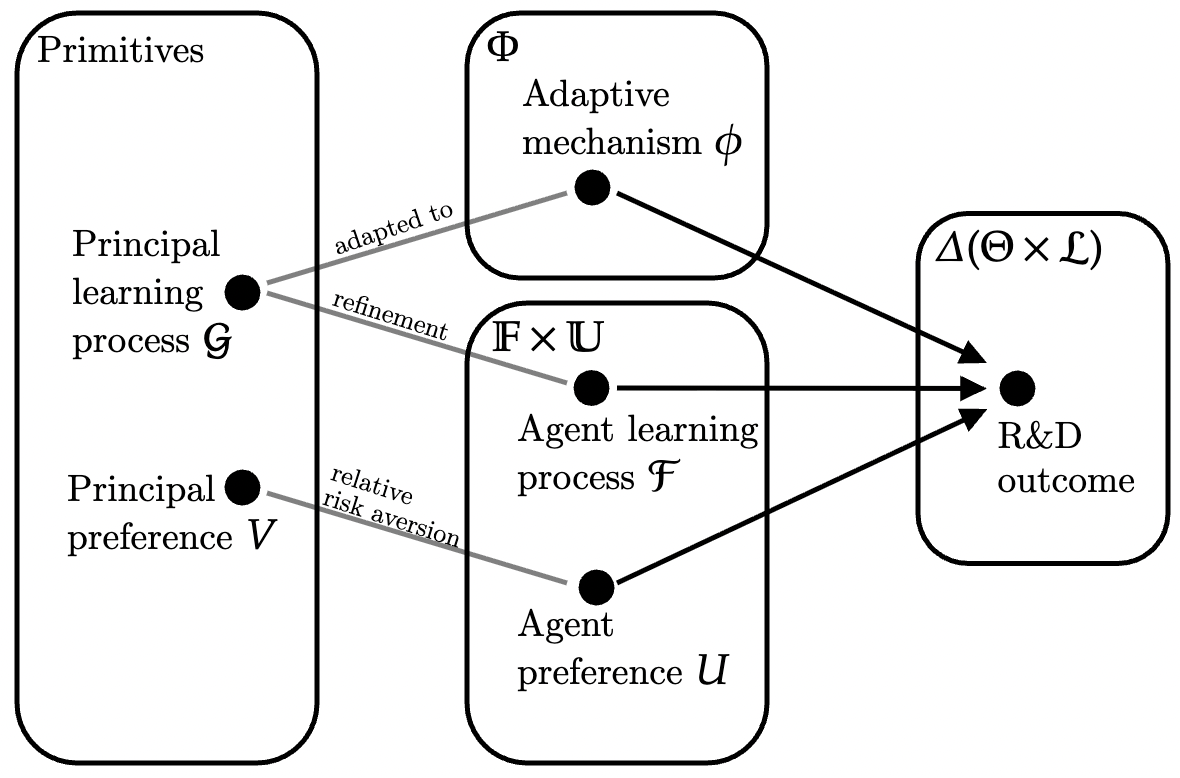}}
    \label{fig:adaptive_outline}
\end{figure}

The agent with preferences $U$, facing mechanism $\phi$ and learning process $\mathcal{F}$, first decides whether or not to participate in the mechanism. If she does not participate, the technology is not developed and payoffs are zero for principal and agent. If she does, the agent solves the following optimal stopping problem: 
\[
     \sup_{\el} \, \Ex\Big[U(\mu_{\el},\el) - \phi_{\el}\Big]
     \quad \text{ s.t. $\el$ is $\mathcal{F}$-adapted.} 
     \tag{O}\label{prob:OSP}
\]
where the agent's technology choices depend jointly on the adaptive mechanism $\phi$, the learning process $\mathcal{F}$, and her preferences $U$. Let $\ell^*(\phi,\mathcal{F},U)$ denote the largest stopping level that solves \eqref{prob:OSP}.\footnote{Existence of solutions to \eqref{prob:OSP} follows from (i) pathwise lower semi-continuity of $\phi$; (ii) uniform integrability of the stopped process $(\phi_l)_l$; and (iii) compactness of the technology space $\mathcal{L}$. Among these optimizers, a largest one exists because $U - \phi$ is upper-semicontinuous in expectation along stopping times \citep{kobylanski2012optimal,el1979aspects}; see Online Appendix \ref{appendix:existence_selection} for a formal treatment.}
All our results hold for a more general selections of optimal stopping levels; we discuss this in Online Appendix \ref{appendix:existence_selection}. \cref{fig:adaptive_outline} takes stock of how different elements of our model interact to jointly shape technology outcomes.  

\subsection{Desiderata} We outline three desiderata for adaptive mechanisms.

\textbf{Robustness.} Suppose our principal evaluates mechanisms according to their payoff guarantee. If the agent's preference $U$ is known but the principal is uncertain about what the agent might learn in the interim, she faces the following learning-robustness problem: 
\[
\sup_{\phi \in \Phi} \, \inf_{\mathcal{F} \in \mathbb{F}} 
\Ex\Big[ V\big(\mu_{\el^*}, \el^*(\phi, \mathcal{F}, U)\big)
\Big] 
\tag{L} \label{eqn:ADV}
\]
Call mechanisms that achieve \eqref{eqn:ADV} \emph{learning-robust}. 

If the principal is uncertain both about the agent's preferences ($U$) and learning process $(\mathcal{F})$, she faces the dual-robustness problem: 
\[
\sup_{\phi \in \Phi} \, \inf_{\substack{{\mathcal{F} \in \mathbb{F}}\\
{U \in \mathbb{U}}}} 
\Ex\Big[ V\big(\mu_{\el^*}, \el^*(\phi, \mathcal{F}, U)\big)
\Big] 
\tag{D} \label{eqn:DR}
\]
Call mechanisms that achieve \eqref{eqn:DR} \emph{dually-robust}.

Our robustness notions recover, as special cases, some of those studied in the literature. When the principal does not learn (i.e., $\mathcal{G}_l = \mathcal{G}_{0-}$ a.s.) and knows the agent's preference but not her learning process, learning-robustness coincides with informational robustness for static \citep{bergemann2005robust} and dynamic \citep{libgober2021informational} mechanisms. When the principal does not learn but is certain that the agent knows the state (i.e., $\theta$ is $\mathcal{F}_0$-measurable), robustness to the agent's preferences is similar to that studied by \citep{armstrong1995delegating,frankel2014aligned}. We think learning- and dual-robustness are apt for new technologies rife with substantial disagreement, not just about the state ($\theta \in \Theta$), but also about what might be learnt over the course more R\&D or deployment of the technology $(\mathcal{F} \in \mathbb{F})$\footnote{For instance, computer scientists disagree about how much we can learn about the safety of AI from empirical testing (see, e.g., \cite{cohen2024regulating} and \cite{anthropic2025}).} and what preference the private sector has $(U \in \mathbb{U})$. Conceptually, our robustness notion is novel and requires mechanisms to learn under ambiguity, continually adapting to interim information while safeguarding against the worst-case learning-process and/or preference.

\paragraph{Dominance and undominated.} We will be interested in assessing mechanisms not just according to their payoff guarantee, but also evaluated away from the worst case. For two different adaptive mechanisms $\phi, \phi' \in \Phi$, 
say $\phi$ \emph{dominates} $\phi'$ if, for any learning process $\mathcal{F} \in \mathbb{F}$ and any agent preference $U \in \mathbb{U}$,  
\[
\Ex \Big[ V \big(\mu_{\el^*},\el^*(\phi,\mathcal{F},U)\big) \Big] \geq 
\Ex \Big[ V \big(\mu_{\el^*},\el^*(\phi',\mathcal{F},U)\big) \Big] 
\]
with strict inequality for some $(\mathcal{F},U) \in \mathbb{F} \times \mathbb{U}$. Our dominance criteria allows the principal to entertain the possibility that the agent's learning process and/or preferences are not adversarially chosen. This in similar spirit to work by \cite{wald1950,savage1972foundations}. The mechanism $\phi$ is \emph{undominated} if it is not dominated by another adaptive mechanism. This notion has been recently analyzed by \cite*{borgers2024undominated} in the context of static mechanisms.\footnote{Moreover, if we require the strict inequality to be over a positive Lebesgue measure set, this has been recently defined as `strong dominance' by \cite*{borgers2024undominated}. While there is no analog of `positive Lebesgue measure' on the space of learning processes,  all examples of dominance in this paper are `strong dominance' with respect to the Wiener measure on the space of c\`adl\`ag paths of the agent's `first-order' belief $\Ex[\theta \mid \mathcal{F}_l]$.}

\textbf{Time-consistency.} A final desideratum is for the regulator to have incentives to follow-through with the mechanism, both on and off the equilibrium path. This matters for feasibility rather than optimality---time consistency determines whether a mechanism can be credibly run. Indeed, such problems looms large in financial regulation \citep{brunnermeier2009fundamental} so we think it is worth analyzing whether the same is true of technology regulation. We defer our definition and analysis of time-consistency to \cref{subsec:time_con}.

\section{Robust, dominant, and time-consistent regulation}\label{sec:robustness}

We begin by defining a simple family of adaptive mechanisms.

\begin{definition}[Adaptive sandboxes]
    $\phi \in \Phi$ is an adaptive sandbox induced by the $\mathcal{G}$-adapted stopping level $\ell$ if:
    \[
    \phi_s = 
        \begin{cases}
            0 \quad &\text{if $s \leq \ell$} \\
            +\infty \quad &\text{if $s > \ell$.} 
        \end{cases}
    \]
\end{definition} 
Adaptive sandboxes prescribe a zero marginal tax on the technology until the stopping level $\el$, beyond which the agent is not allowed to push the technology further. Sandboxes are adaptive mechanisms since the location of the limit $\ell$ is itself a $\mathcal{G}$-adapted stopping level.

\begin{figure}[H]
\begin{minipage}[t]{0.45\linewidth}
Let $\phi^*$ be the adaptive sandbox induced by stopping level $\overline{\el}$ solving
\begin{align*}
\sup_{\el}\, &\Ex\big[V(\mu_\el,\el)\big]     \\
&\text{ s.t. $\el$ is $\mathcal{G}$-adapted.}
\end{align*}
$\phi^*$ is illustrated in \cref{fig:mechanism_illust}. To ease exposition, we will assume $\overline{\el}$ is the unique solution to the above (which holds generically).
\end{minipage}%
\hfill%
\begin{minipage}[t]{0.55\textwidth}
\vspace{-1em}
\hfill 
{\includegraphics[width=0.95\textwidth]{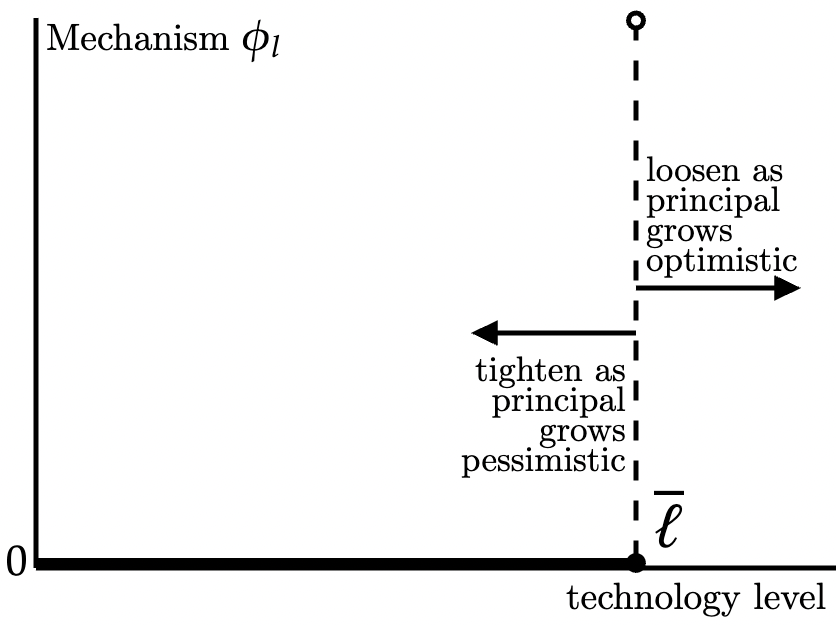}}
    \vspace{-1em}
    \caption{Adaptive sandbox}\label{fig:mechanism_illust}
\end{minipage}
\end{figure}


\subsection{Robustness}  
\begin{theorem}[Robustness]\label{thrm:robustness}
The adaptive sandbox $\phi^*$ is learning-robust and dually-robust i.e., it solves \eqref{eqn:ADV} and \eqref{eqn:DR}. 
\end{theorem}

\cref{thrm:robustness} establishes the learning- and dual-robustness of $\phi^*$. The proof is quite simple and instructive so we will sketch the main steps here. 

\underline{\smash{Step 1: Robustness problem upper-bounded by direct control.}} A basic observation is that the value of learning-robust problem is upper-bounded by `direct control' in which the principal chooses the technology level directly: 
\begin{align*}
\sup_{\phi \in \Phi} \, \inf_{\mathcal{F} \in \mathbb{F}} 
\Ex\Big[ V\big(\mu_{\el^*}, \el^*(\phi, \mathcal{F}, U)\big)
\Big] 
&\leq 
\sup_{\phi \in \Phi} \, 
\Ex\Big[ V\big(\mu_{\el^*}, \el^*(\phi, \mathcal{G}, U)\big)
\Big] 
\\
&\leq \begin{array}[t]{l}
     \sup_{s} \Ex\big[V(\mu_s,s) \big]  
     \\
    \text{s.t. $s$ is $\mathcal{G}$-adapted}
    \end{array} 
\end{align*}
where the first inequality is because $\mathcal{G} \in \mathbb{F}$ is \emph{a} possible learning process, and the second is because the agent's technology choice $\el^*$ is \emph{a} $\mathcal{G}$-adapted stopping level so the principal might just as well control experimentation directly. It suffices to show that our adaptive sandbox $\phi^*$ achieves this upper-bound over all learning processes $\mathcal{F} \in \mathbb{F}$.  

\underline{\smash{Step 2: $\phi^*$ achieves the upper-bound.}} First and most immediately, the adaptive sandbox $\phi^*$ delivers a \emph{participation guarantee}: pushing the technology up to the sandbox limit---always a feasible strategy under any $\mathcal{F} \in \mathbb{F}$---is always weakly better than not participating. Next, we develop a simple but powerful comparative static result relating risk aversion to optimal experimentation: 
\begin{lemma}[Experimentation comparative static] \label{lem:compstat}
    Fix any learning and preference pair $(\mathcal{F},U) \in \mathbb{F} \times \mathbb{U}$ and any pair of $\mathcal{F}$-stopping levels $\el_1 \leq\el_2$. The set of $\mathcal{F}$-stopping levels that solve $\sup_{\el \in [\el_1,\el_2]} \Ex[V(\mu_\el,\el)]$ is dominated by the set of $\mathcal{F}$-stopping levels that solve $\sup_{\el \in [\el_1,\el_2]} \Ex[U(\mu_{\el},\el)]$ in the strong set order.\footnote{Recall that for a lattice $L$ and any $A, B \subseteq L$, $A$ dominates $B$ in the strong set order if for any $a \in A$ and $b \in B$, $a \vee b \in A$ and $a \wedge b \in B$. } 
\end{lemma}

\cref{lem:compstat} nests and unifies previous work analyzing how different kinds of risk-aversion can either prolong or shorten experimentation \citep{keller2019note,chancelier2009risk}. \cref{lem:compstat} reconciles these results, and holds path-by-path for any learning process, including those that induce non-Markov and/or non-continuous beliefs, without requiring explicit solutions to the optimal stopping problem.\footnote{The connection between risk-aversion, patience, and stopping has a long history in economics \citep{keller2019note,chancelier2009risk,quah2013discounting}. In Online Appendix \ref{app:experimentation discussion} we discuss how \cref{lem:compstat} fully or partially generalizes these results, and allows the analyst to make predictions about the duration of experimentation when risk- and time-preferences are both changing.}

Now suppose that under the sandbox $\phi^*$, the agent stops strictly before the sandbox limit i.e., $\el^* < \overline{\el}$. Since the adaptive sandbox imposes a zero-marginal tax, in the absence of any regulation the agent would continue to stop at $\el^*$ facing the truncated technology space $[0,\overline{\el}]$. By \cref{lem:compstat}, this means that if the principal had access to the agent's information, she would also prefer to stop at $\mathcal{F}_{\el^*}$. 

\begin{figure}[H]
\begin{minipage}[t]{0.45\linewidth}  
This \emph{aligned sensitivity} is depicted in \cref{fig:aligned_illust} where each of the \textcolor{red!80!black}{red paths} depict sample paths of the agent's stopped beliefs. By minimally distorting the shape of relative incentives, the adaptive sandbox generates positive correlation between the events that both principal and agent prefer to stop, thereby guaranteeing that any deviation by nature $\mathcal{G} \to \mathcal{F}$ must improve the principal's payoff. 
\end{minipage}%
\hfill%
\begin{minipage}[t]{0.55\textwidth}\vspace{0pt}
\vspace{-1em}
\centering
{\includegraphics[width=0.9\textwidth]{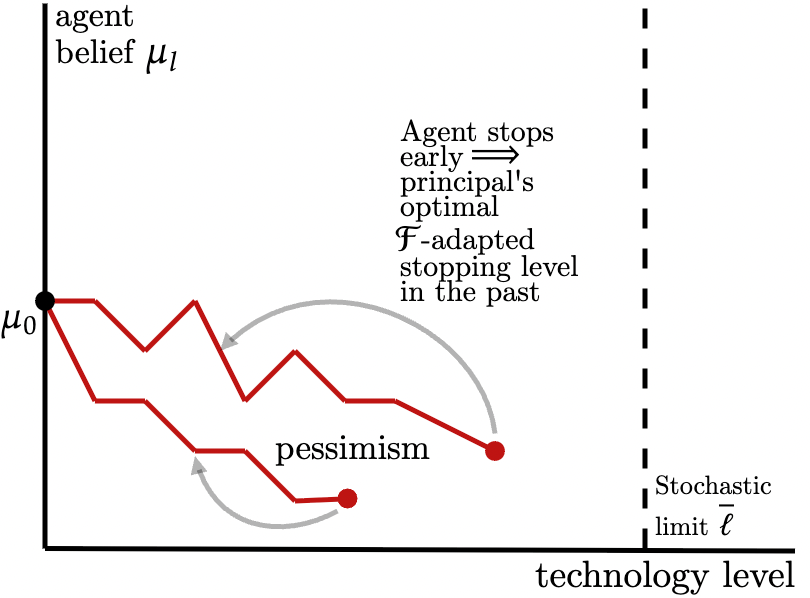}}
    \caption{\small Aligned sensitivity before limit}\label{fig:aligned_illust}
\end{minipage}
\end{figure}

Finally suppose that under the sandbox $\phi^*$, the agent stops at the sandbox limit i.e., $\el^* = \overline{\el}$. But the principal's payoff from such paths coincide with that of `direct experimentation'. Then partitioning up the set of paths into those in which the agent either stops early $(\el^* < \overline{\el})$ or pushes until the sandbox limit $(\el^* = \overline{\el})$ and taking total expectation implies that the principal's payoff under the adaptive sandbox $\phi^*$ and over all learning process $\mathcal{F} \in \mathbb{F}$ achieves the upper-bound constructed in Step 1, yielding learning-robustness. Dual-robustness follows by noticing that the adaptive sandbox $\phi^*$ does not depend on the agent's preference $U$. $\hfill \diamondsuit$

\begin{figure}[H]
\begin{minipage}[t]{0.45\linewidth} 
The hard limit induces \emph{minimal sensitivity} to information---whatever the agent's interim beliefs, she is not allowed to push past it. Although this is stringent, there is a sense in which this is exactly what robustness demands. Consider, for instance, the realization of the \textcolor{blue}{blue path} (`strong optimism') in \cref{fig:minimal_illust} in which the agent has privately become pretty optimistic and the principal, if she knew this, would like to slacken the sandbox limit. 
\end{minipage}%
\hfill%
\begin{minipage}[t]{0.55\textwidth}\vspace{0pt}
\vspace{0em}
\centering
{\includegraphics[width=0.9\textwidth]{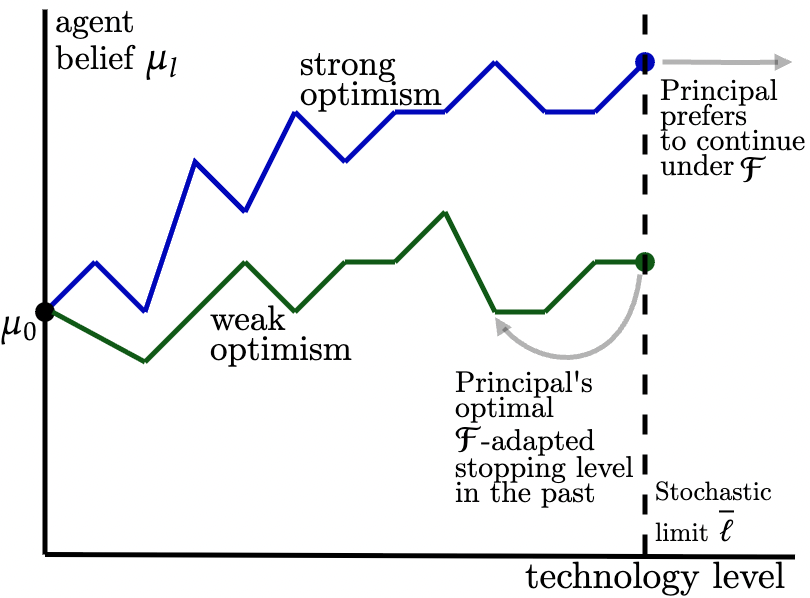}}
    \caption{\small Minimal sensitivity at limit}\label{fig:minimal_illust}
\end{minipage}
\end{figure}
Under the \textcolor{green!40!black}{green path} (`weak optimism'), however, the principal prefers stopping although the agent is willing to take on more risk.  It is precisely because the agent's learning process is adversarially chosen that adaptive mechanisms cannot distinguish between these kinds of optimism to guarantee positive correlation between the events that both agent and principal prefer a higher technology level. 
Indeed, we will see in \cref{sec:importance} that in the absence of robust regulation, the worst-case learning process negatively correlates these events by inducing a specific kind of weak optimism that maximizes risk-taking---hard limits safeguard against exactly this.

The adaptive sandbox $\phi^*$ is relatively simple. This echoes a broader theme from the literature on robust mechanism design \citep{bergemann2005robust,carroll2019robustness} that robustness is particularly appealing when it prescribes mechanisms that are simple, intuitive, and mirror those used in practice. By contrast, the Bayesian problem is intractable and sensitive to the prior with two exceptions: 
\begin{itemize}
    \item \emph{Unknown but static learning.} Suppose that the agent only learns information at level-$0$,\footnote{That is, we can have `learning at zero' with $\mathcal{F}_{0-} \neq \mathcal{F}_0$, but $\mathcal{F}_{0} = \mathcal{F}_l$ for all $l \in \mathcal{L}$.} but the regulator does not know the law of the learning process. We show (see \cref{prop:static_firstbest} in Online Appendix \ref{appendix:static_vs_dynamic}) that a nonlinear tax can be carefully constructed to correct the externality belief-by-belief i.e., it implements the first best ex post. This echoes recent work highlighting the power of nonlinear instruments under uncertainty in market design \citep{celebi2025adaptive} and networks \citep{koh2025prices}.
    \item \emph{Known and Markov learning.} Suppose the principal does not learn and the agent's learning process $\mathcal{F}$ is (i) known; and (ii) the agent's beliefs $\Ex[\theta |\mathcal{F}_l]$ is Markov with continuous sample paths. Then, results from \cite{kruse2015optimal,kruse2019inverse} can be adapted to solve for the optimal mechanism---this also takes the form of a nonlinear tax whose form is quite sensitive to the agent's learning process. 
\end{itemize}

But the specificity of these exceptions suggest that the Bayesian problem in which the agent's learning process $\mathcal{F} \in \mathbb{F}$ is unknown is hard and detail-dependent. On this view, \cref{thrm:robustness} represents the best a boundedly rational regulator can do.

Indeed, qualitative features of our adaptive sandbox---zero marginal tax, quantity limits, and adaptivity---broadly resembles real-world policies that have been deployed by policymakers to manage uncertain technologies e.g., real-world `regulatory sandboxes' for financial technology \citep{fca2015sandfbox} and AI \citep{EU2024AI}, or phases of FDA drug trials \citep{fda2008guidance}. We detail these connections at the end of this section.

\subsection{Undominated and dominance} \cref{thrm:robustness} establishes that the adaptive sandbox $\phi^{*}$ is both learning- and dually-robust. This leaves open the question as to whether other mechanisms might also be robust and, if so, why might we prefer one mechanism over another? We will offer a simple but, we think, compelling reason: $\phi^*$ is dominant within a natural class of mechanisms we now define. 

\begin{definition}[Regularity]
    An adaptive mechanism $\phi \in \Phi$ is \emph{regular} if it is almost-surely single-peaked or single-dipped. 
\end{definition}

\begin{figure}[H]  
\centering
    \caption{Illustration of regularity} 
    \subfloat[$\substack{\text{\footnotesize Nonlinear tax} \\ \text{\footnotesize \textcolor{blue!70!black}{Regular}}}$]
    {\includegraphics[width=0.24\linewidth]{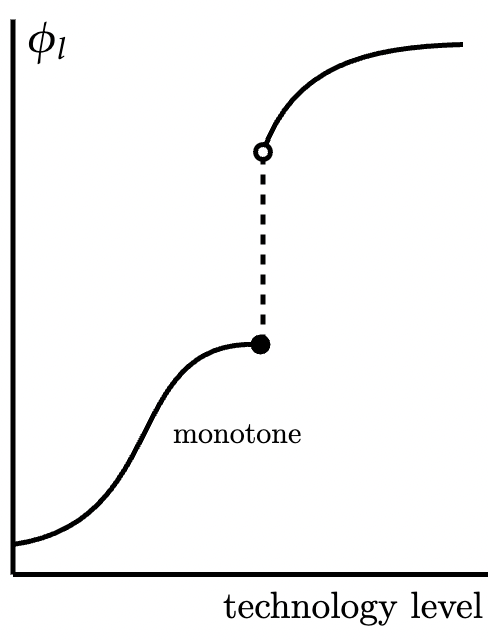}}
    \subfloat[$\substack{\text{\footnotesize Tax$\to$subsidize} \\ \text{\footnotesize \textcolor{blue!70!black}{Regular}}}$]
    {\includegraphics[width=0.24\linewidth]{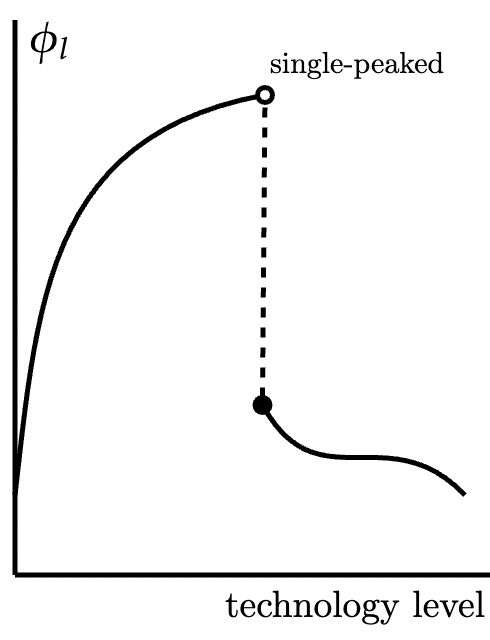}}
    \subfloat[$\substack{\text{\footnotesize Quota} \\ \text{\footnotesize \textcolor{blue!70!black}{Regular}}}$]
    {\includegraphics[width=0.24\linewidth]{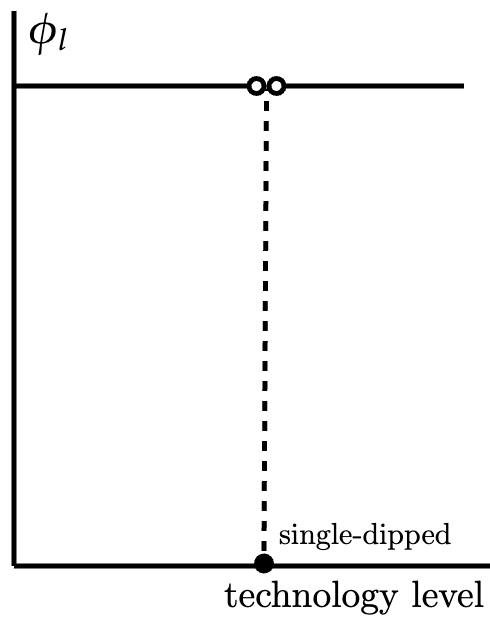}}
    \subfloat[$\substack{\text{\footnotesize Oscillatory} \\ \text{\footnotesize \textcolor{red!70!black}{Not regular}}}$]{\includegraphics[width=0.24\linewidth]{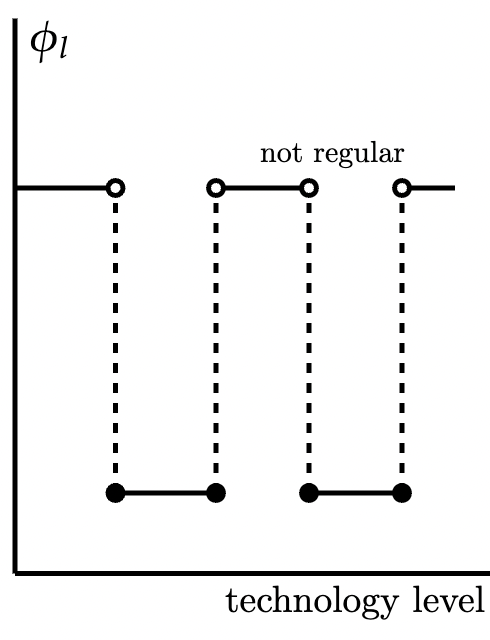}}
    \label{fig:regularity}
\end{figure}
 
 Regularity requires mechanisms to not vacillate between penalizing and rewarding the agent for pushing the technology. This is, in part, motivated by practical considerations: regular mechanisms are "not too complex" \citep{diamond2011case}.\footnote{\cite{diamond2011case} argue for this in the context of income taxation but we think this applies equally to taxes on R\&D or deployment.} At the same time, regularity is rich enough to capture a wide suite of instruments such as (potentially stochastic) nonlinear taxes and/or subsidies, quotas, and mixtures thereof. Sample paths of regular mechanisms are illustrated in panels (a)-(c) of Figure \ref{fig:regularity}. Panel (d) illustrates an irregular mechanism which oscillates between a low and a high tax.\footnote{Regularity is more permissive than the monotonicity condition typically imposed in security design \citep{nachman1994optimal,demarzo1999liquidity}  since any tax schedule that penalizes R\&D or deployment is increasing and thus regular.}

\begin{theorem}\label{thrm:dominance}
The adaptive sandbox $\phi^*$ is undominated. If the regulator chooses among regular mechanisms, all other dually-robust mechanisms are dominated by $\phi^*$.
\end{theorem}

The first part of \cref{thrm:dominance} establishes that $\phi^*$ is undominated and can thereby be rationalized over any other adaptive mechanism---for any alternate mechanism $\phi$, we can find some learning process $\mathcal{F}$ and preference $U$ so that $\phi^*$ delivers strictly higher regulator payoffs than $\phi$. The second part of \cref{thrm:dominance} states that $\phi^*$ dominates all other mechanisms $\phi$ which are regular and dually-robust.\footnote{Our focus on dominance is in the spirit of \cite{borgers2017no} who argue that simply focusing on robustness may be dominated; \cref{thrm:dominance} makes precise that $\phi^*$ is the unique undominated, robust, and regular mechanism.} Thus, if our designer has lexicographic attitudes toward uncertainty about the agent's learning process and preferences and, perhaps for practical reasons, is restricted to regular mechanisms which do not vacillate between punishing and rewarding pushing the technology, then $\phi^*$ is uniquely optimal.

\begin{figure}[H]
\begin{minipage}[t]{0.45\linewidth}
This is illustrated in \cref{fig:dominance_illust}: holding fixed the agent's preference $U \in \mathbb{U}$, the figure illustrates how the principal's payoff varies with the learning process $\mathcal{F} \in \mathbb{F}$. The principal's payoff under the adaptive sandbox $\phi^{*}$ (\textcolor{blue}{blue}) is minimized when the agent has no additional information ($\mathcal{F}_l = \mathcal{G}_l$ a.s.). But away from the worst-case, it performs better than other robust mechanisms e.g., an adaptive quota (\textcolor{red}{red}). 

\end{minipage}%
\hfill%
\begin{minipage}[t]{0.55\textwidth}\vspace{0pt}
\vspace{-1.5em}
\hfill 
{\includegraphics[width=0.9\textwidth]{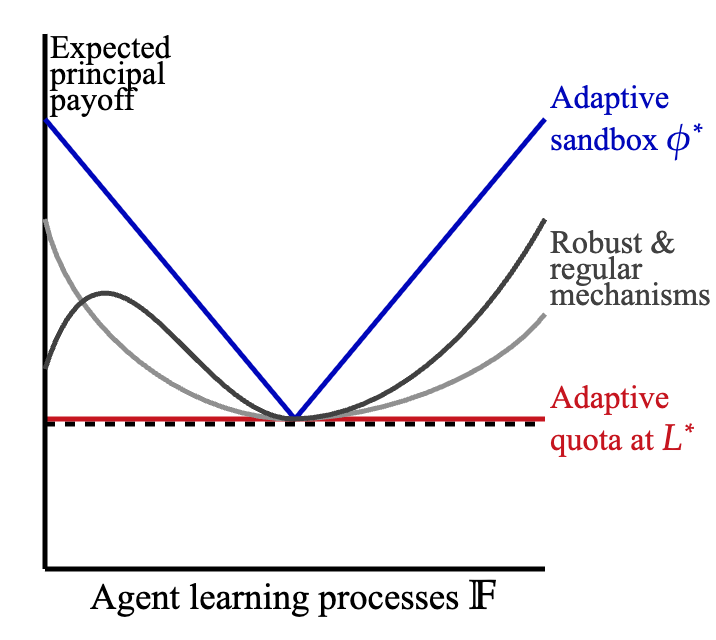}}
    \vspace{-0.5em}
    \caption{\small{Illust. of dominance}}\label{fig:dominance_illust}
\end{minipage}
\end{figure}

What drives the dominance of $\phi^{*}$? We have already highlighted how the zero marginal tax induces aligned sensitivity to new information by positively correlating the events that the agent and principal both prefer to stop. Among mechanisms with this property, the zero marginal tax within the sandbox induces \emph{maximal sensitivity} to new adverse information---we will briefly sketch the key ideas behind the argument for dominance here. The complete proof of \cref{thrm:dominance} is quite involved and is in \cref{app:proofs}.

\underline{\smash{Step 1: Characterization of dual-robustness.}} We begin with a complete characterization of dually-robust mechanisms. 

\textbf{Limit with option mechanisms.} $\phi$ is a \emph{limit with option mechanism} induced by the $\mathcal{G}$-adapted level $\ell$ if: 
\begin{itemize}
    \item[(i)] \emph{Limit:} $\phi_{s} = +\infty$ a.s. for all $s > \ell$; and 
    \item[(ii)] \emph{Option:} $\phi_s \geq \phi_\ell$ a.s. for all $\mathcal{G}$-adapted stopping levels $s \leq \ell$.
\end{itemize}  
Such mechanisms impose a quantity limit on the technology at some $\mathcal{G}$-adapted level $\ell$ and, in this regard, resemble sandboxes. Differently, they only demand that for any $\mathcal{G}$-stopping level $s \leq \ell$, the agent has the option of pushing the technology until $\ell$ to enjoy a \emph{guaranteed subsidy} of $\phi_s - \phi_{\ell} \geq 0$ relative to stopping at $s$. Since a zero marginal tax is one such option, all sandboxes are limit with option mechanisms. Limit with option mechanisms completely characterize dual-robustness.

\begin{proposition}\label{prop:dualrobust_characterization}
    $\phi$ is dually-robust if and only if it is a limit with option mechanism induced by $\overline \ell$. 
\end{proposition}

The proof of \cref{prop:dualrobust_characterization} is fairly technical and deferred to \cref{app:proofs}. We will sketch the broad intuition here. The limit at $\overline \ell$---which, recall, is the solution to the principal's direct experimentation problem $\sup_\el \Ex[V(\mu_\el,\el)]$---is driven by concerns that the agent might grow (with high probability) just a little more optimistic, but also turn out to be quite risk-seeking and so take on excessive risk. Limits safeguard against the prospect that equilibrium technology levels are distorted upwards in this manner, and is why `quantity-like' instruments must emerge under robustness concerns. By contrast, the option of receiving a guaranteed subsidy if the agent were to push the technology till $\ell$ (though she might be heavily taxed for stopping in the interim) safeguards against the prospect that regulation might stifle innovation by distorting the technology downward. Limit with option mechanisms simultaneously safeguard against both kinds of distortions which is exactly what dual-robustness demands. 

\begin{figure}[H]
\begin{minipage}[t]{0.45\linewidth}
\cref{prop:dualrobust_characterization} implies that if a dually-robust mechanism is also regular, it must be almost surely decreasing until the hard limit at $\overline{\el}$. That is, at any interim history, the agent knows that she will \emph{always} receive a subsidy for pushing the technology within the sandbox limit. This is illustrated by \cref{fig:subsidy_sandbox} which depicts three possible sample paths continuing from some interim history.   
\end{minipage}%
\hfill%
\begin{minipage}[t]{0.55\textwidth}\vspace{0pt}
\vspace{-0.5em}
\centering
{\includegraphics[width=0.99\textwidth]{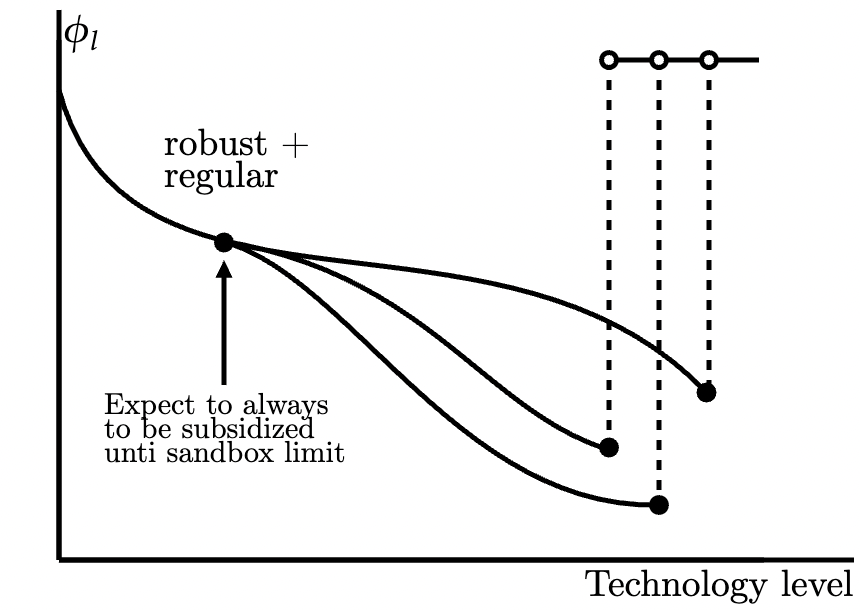}}
    \caption{Subsidy within sandbox}\label{fig:subsidy_sandbox}
\end{minipage}
\end{figure}

\underline{\smash{Step 2:  Coupling.}} Take any dually-robust and regular mechanism $\phi$. We have already observed that $\phi$ must subsidize the technology until the hard limit at $\overline{\el}$. Now for any learning process $\mathcal{F} \in \mathbb{F}$ and any agent preference $U \in \mathbb{U}$, we construct the simplest coupling of the optimal stopping problems induced by $\phi$ and $\phi^*$ on an enlarged probability space such that (i) the learning processes is perfectly correlated across problems; and (ii) the marginal distribution of learning processes for each individual problem coincides with $\mathcal{G}$ and $\mathcal{F}$. 

\begin{figure}[H]
\begin{minipage}[t]{0.45\linewidth}
On this coupled problem, the agent always stops sooner facing the adaptive sandbox $\phi^*$ than under the alternative $\phi$: 
\[
\el^*(\phi^*,\mathcal{F},U) \leq \el^*(\phi,\mathcal{F},U)\text{ a.s.}
\]
precisely because $\phi$ imposes a marginal subsidy on the technology. This is depicted by \cref{fig:pathwise_compstat} which illustrates two sample paths of beliefs (\textcolor{red}{red} and \textcolor{red!50!black}{dark red}). On each of these paths, the agent stops sooner facing $\phi^*$ than under the alternate mechanism $\phi$. 
\end{minipage}%
\hfill%
\begin{minipage}[t]{0.55\textwidth}\vspace{0pt}
\vspace{0em}
\centering
{\includegraphics[width=0.99\textwidth]{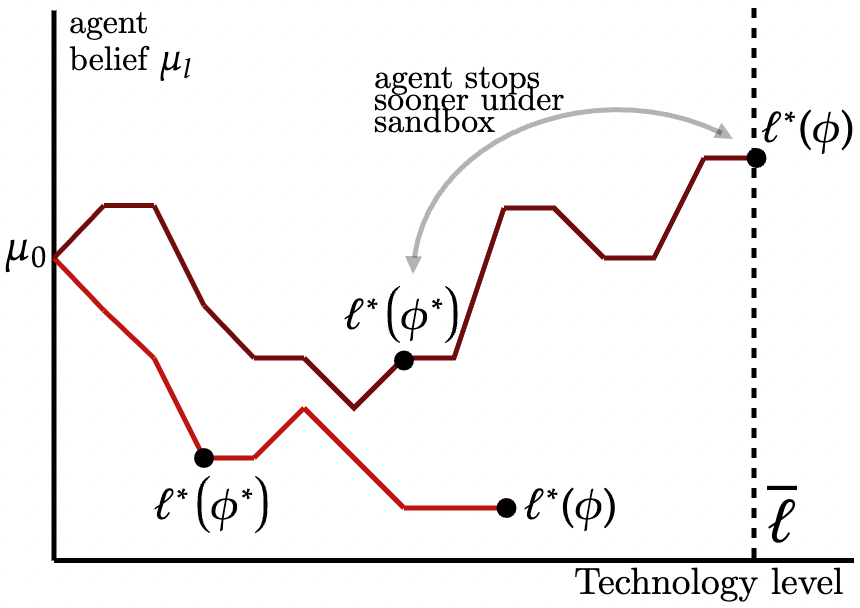}}
    \caption{Pathwise comparison}\label{fig:pathwise_compstat}
\end{minipage}
\end{figure}
 
\vspace{-1em}This is the precise sense in which the regulatory sandbox induces \emph{maximal sensitivity} to information. 

\underline{\smash{Step 3: Maximal + aligned sensitivity $\implies$ dominance.}} To finish, recall that our argument from \cref{thrm:robustness} and our comparative static \cref{lem:compstat} implied that the adaptive sandbox $\phi^*$ induced aligned sensitivity---whenever the agent prefers to stop, so does the principal as evaluated through the lens of the agent's private information:
\[
 \Bigg\{\substack{\text{\normalsize Agent prefers to stop facing} \\ \text{\normalsize mechanism $\phi^*$ and information $\mathcal{F}_{\el^*}$}} \Bigg\} \subseteq 
\Bigg\{\substack{\text{\normalsize Principal prefers to stop facing} \\ \text{\normalsize information $\mathcal{F}_{\el^*}$}} \Bigg\}. 
\]
But if this happens it is already too late---we are already past the principal's optimal technology level and the best thing to do is to stop the technology immediately. This implies that under this coupling, $\phi^*$ delivers a higher principal payoff path-by-path and so too in expectation. $\hfill \diamondsuit$

\cref{thrm:dominance} speaks directly to current debates around how regulatory sandboxes should be implemented in practice. In a 10-year legal retrospective on regulatory sandboxes, \cite{allen2025sandbox} notes the \emph{"transformation of financial regulators into cheerleaders and sponsors for the innovations they’ve selected for their sandboxes"} and, as a result, facilitated the proliferation of harmful technologies like predatory lending. On this view, real-world sandboxes were implemented with a positive marginal subsidy within its boundaries which make tech firms less sensitive to interim information---they are not incentivized to halt even if they turn pessimistic, thereby leading to the over-proliferation of harmful technologies. \cref{thrm:dominance} refines and unifies such different intuitions about why real-world sandboxes might not have worked well, as well as guidance for how they should be operationalized.

\subsection{Time consistency} \label{subsec:time_con} Robustness and dominance are ex ante criteria for assessing mechanisms before they are run. Might regulators be tempted to deviate to a different mechanism at interim histories? If so, this can render such mechanisms infeasible. Motivated by these concerns, we develop an analysis of time-consistency.

\textbf{Time-consistency.} $\phi$ is \emph{time-consistent} if for each $\mathcal{G}$-stopping level $\ell$: 
    \begin{itemize}
        \item[(i)] \textbf{Stopping.} If the agent stops at $\el$, $\phi$ is conditionally undominated i.e., there is no mechanism ${\phi'} \in \Phi$ such that 
        \begin{align*}
        \Ex\big[V\big(\mu_{\ell^*},\ell^*(\phi, \mathcal{F},U) \big) &\big| \mathcal{G}_{\el}, \ell^* = \ell \big] \leq 
        \Ex\big[V\big(\mu_{\el'},\ell'\big)\big| \mathcal{G}_{\el}, \ell^* = \ell\big] \\
            &\text{where $\ell'$ is the largest solution to $\sup_{s \geq \ell} \Ex[U^{\phi'}(\mu_s,s)]$}
        \end{align*}
        with strict inequality for some $(\mathcal{F},U) \in \mathbb{F} \times \mathbb{U}$.
    \item[(ii)] \textbf{Continuing.} If the agent continues at $\el$, $\phi$ is undominated i.e.,  there is no mechanism ${\phi'} \in \Phi$ such that 
        \begin{align*}
        \Ex\big[V\big(\mu_{\ell^*},\ell^*(\phi, \mathcal{F},U) \big) &\big| \mathcal{G}_{\el}, \ell^* > \ell \big] \leq 
        \Ex\big[V\big(\mu_{\el'},\ell'\big)\big| \mathcal{G}_{\el}, \ell^* > \ell\big] \\
            &\text{where $\ell'$ is the largest solution to $\sup_{s \geq \el} \Ex[U^{\phi'}(\mu_s,s)]$}
        \end{align*}
        with strict inequality for some $(\mathcal{F},U) \in \mathbb{F} \times \mathbb{U}$.
    \end{itemize}

Time-consistency requires mechanisms to not be dominated following from at any history. There are three distinct ways it can fail. The first resembles the logic of the Coase conjecture: the agent might, knowing that the regulator lacks commitment, deviate from continuing to stopping, or from stopping to continuing such as to influence future regulation.\footnote{Here, the agent believes that the regulator will deviate to some dominant mechanism.} The second is because the set of learning processes and preferences the regulator regards as possible shrinks in the interim so a mechanism that might have been previously rationalized for performing well at some learning-preference pair $(\mathcal{F},U) \in \mathbb{F} \times \mathbb{U}$ might become dominated.\footnote{The intuition here is similar to time-inconsistency generated by learning under ambiguity \citep{epstein2003recursive}.} Finally, as the technology level increases irreversibly, regulators might simply lose incentives to follow-through at interim histories. Time-consistency rules out all these possibilities.

Our notion of time-consistency is quite weak, and does not require us to make strong positive assumptions on the objective the regulator running the mechanism `actually has' e.g., she might have the robust objective, maximize expected utility, have a concern for misspecification, and so on.\footnote{We might, instead, demand that under $\phi$ the mechanism is `conditionally robust' at every history so that if the regulator actually had a robust objective she has no incentives to deviate. This notion is related to learning under ambiguity \citep{epstein2003recursive} and a variant (with no principal learning) has been studied in the context of monopoly pricing \citep*{li2022sequentially}. Our adaptive sandbox $\phi^*$ is also time-consistent under this alternate notion.} Nonetheless, this minimal notion of time-consistency is enough to refine the set of robust mechanisms:

\begin{theorem}[Time-consistency]\label{thrm:time-consistency}
The adaptive sandbox $\phi^*$ is time-consistent. If the technology space $\mathcal{L}$ is finite then $\phi^*$ is the only time-consistent and dually-robust mechanism.
\end{theorem}


We first sketch why the sandbox is time-consistent. Then, we sketch why it is the only dually-robust mechanism that is time-consistent. 

\textbf{Why is the sandbox time-consistent?} There are three kinds of histories: 
\begin{enumerate}[leftmargin = *]
    \item \underline{\smash{Stopping before sandbox limit.}} If the agent stops strictly before the sandbox i.e., at technology level $\el$ where $\el = \el^* < \overline{\el}$, then by aligned sensitivity, the principal is certain that her own optimal stopping level (viewed through the agent's filtration $\mathcal{F}$) has already passed 
    so the best thing continuing from that history is to let the agent stop. In fact, the sandbox $\phi^*$ is not only undominated, but it also \emph{dominates} all other dually-robust mechanisms from this history. 

\item \underline{\smash{Continuing before sandbox limit.}}  If the agent chooses to continue pushing the technology before the sandbox i.e., at technology level $\el < \overline{\el}$, then the regulator can rule out some possibilities e.g., that the agent has conclusively learnt that the technology is harmful, but her ambiguity set over $\mathbb{F} \times \mathbb{U}$ remains large so maintaining the sandbox is conditionally undominated.\footnote{This can be thought of as a kind of (non-Bayesian) positive selection \citep{tirole2016bottom}.}  

\item \underline{\smash{Stopping at sandbox limit.}} If the agent pushes the technology until the sandbox limit i.e., $\el^* = \overline{\el}$, the regulator cannot rule out the possibility that the agent's information at $\overline{\el}$ is such that if the regulator observed the agent's information she would prefer to stop. That is, the fear that the agent might only be weakly optimistic rationalizes keeping the hard limit in place.
\end{enumerate}

\textbf{Why are other robust mechanisms not time-consistent?}
Consider an alternate mechanism $\phi$ and recall dual-robustness which requires $\phi_\el \geq \phi_{\overline{\el}}$ a.s. for all $\el \leq \overline{\el}$ (via our characterization from \cref{prop:dualrobust_characterization} above). But suppose that with positive probability this inequality is strict at some $\mathcal{G}$-stopping level $\ell$ i.e., $\phi$ almost surely decreases until $\overline{\el}$. 

\begin{figure}[H]  
\centering
    \caption{Why are other robust mechanisms not time-consistent?} 
    \subfloat[Original mechanism $\phi$]{\includegraphics[width=0.47\linewidth]{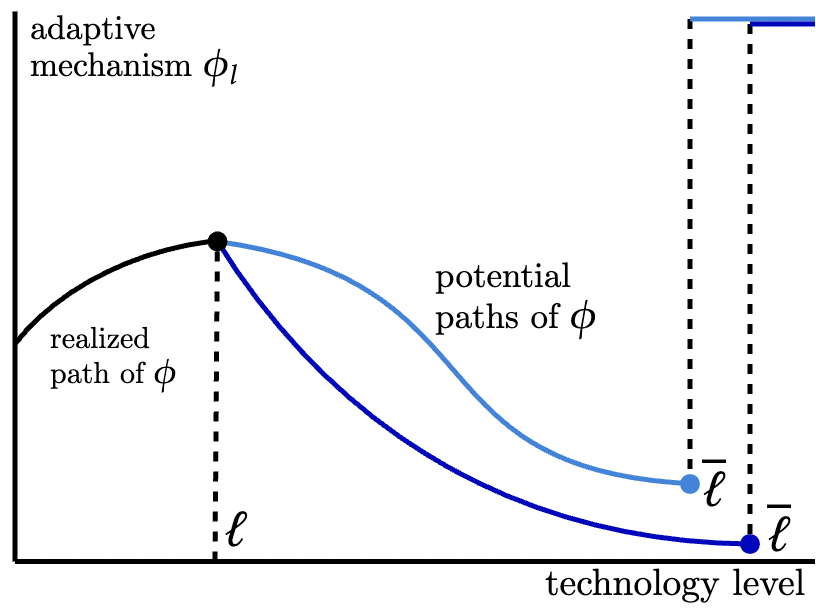}}
    \hfill 
    \subfloat[Dominant deviation at $\ell$]
    {\includegraphics[width=0.47\linewidth]{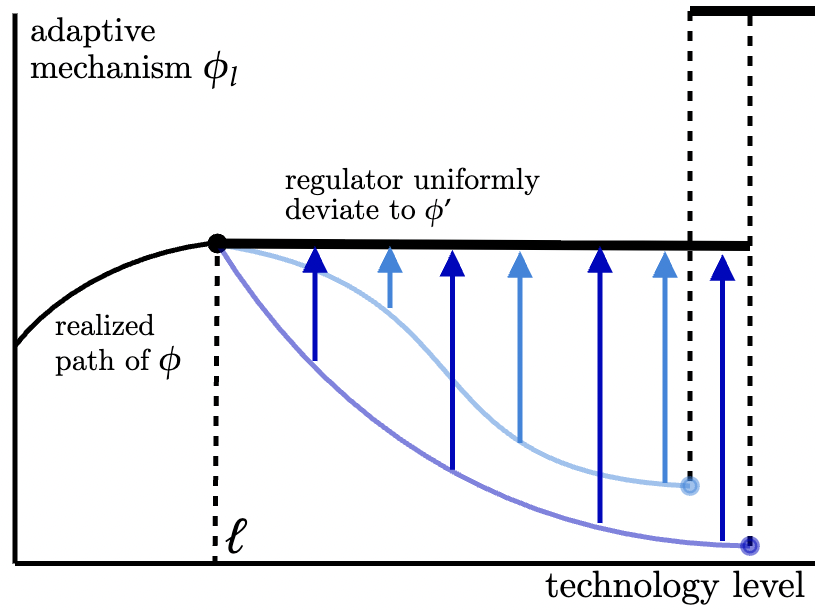}}
    \label{fig:pathwise_ironing}
\end{figure}

This candidate mechanism $\phi$ is depicted in \cref{fig:pathwise_ironing} (a) where the \textcolor{blue}{blue paths} reward pushing the technology between $\el$ and $\overline{\el}$. Now consider the uniform deviation to an alternate mechanism $\phi'$ continuing from $\el$ which `irons out' the mechanism along each path until the random level $\overline{\el}$. This deviation by the \textbf{bold black paths} in \cref{fig:pathwise_ironing} (b), noticing that (i) this uniform deviation is feasible i.e., the resultant mechanism $\phi'$ remains $\mathcal{G}$-adapted; and (ii) $\phi'$ is simply an adaptive sandbox continuing from $\el$. If the agent stops pushing the technology following this deviation to $\phi'$, then since the sandbox is aligned sensitive, this means that the principal's optimal technology level has already passed so this deviation delivers a strict improvement to the regulator's payoff. 
It is precisely this temptation to screen out weakly optimistic agents that is in tension with time-consistency.

\subsection{Connection to policy} We have thus far offered a rather abstract treatment of adaptive sandboxes. We now draw connections to real-world policy. Then, we will explain how our results might offer guidance for how policy can be improved. 

Over the past century, regulators have gradually converged on various kinds of `dynamic quantity limits' to regulate uncertain technologies. Drugs are the most paradigmatic example of this. In 1935, the Michigan state Supreme Court recognized that ``the fact that if the general practice of medicine and surgery is to progress, there must be a certain amount of experimentation carried on.'' But it was not until the 1960s with the Kefauver-Harris Drug Amendments that strengthened FDA oversight over drug experimentation, requiring pharmaceutical companies to demonstrate that it was safe to conduct clinical trails on humans. Regulation continued evolving into the present phased system in which the firm is progressively allowed to experiment on a growing number of humans in each phase \citep{junod2008fda}. 

A similar progression has taken place in financial and AI regulation. In 2015, the UK's Finanical Conduct Authority (FCA) proposed `regulatory sandboxes' to regulate new financial technologies in which participating firms could experiment with otherwise untested financial products e.g., insurance tech, peer-to-peer lending, distributed ledgers etc. for six months, upon which they would seek approval to be released into the wider market. As of 2020 over 50 jurisdictions have implemented versions of `regulatory sandboxes' to regulate new financial technologies \citep{appaya2020global}. Such `regulatory sandboxes' are increasingly enacted to regulate artificial intelligence across Europe \citep{EU2024AI} and the US \citep{sandboxact2025}. Our mechanism $\phi^*$ reflects the important features common across these policies: 

\begin{enumerate}[leftmargin =*]
    \item \emph{Zero marginal tax as fostering innovation}. In the first use of regulatory sandboxes, UK regulators described it as a {```safe space’ in which businesses can test innovative products... without immediately incurring all the normal regulatory consequences''} \citep{fca2015sandfbox}. This was implemented in 2016, and has been credited for making London a fintech hub.\footnote{\cite{cornelli2024regulatory} estimate that the UK fintech sandbox led new financial technology firms to raise 15\% more capital.} Within our framework, the zero marginal tax before the limit corresponds to relaxing (monetary and non-monetary) regulatory burdens e.g., lengthy and costly approval processes. \cref{thrm:robustness} showed that this zero marginal tax was sufficient to induce \emph{aligned sensitivity}. \cref{thrm:dominance} showed that subsidies, by contrast, are dominated because it dulls the firm's sensitivity to interim adverse information. This makes precise the intution for why sandboxes have not always worked well \citep{allen2025sandbox,sarro2025sandbox}, and guidance for how they might be improved. 
    
    \item \emph{Limits as safeguards}. Real-world implementation of sandboxes have imposed limits on different aspects of the technology. With fintech sandboxes, this has differed across jurisdictions e.g., (i) time limits for firms to roll-out new financial products (UK, Australia, Japan); (ii) {scale} limits which impose caps on the number of customers and/or deposit volumes (UK and Switzerland); or (iii) {scope} limits which fences off certain kinds of applications (virtually all jurisdictions).\footnote{For instance, the Financial Conduct Authority in the UK have excluded cryptocurrency-based derivative products from their sandbox. Similarly, the Monetary Authority in Singapore have excluded cryptocurrency exchanges and `initial coin offerings' from their sandbox.} More recently, the EU AI Act imposes a sandbox to "facilitate the development, training, testing and validation of innovative AI systems for a limited level" \citep{EU2024AI}, and more broadly regulators in the EU and US have imposed thresholds on training compute \citep{heim2024training}. A common theme across such policies is some form of `quantity limits' on research, development, or deployment. 
    \item \emph{Adaptivity.} Our sandbox limit $\overline \ell$ loosens and tightens with the principal's interim belief. This reflects the practice of regulatory sandboxes in fintech---to push beyond the boundary e.g., to scale up the customer base of new producers---firms must convince the regulator to extend the sandbox by producing evidence that the technology is beneficial for consumers\footnote{At the sandbox limit, firms can `apply to remove the restrictions on... permissions to become a fully authorised firm.' \citep{fca_apply_sandbox_2023}. A similar policy is in place in Australia \citep{asic_info248}.}---failing which the firm is forced to halt roll-out of the technology. Similar kinds of `step-by-step' quantity limits predates regulatory sandboxes: the FDAs regulations on clinical research progressively rolls out a new drug in phases to a progressively larger number of human participants, halting experimentation (failing) if interim results make regulators pessimistic that the drug is safe and efficient.\footnote{For an outline of current regulation FDA clinical trials see \url{https://www.fda.gov/patients/drug-development-process/step-3-clinical-research}.}
    \item \emph{Encouraging learning.} A common theme articulated across regulatory agencies is that real-world sandboxes are {``intended to foster business learning, i.e., the development and testing of innovations in a real-world environment''} \citep{euparliament2024sandbox}. Our results crystalize this ambition: by loosening the limit whenever the principal turns more optimistic, this carves out extra room for the agent to experiment. But if the agent then halts experimentation prematurely, aligned sensitivity to information ensures that the principal also prefers to halt. 
\end{enumerate}

We emphasize that our results should not be taken as an endorsement of quantity limits \emph{tout court}. Policymakers can, of course, get things wrong---they might be too stringent with these limits (as has been argued in the case of drug trials), or too lax (as has been argued for AI). Instead, our results formalize the intuition for why certain \emph{kinds} of regulatory instruments should be preferred: adaptive quantity limits are simple, transparent, and delivers optimal payoff guarantees (\cref{thrm:robustness}) in an otherwise complicated environment (what beliefs do you have over $\mathbb{F} \times \mathbb{U}$?) Moreover, we offered sharp refinements over the universe of quantity instruments via dominance and time-consistency (\cref{thrm:dominance,thrm:time-consistency}). Taken together, our results imply that \emph{if} we took robustness seriously, the adaptive sandbox $\phi^*$ is, in effect, the uniquely optimal mechanism. Should we?

\section{The Importance of Robustness}\label{sec:importance}
We will argue that in the context of technology regulation, robustness is \emph{important} in two regards. First, the worst-case learning process under non-robust mechanisms is natural---that is, robust mechanisms safeguard against something real. Second, the worst-case payoff under non-robust mechanisms can be arbitrarily poor, even when the regulator and firms' risk preference differ slightly. Taken together, we think this offers a strong, if not indefeasible, reason to care about robust technology regulation.\footnote{There are, of course, more `standard' reasons for robustness: simplicity, difficulty of forming beliefs over the space of learning processes, axioms for ambiguity etc. that we will not belabor. For discussions of simplicity see \cite{carroll2019robustness}; for informational robustness see \cite{bergemann2005robust}; for axiomatic approaches to learning under ambiguity see \cite{epstein2003recursive}.}

We start by characterizing the worst-case learning process under non-robust mechanisms. We will focus on the case in which the principal does not learn i.e., $\mathcal{G}_l = \mathcal{G}_0$ a.s. for all $l \in \mathcal{L}$. This approximates the `short run' which we view as a conceptual shorthand for the period of analysis in which the principal's beliefs are fixed,\footnote{Analogous to the Marshallian short run \citep{marshall1890}.} or for technologies with large `regulator lags'. Let $\overline{\mathbb{F}}$ denote the set of all agent learning processes and let $\underline{\Phi}$ denote the set of mechanisms under no principal learning. A few definitions are in order.  

\begin{definition}[Bad news process] The belief process induced by $\mathcal{F} \in \overline{\mathbb{F}}$ is a \emph{bad news process} with \emph{continuation belief path} $\gamma: \mathcal{L} \to [0,1]$ if  
\[\text{supp} \, \Ex\big[\theta \big | \mathcal{F}_l\big] \subseteq \Big\{0, \gamma(l) \Big\}\]
where $\gamma$ is nondecreasing and right-continuous. Note that since $\Ex[\theta \mid \mathcal{F}_l]$ is an $\mathcal{F}$-martingale, $\gamma$ fully specifies the law of the belief process.
\end{definition}

\begin{figure}[H]
\begin{minipage}[t]{0.45\linewidth}
Bad news processes are such that, at each technology level, bad news arrives at an inhomogeneous rate so the agent either conclusively learns that the technology is harmful or, in the absence of such news, becomes increasingly more optimistic and follow the path $\gamma$.\footnotemark This is illustrated in \cref{fig:badnews} where the gray lines show different belief paths and the \textcolor{red}{red} region denotes the support of the equilibrium joint distribution over technology levels and beliefs.
\end{minipage}%
\hfill%
\begin{minipage}[t]{0.55\textwidth}\vspace{0pt}
\vspace{-0.5em}
\centering
{\includegraphics[width=0.97\textwidth]{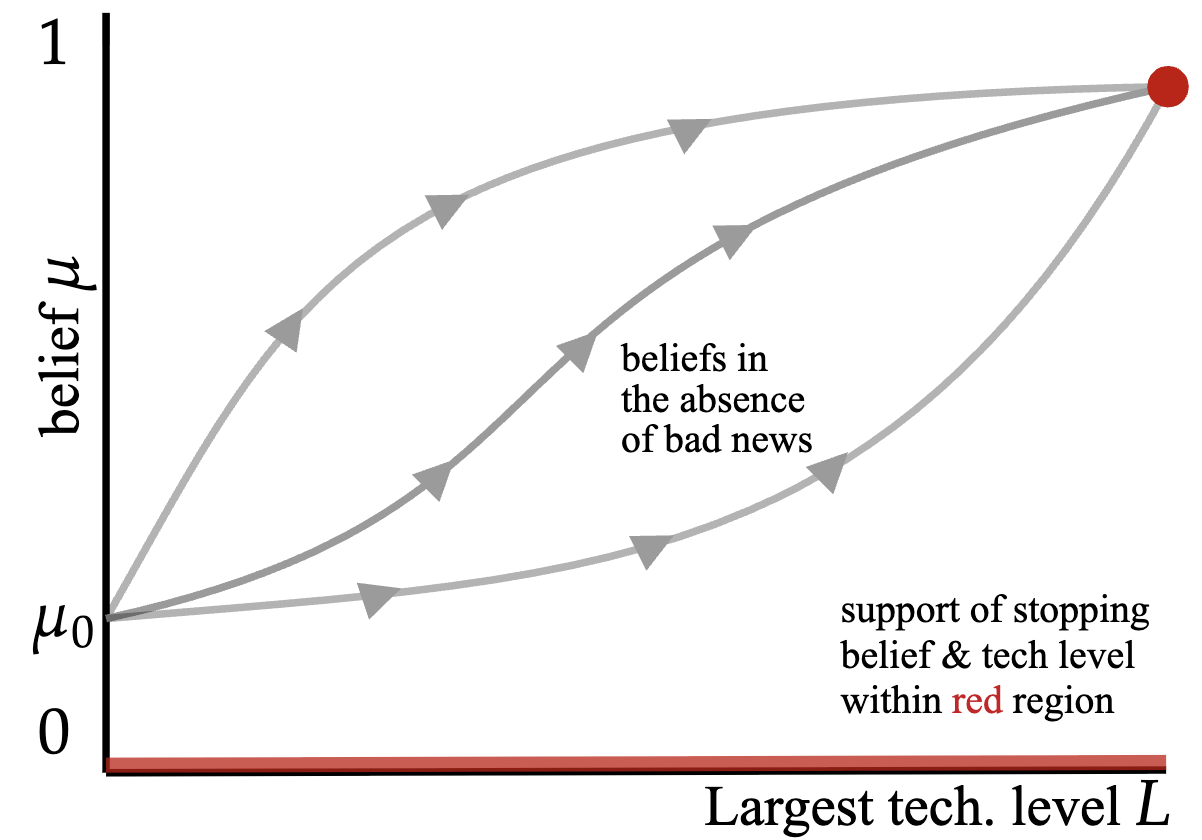}}
    \caption{Bad news processes}\label{fig:badnews}
\end{minipage}
\end{figure}

\footnotetext{The connection between this arrival rate and beliefs is $\gamma(l) = \mu_0/(1 - F(l))$ where $F$ is the (unnormalized) CDF of the arrival level of bad news.}

The set of bad news processes is large. Let us construct a specific one. For the indirect utility $f: \Delta(\Theta) \times \mathcal{L} \to \mathbb{R}$, define: 
\[
\underbrace{l^*_f(\mu) := \max \, \argmax_{l \in \mathcal{L}} f(\mu,l)}_{\text{Static optimal technology level at belief $\mu$}} \quad \text{and} \quad \underbrace{\mu^*_{f}(l) := \min \Big\{ \mu \in \Delta(\Theta): l_f^*(\mu) \geq l \Big\}}_{\text{Smallest belief rationalizing $l$}}. 
\]
$l^*_f(\mu)$ is the largest optimal `one-shot' technology level under belief $\mu$ in the absence of further information; $\mu^*_f(l)$ is the belief threshold---the most pessimistic belief that can rationalize pushing the technology until level $l$.

\begin{definition}[Weak optimism]
    The bad news process $\mathcal{F} \in \overline{\mathbb{F}}$ induces \emph{weak optimism} for indirect utility $f$ if it has continuation belief path
    \[
    \gamma(l) = 
    \begin{cases}
        \mu_0 \quad &\text{if $l < l^*_f(\mu_0)$;} \\
        \mu^*_f(l)
        &\text{if $l \geq l^*_f(\mu_0)$.}
    \end{cases}
    \]
\end{definition}

Bad news processes which induce weak optimism deliver conclusive evidence of bad news that the technology is harmful at a rate so that the agent's interim belief $\gamma(l)$ in the absence of bad news is exactly equal to the minimal belief $\mu^*_f(l)$ that can rationalize pushing the technology until level $l$ in the absence of further information. The following result establishes that this is indeed the worst-case in the absence of robust regulation:  

\begin{theorem}\label{thrm:worstcase} The form of the worst-case learning process and value of robustness is as follows: 
\begin{itemize}
    \item[(i)]  \textbf{Form.} If the agent's preference $U$ is known, then for any mechanism $\phi$ such that $U^{\phi} = g \circ V$ for some increasing convex function $g$, 
the worst-case learning process is a bad news process that induces {weak optimism} for $U^{\phi}$.
    \item[(ii)] \textbf{Importance.} 
If the agent's preference $U$ is also chosen adversarially, then for any mechanism $\phi$ which does not impose a hard limit, dual-robustness is arbitrarily important:
\[
\lim_{L \to \infty}  \inf_{\substack{\mathcal{F} \in \overline{\mathbb{F}} \\ U \in \mathbb{U}}} \Ex\Big[V( \mu_{\el^*}, \el^*(\phi, \mathcal{F}, U)) \Big] = -\infty 
\quad \text{if} \quad \lim_{L \to +\infty} \Bigg|\cfrac{V(0,L)}{V(1,L)} \Bigg| = +\infty. 
\]
\end{itemize}
\end{theorem}

\cref{thrm:worstcase} pins down the learning processes which minimizes the principal's payoff under any mechanism $\phi$ that preserves the order of risk preferences.\footnote{This is satisfied by a large class of mechanisms and payoff functions. A previous version of this paper established that if $l^*_{U^{\phi}}$ and $l^*_V$ are continuous the worst-case learning process continues to be some some bad-news process, though not necessarily binding the agent's continuation incentive at every history.} The proof proceeds by recasting nature's problem as a dynamic information design problem to maximize $-V$, noting that here nature has state-dependent preferences which is novel to the literature\footnote{To handle this, we modestly extend techniques developed in some of our previous work \citep*{koh2024persuasion} that established a general toolkit for such problems.}---this is fairly technical and we defer it to \cref{app:proofs} for we have quite a lot to say about the economics. 

\textbf{The form and functioning of weak optimism.}  
Weak optimism hurts the regulator by (stochastically) maximizing the technology level when it is harmful, subject to the constraint that the agent does, in fact, want to push on with the technology---i.e., by \emph{magnifying false positives}. 
\begin{figure}[H]
\begin{minipage}[t]{0.45\linewidth}

This is illustrated by the \textcolor{red}{red} curve in \cref{fig:false_positive} which shows the CDF of technology levels conditional on $\theta = 0$. This is chosen so that, in the absence of bad news, the agent grows progressively more emboldened to develop the technology. Conversely, the technology always fully developed when the technology is safe $(\theta = 1)$ i.e., there are no false negatives as illustrated in \textcolor{blue}{blue}.
\end{minipage}%
\hfill%
\begin{minipage}[t]{0.55\textwidth}\vspace{0pt}
\vspace{-1em}
\centering
{\includegraphics[width=0.9\textwidth]{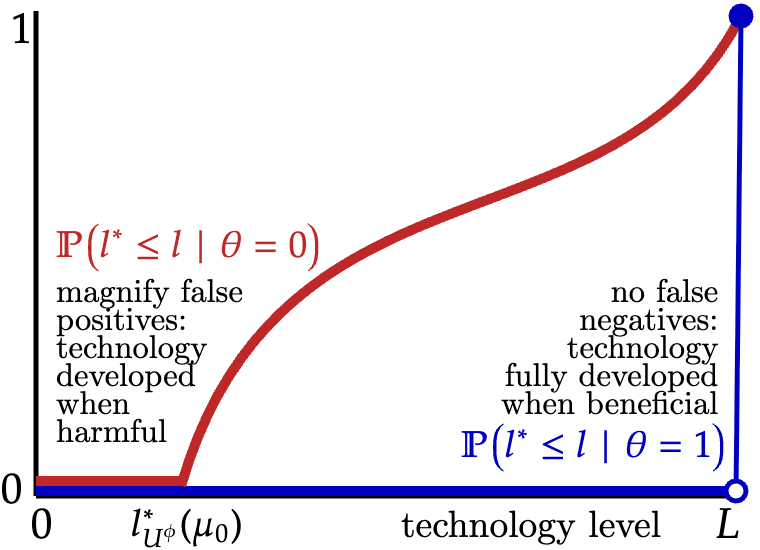}}
    \caption{Worst-case joint dist.}\label{fig:false_positive}
\end{minipage}
\end{figure}

Indeed, a distinctive feature of the worst-case learning process is that it progressively exploits \emph{only} the static wedge between the principal's and agent's preferences. The agent's payoff from incrementally increasing the technology level can be heuristically decomposed as: 
\[
\underbrace{\substack{\text{\normalsize Value of more technology} \\ \text{\normalsize at current belief}}}_{\substack{\text{$\gtrless 0$ under arbitrary} \\ \text{learning process} \\
\text{\textcolor{red!70!black}{Worst-case: $0$}} 
}
} 
\, + \,
\underbrace{\substack{\text{\normalsize  Value of future info} \\ \text{\normalsize from more technology}}}_{\substack{\text{always $\geq 0$} \\ \text{(free disposal)} \\
\text{\textcolor{red!70!black}{Worst-case: $0$}}
}} 
\quad 
\]

where the first term gives the value of additional technology levels holding fixed the current belief, while the second gives the value of learning more about the technology which might inform future technology choices. For a typical learning process the agent has positive value for future information which incentivizes risk-taking e.g., the first term can be negative while the second positive. Weak optimism essentially holds both terms down to zero:\footnote{The first term is strictly positive before any information arrives i.e., up until level $l^*_{U^{\phi}}(\mu_0)$ and zero thereafter.} the agent's beliefs are dynamically steered such that she is continually kept indifferent between pushing the technology and stopping, while her value from future information is kept at zero at all histories. 

The form of weak optimism carries two further implications. First, fixing the mechanism $\phi$, and agent preference $U$, the worst-case learning process and attendant regulator payoff can be straightforwardly computed. We leverage this in \cref{eg:CRRA_worstcase} below to quantify the worst-case under CRRA utility. Second, the worst-case payoff remains possible whenever the agent employs any \emph{undominated stopping rule}. This includes agents who are misspecified or ambiguity-averse over what they might learn in the future \citep{riedel2009optimal}, precisely because weak optimism exploits only the static wedge in risk-preferences. Indeed, computer scientists have raised worries that they might not learn much about the dangers of artificially intelligent agents as they continue to be developed.\footnote{This is formalized in Online Appendix \ref{app:undominated_stopping}. Here our agent correctly interprets the information that arrives and updates via Bayes' rule. Her ambiguity and/or misspecification is over the law of future information which shapes choices in the present. This notion is closely related to that of \cite{riedel2009optimal} who studies a stopping problem where the agent perfectly observes realizations of the payoff process, but faces ambiguity over the future law.} \cref{thrm:worstcase} implies that tech firms' conservatism about future information per se is not enough to protect society writ large.

\begin{figure}[H]  
\centering
    \caption{Weak optimism under linear taxes} 
    \subfloat[Increasing the linear tax]{\includegraphics[width=0.4\linewidth]{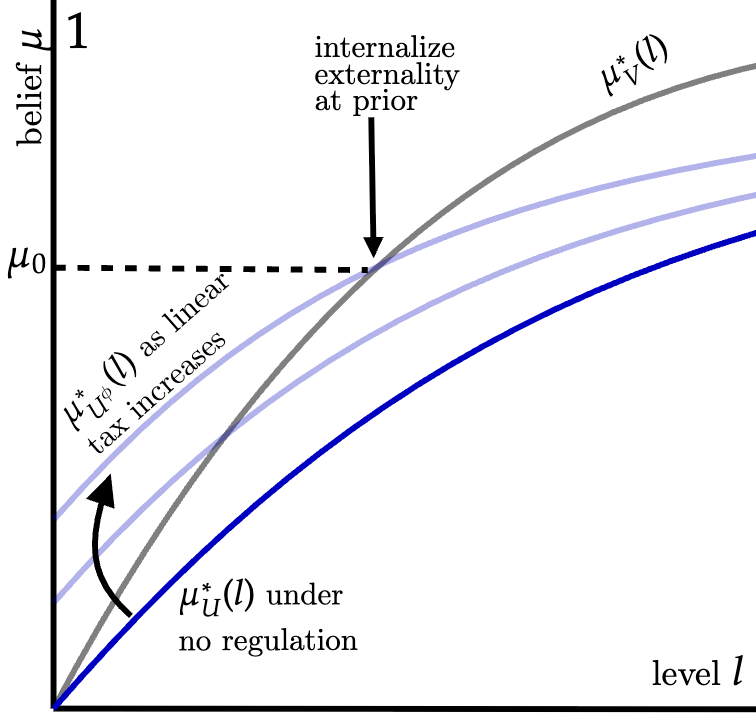}}
    \subfloat[Worst-case learning process]
    {\includegraphics[width=0.4\linewidth]{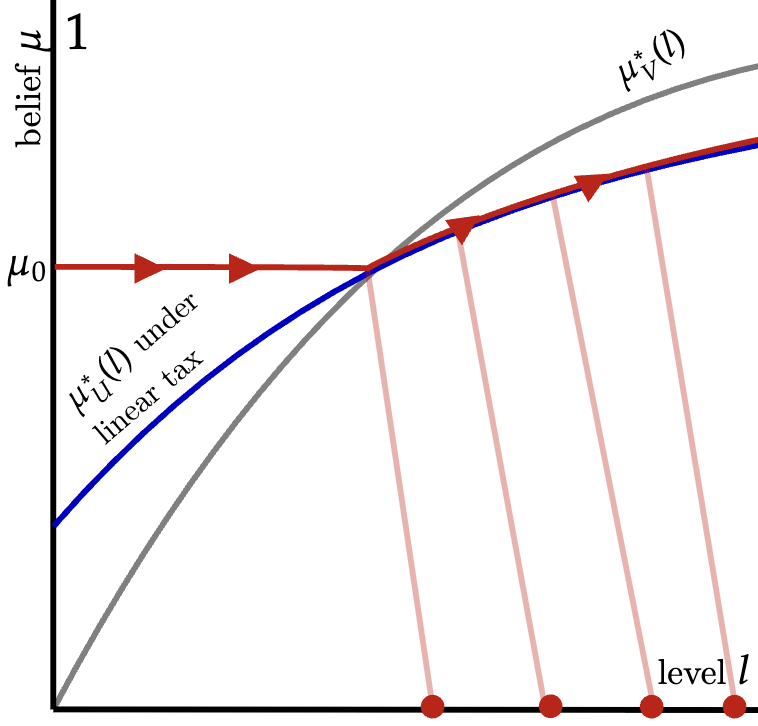}}
    \label{fig:worstcase_linear}
\end{figure}

\textbf{Policy implications of weak optimism.} \cref{thrm:worstcase} warns against correcting externalities only `locally'. Consider, for instance, the linear Pigouvian tax $\phi(l) = \alpha \cdot l$ with slope $\alpha > 0$ chosen to internalize externality under the prior. Panel (a) of \cref{fig:worstcase_linear} illustrates the belief thresholds $\mu^*_{V}$ (\textcolor{gray!30!black}{gray}) and $\mu^*_{U^{\phi}}$ (\textcolor{blue}{blue}). As the Pigouvian tax increases, the belief threshold for the agent increases accordingly, and the slope can be chosen such that the externality is perfectly internalized under the prior: $l^*_{U^{\phi}}(\mu_0) = l^*_{V}(\mu_0) = \overline{\el}$.
Panel (b) of \cref{fig:worstcase_linear} depicts the worst-case learning process under this tax: the agent learns nothing about the safety of the technology until $\overline{\el}$, then bad news that the technology is unsafe arrives at a rate such that in its absence, the agent becomes progressively more confident in the technology (\textcolor{red!70!black}{red arrow}) which encourages---from the principal's point-of-view---excess risk-taking.

\begin{figure}[H]  
\centering
    \caption{How convexity safeguards against weak optimism} 
    \subfloat[Increasing tax convexity]{\includegraphics[width=0.4\linewidth]{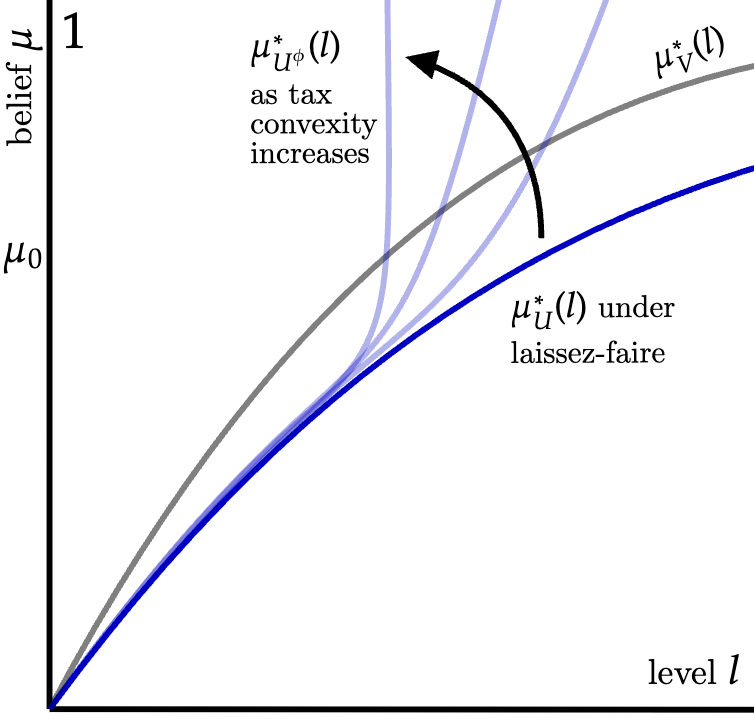}}
    \subfloat[Worst-case learning process]
    {\includegraphics[width=0.4\linewidth]{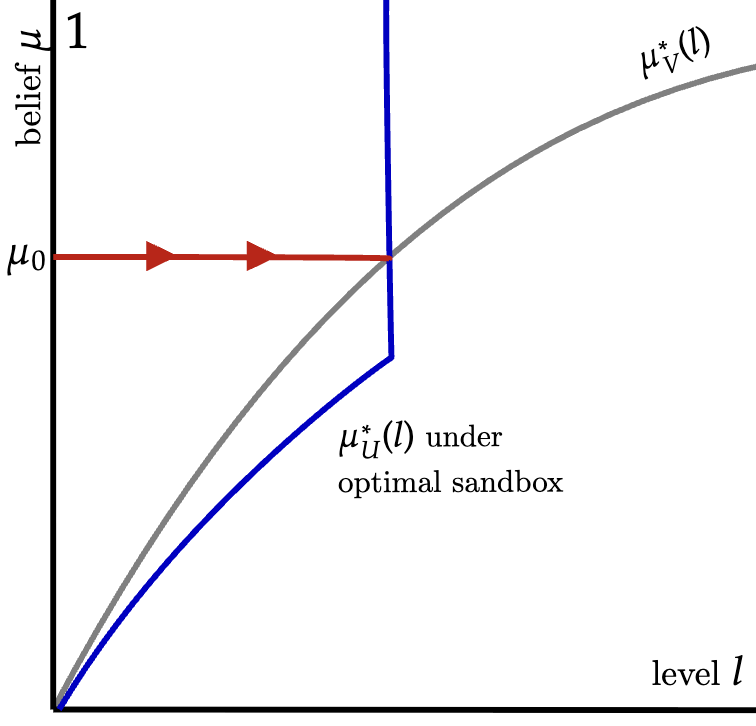}}
    \label{fig:worstcase_convex}
\end{figure}

Now consider panel (a) of \cref{fig:worstcase_convex} which depicts the agent's belief thresholds under a sequence of mechanisms of `increasing convexity'. As before, this can be chosen such as to internalize the externality locally at the prior. But differently, at more optimistic beliefs the agent now prefers a lower technology level than the principal, while at less optimistic beliefs the agent prefers more. Panel (b) of \cref{fig:worstcase_convex} depicts a regulatory sandbox as the limit of such mechanisms. From \cref{thrm:robustness} (i), aligned sensitivity ensures that any deviation by nature to a more informative learning process must benefit the principal so the worst-case learning process is, in fact, no learning.  

\textbf{The plausibility of weak optimism.} Weak optimism identified by \cref{thrm:worstcase} (i) correspond to technologies that pose rare disaster risk: financial crises induced by the proliferation of new financial instruments, pandemic induced by gain-of-function research, or AI-induced catastrophes induced by misalignment. For instance, prominent computer scientists have noted that \emph{"sufficiently capable LTPA, safety testing is likely to be either dangerous or uninformative"} \citep{cohen2024regulating}. This corresponds to a bad news process where the lack of a disaster makes the agent (financial engineer, biologist, AI scientist)---via Bayes' rule---progressively more optimistic that the technology is safe. But in the absence of stringent regulation, our result shows that learning per se is not enough to safeguard society---in fact, a little learning can be much worse than no learning at all. Indeed, with AI development there is substantial concerns among AI scientists at the technological frontier on dangers of powerful AI systems \citep{greenblatt2024alignment}---but not enough to induce stopping. We think this is consistent with our notion of `weak optimism'.   

\textbf{The quantitative importance of robustness.} \cref{thrm:worstcase} (ii) establishes the quantitative importance of robustness: when both the agent's preference and learning process are chosen adversarially, then without a hard limit payoffs are arbitrarily poor if, in the large technology limit (with `superhuman' AI systems, say) the loss to payoffs when the technology is dangerous ($\theta = 0$) dwarfs gains when the technology is beneficial ($\theta = 1$). 
We emphasize, however, that learning \emph{per se} can magnify even small wedges in differential risk-aversion as the following example illustrates: 
\begin{example}[Worst-case under CRRA utility]\label{eg:CRRA_worstcase} Suppose, following \cite{jones2024ai} that both agent and principal have CRRA preferences over consumption with risk-aversion parameter $\gamma_P > \gamma_A \geq 1$. Consumption is given by 
\[
c(\theta,l) = \begin{cases}
    \exp\big(a\cdot l\big) \quad &\text{if $\theta = 1$}\\ 
    \exp\big(-b\cdot l\big) \quad &\text{if $\theta = 0$}
\end{cases}
\]
with $a, b > 0$. Then for any linear Pigouvian tax $\phi_l = \alpha \cdot l$ that satisfies the condition in \cref{thrm:worstcase},  
\[
\lim_{L \to \infty}  \inf_{\mathcal{F} \in \overline{\mathbb{F}}} \Ex\Big[V( \mu_{\el^*}, \el^*(\phi, \mathcal{F}, U)) \Big] = -\infty  \iff \text{$U,V$ s.t. $\cfrac{\gamma_P-1}{\gamma_A - 1} > 1 + \cfrac{a}{b}$.}
\]
\cref{fig:numerics} illustrates worst-case payoffs as the principal and agent's risk-aversion coefficients vary. Panel (a) illustrates parameter regions under which the principal's payoffs under weak optimism is unboundedly poor. Panel (b) illustrates the principal's payoff under weak optimism when $\gamma_P = 2$ and $\gamma_A = 1.6$, noting that it asymptotes to approximately $-3$; by contrast, when the agent is just a little less risk-averse with $\gamma_A = 1.4$ the principal's payoff under weak-optimism diverges to $-\infty$ exponentially quickly with size of the technology space $L$. 
\begin{figure}[H]  
\centering
    \caption{Worst case (weak optimism) payoffs under CRRA utility \\ \footnotesize $a = b = 1$} 
    \subfloat[]{\includegraphics[width=0.45\linewidth]{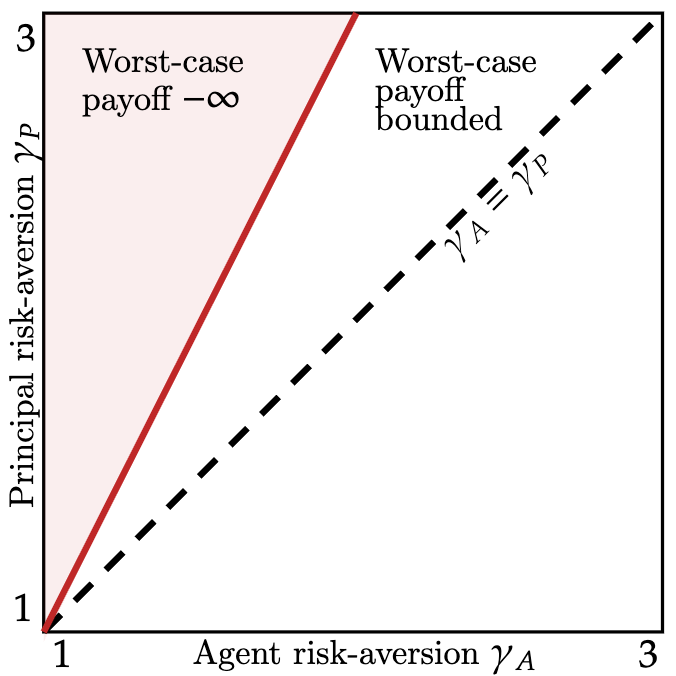}}
    \subfloat[]
    {\includegraphics[width=0.448\linewidth]{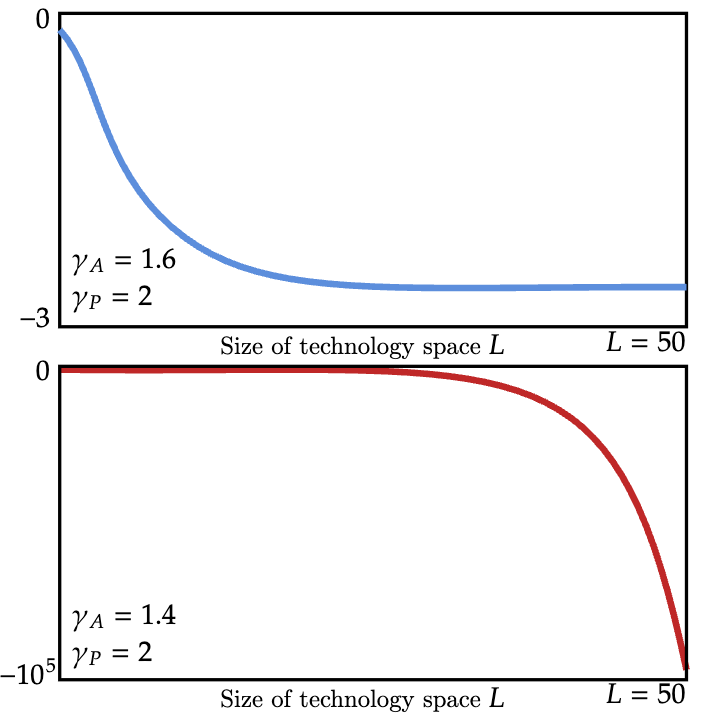}}
    \label{fig:numerics}
\end{figure}
We take this example to suggest that with \emph{transformative} technologies---those that can drastically shape consumption-equivalents---even small wedges in risk preferences can translate into vastly different welfare outcomes under the worst-case learning process of weak optimism. We think this is reason for caution as reflected by our robustness notions. $\hfill \diamondsuit$
\end{example}

Indeed, that robustness is both quantitatively and qualitatively important allows us to partially bridge robustness and Bayesian optimality: 

\begin{proposition}[Informal]\label{prop:Bayesian} Suppose that the regulator's prior over learning processes and preferences $f \in \Delta(\mathbb{F} \times \mathbb{U})$ is `diffuse'. Then as $L \to +\infty$, all mechanisms that attain  
    \[
    \underset{\phi \in \Phi}{\sup}\,
    \int_{\substack{(U,\mathcal{F}) \\ \in \mathbb{U} \times \mathbb{F}}} \Ex \Big[ V\big(\mu_{\el^*}, \el^*(\phi, \mathcal{F}, U) \big) \Big] df  
    \]
    imposes a hard limit at a finite level. 
\end{proposition}

\cref{prop:Bayesian} states that when the principal's prior $f$ is sufficiently `diffuse' over a simple parametric class of L\`evy signal processes and CRRA agent preferences, Bayesian optimality requires hard limits. A precise statement is in Online Appendix \ref{appendix:Bayes}. The intuition is straightforward: without hard limits, the richness of the regulator's prior implies the existence of a positive probability set $\mathbb{S} \subseteq \mathbb{F} \times \mathbb{U}$ under which payoffs are unboundedly poor. Thus, our results imply that not only do hard limits perform well at \emph{a} prior \citep{wald1950,lecam1955extension}, but that it is optimal under \emph{flat} priors---sufficient uncertainty of the Bayesian kind is another reason for robust technology regulation.

\textbf{Ranking payoff guarantees.} \cref{thrm:worstcase} can be deployed to rank non-robust mechanisms according to their payoff-guarantees. This offers guidance for a regulator who might be concerned about robustness but faces political or other constraints and must choose the `least bad' policy from a restricted set. In particular, if one mechanism induces more agent risk-aversion than another, it delivers better payoff guarantees when the agent's learning process is chosen adversarially:  

\begin{proposition}\label{prop:worstcasecompstat}
    Fix two mechanisms $\phi$ and $\phi'$ such that there exists an increasing convex function $g: \mathbb{R} \to \mathbb{R}$ such that $U^{\phi'} (\theta,l) = g \circ U^{\phi}(\theta,l)$ for every $\theta \in \Theta$ and $l \in \mathcal{L}.$ Then worst-case payoffs are ordered as follows: 
    \[
    \inf_{\mathcal{F} \in \overline{\mathbb{F}}} \Ex\Big[V( \mu_{\el^*}, \el^*(\phi, \mathcal{F}, U)) \Big] \geq  
    \inf_{\mathcal{F} \in \overline{\mathbb{F}}} \Ex\Big[V( \mu_{\el^*}, \el^*(\phi', \mathcal{F}, U)) \Big].
    \]
\end{proposition}

\textbf{The worst-case under principal learning.} We have focused on the case in which the principal does not learn, which gave nature the most room to choose the agent's learning process adversarially. If the principal learns, nature becomes more constrained since the agent has superior information i.e., $\mathcal{F} \supseteq \mathcal{G}$. Intuitively if the principal's learning process is `fast' then the agent cannot be weakly optimistic (in the sense constructed above). Is robustness still important in the presence of principal learning? Yes---as long as the principal's learning process is `slow'. 

Suppose, for instance, that the $\mathcal{G}$ is a perfect bad news process that arrives at some low rate. This might corresponds to the public realization of a `disaster' from developing or deploying the technology. Weak optimism remains possible here: the agent also observes the public realization but, in addition, is privy to extra information (e.g., through testing) that can also yield conclusive evidence of harm. This rate can be chosen such that, combined with the public process, the agent's learning process is exactly weak optimism. More broadly, for a fixed principal learning process $\mathcal{G}$, the agent's worst-case learning process $\mathcal{F}$ might correlate the realizations of the agent's past, present, and future signals with realizations of the principal's \emph{present} signals. This can be leveraged to great effect to induce excessive risk-taking.

\section{Policy Implications and Extensions} \label{sec:policy}

We have argued that a regulator concerned about robustness should adapt policy to evolving information by relaxing and/or tightening a quantity limit on risky experimentation (\cref{thrm:robustness,thrm:dominance,thrm:time-consistency}). We now distill several policy implications before briefly discussing how our framework and results can be extended in various directions.

\textbf{A taxonomy of learnability for technologies.} Our results offer simple guidance for how stringently different technologies should be regulated as a function of their learnability---how easily and predictably they can be learnt about---as well as the kinds of information regulators should seek out.  

Some technologies are \emph{transparently learnable}. These include clinical FDA drug trials, randomized lab or field interventions, and more broadly technologies for which the methods of modern science---double-blind randomization, placebo controls, and significance testing---can be brought to bear. Within our framework, the technology level $l$ might correspond to the number of humans that have taken the drug so the regulator's learning process $\mathcal{G}$ comprises of iid signals resembling the Wald problem \citep{wald1947foundations}. Indeed, this is in essence what the FDA already does for clinical trials in which a drug of uncertain safety and efficacy is rolled out to a progressively larger number of participants, with the option for either party to halt experimentation: \emph{within} each phase, the participating company might voluntarily choose to halt the trial; \emph{at the boundary} of each phase, the FDA might stop the trial from proceeding. This practice mirrors our adaptive sandbox quite closely\footnote{See also \cite{mcclellan2022experimentation} for a distinct analysis of `approval mechanisms' that can be viewed as a once-and-for-all decision on whether a drug is approved---in which case it can be deployed limitlessly.
} 
and for such transparently learnable technologies, the challenge regulators face is to monitor the agent's experimentation such as to close the gap between $\mathcal{G}$ and $\mathcal{F}$ e.g., through enforcing the reporting of clinical trials, pre-analysis plans, and detecting fraud.\footnote{There has been complaints that regulatory agencies like the FDA and FAA have been too conservative. We emphasize that our framework and results are consistent with this view since they are about the form of instrument choice---dynamic quantity limits---rather than the specifics around where these limits should be.} 

Other technologies are \emph{opaque}, beset by difficulties not just around learning about the technology, but also uncertainty about the learning process itself. We think transformative technologies like artificial intelligence often fall into this category---what, if anything, will the next generation of models teach us about whether powerful AI systems further out into the future can be aligned to human values? What kinds of biosecurity or labor market risks will such models pose? For such technologies, regulators might not learn \emph{by default} as they are developed, but might nonetheless invest in \emph{active learning} i.e., choosing $\mathcal{G}$ at some cost.  Indeed, computer scientists have recently advocated for these kinds of evidence-seeking AI policy that ``proactively helps to produce more information.'' \citep*{casper2025pitfalls}. But this raises further questions: what kinds of information should regulators acquire, and how much resources should they expand to do so? Our framework is able to speak directly to such questions since our analysis can be viewed as the `inner problem' that fixes the principal's learning process $\mathcal{G}$ and optimizes over $\mathcal{G}$-adapted mechanisms. This paves the way to analyze the `outer problem' of how $\mathcal{G}$ should be chosen subject to constraints and/or costs---a rich toolkit from dynamic information acquisition \citep{moscarini2001optimal,zhong2022optimal,hebert2023rational,sannikov2024exploration} can be drawn upon here.

\textbf{How much are we giving up for robustness?} A natural worry is that robustness demands \emph{too much}---that is, the payoff difference between robust and Bayes-optimal mechanisms can be large and, if so, our adaptive sandbox might be excessively stringent. We resist this claim for two distinct reasons.  

If regulator learning is fast i.e., $\mathcal{G}$ is informative about the technology, then the difference between robustness and Bayes-optimal mechanisms must be small since the value of the latter is upper-bounded by the regulator directly observing the agent's learning $\mathcal{F}$ which, in turn, is upper-bounded by instantaneously learning the truth. In such cases, the adaptivity of our sandbox paired with fast regulator learning---e.g., for what we have called transparently learnable technologies---ensures that we are not giving up too much. 

If regulator learning is slow i.e., $\mathcal{G}$ is not very informative about the technology, then as we have seen from \cref{sec:importance}, non-robust mechanisms are susceptible to worst-case learning processes that are both natural and generate unboundedly poor regulator payoffs. Thus the value of Bayes-optimal mechanisms\footnote{For a sufficiently `rich' prior over $\mathbb{F} \times \mathbb{U}$; see \cref{prop:Bayesian}.} is now upper-bounded by the performance of mechanisms that impose hard limits on the technology (see \cref{prop:Bayesian}) so the payoff gap between robustness and Bayes-optimal mechanisms once again cannot be too large. 

\textbf{Extensions and generalizations.} 
We have deliberately worked within the simplest model in which equilibrium technology choices are jointly shaped by learning, preferences, and policy. But there is quite a lot we can add to enrich the model along various dimensions---we outline some possibilities here and collect details in Online Appendix \ref{app:ext}. 
\begin{enumerate}[leftmargin =*]
    \item 
\textbf{Randomization and elicitation have zero value.} Our definition of mechanisms were those adapted to the principal's filtration which does not allow for additional randomization or type elicitation. Nonetheless, our regulator cannot improve upon her worst-case guarantee by doing either, precisely because under robust mechanisms, the worst-case learning process offers the agent no additional information beyond the principal's own. That our regulator cannot do better via randomization stands in contrast to work on robust pricing and contracting \citep{libgober2021informational}; that our regulator cannot do better via elicitation of type reports is reminiscent of \citep{kruse2019inverse}.
\item 
\textbf{Revenue motives.} Suppose that the principal is interested in raising revenue i.e., we are in a (partially or fully) transferable utility environment e.g., the regulator's payoff from the technology until level $l$ in state $\theta$ is: $v(\theta, l) + \lambda \cdot \phi_l$ for some $\lambda \geq 0$. An adaptive sandbox that solves the experimentation problem for a \emph{weighted sum} of the principal's and agent's payoffs 
\[\sup_{\el} \, \Ex\Big[V(\mu_\el,\el) + \lambda \cdot U(\mu_\el,\el)\Big] \quad \text{s.t. $\el$ is $\mathcal{G}$-adapted}
\]
augmented with a carefully chosen participation tax is learning-robust. 
\item \textbf{Endogenous agent learning.} Suppose that the agent's learning process $\mathcal{F}$ is flexibly chosen at some cost which is increasing in the flow variation of the belief process \citep{moscarini2001optimal,zhong2022optimal,hebert2023rational}. When the principal is uncertain about how difficult it is to acquire information i.e., about the agent's cost function and/or variation measure, the adaptive sandbox $\phi^*$ remains robustly optimal.
\item \textbf{Behavioral robustness.} If the regulator does not learn, the sandbox $\phi^*$ delivers optimal worst-case guarantees whenever the agent employs any \emph{undominated stopping rule}. Undominated stopping rules impose minimal assumptions on the agent's rationality by \emph{only} ruling out stopping at a given pair of beliefs and technology levels $(\mu,l)$ when more technology is strictly better in the absence of further information. These include stopping levels which are optimal, but also allows for agents to be misspecified or ambiguity averse about the future learning process, misoptimize, or be myopic. 
\item \textbf{Technology shapes the state.} Suppose that the state $\theta_l \in \Theta$ evolves as an absorbing Markov process where it transitions from $0$ to $1$ at (inhomogenous) Poisson rate $\lambda_l \geq 0$, and state $1$ is absorbing. This might correspond to a breakthrough in the technology that could make a technology safe. The sandbox mechanism $\phi^*$ remains robustly optimal. In particular, the regulator's beliefs about the state $\Ex[\theta \mid \mathcal{G}_l]$ exhibits a positive drift even in the absence of any information, and the sandbox limit $\overline{\el}$ takes this into account by loosening accordingly.
\item  
\begin{minipage}[t]{0.5\linewidth}  
\textbf{Multidimensional Technology.} Suppose that our technology space $\mathcal{L}^{\mathsf{M}} \subseteq \mathbb{R}^d$ is now a compact subset of the $d-$dimensional reals, and continue to assume that uncertainty $\Theta$ is totally ordered i.e., a higher-state corresponds to higher payoffs for each $\bm{l} \in \mathcal{L}^{\mathsf{M}}$. If the regulator does not learn, a \emph{multidimensional sandbox} that imposes a zero tax on the technology, and a hard limit at a manifold $\mathcal{O}_* \subset \mathcal{L}^{\mathsf{M}}$ (see \cref{fig:multidim}) is robustly optimal, undominated, and time-consistent.  
\end{minipage}%
\hfill%
\begin{minipage}[t]{0.45\textwidth}\vspace{0pt}
\vspace{-1.7em}
\begin{figure}[H]
\centering
{\includegraphics[width=0.9\textwidth]{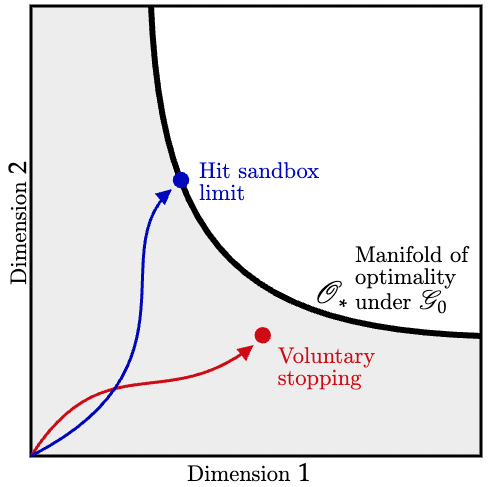}}
    \caption{Multidim. sandbox}\label{fig:multidim}
\end{figure}
\end{minipage}
\end{enumerate}

\section{Conclusion} \label{sec:conclude}

We have developed a parsimonious model in which learning, preferences, and policy jointly shape the equilibrium technology level. We showed that adaptive sandboxes mechanisms comprising of a zero marginal tax on the technology up to an evolving hard limit is robust to the agent's learning process and/or preferences (\cref{thrm:robustness}), undominated and dominant among robust and regular mechanisms (\cref{thrm:dominance}), and uniquely time-consistent among robust mechanisms (\cref{thrm:time-consistency}). Our results clarify when and why dynamic quantity limits on risky experimentation e.g., regulatory sandboxes and FDA clinical trials work well, as well as guidance for how they might be improved.



What do sandboxes safeguard against? We showed that they served as bulwarks against weak optimism: sans robust regulation, the agent's learning process is adversarially chosen to entice excessive risk-taking. This can magnify even small wedges in risk aversion to deliver unboundedly poor principal payoffs (\cref{thrm:worstcase}). The worst-case learning process is natural and arises when new technologies pose tail risks so learning is lumpy e.g., financial crises induced by synthetic instruments, pandemics induced by gain-of-function research, or AI catastrophes. Robustness is not always (or even often!) a good guide to policy choice. But we have showed that it is especially important in the context of technology regulation---this, we think, are reasons to take it seriously. 


Taken together, our results offer concrete guidance for policymakers who, in the coming years, will face difficult choices around how to regulate new and transformative technologies like artificial intelligence whose promises and perils loom large over human life. There are, of course, further questions around political feasibility, international coordination, and so on. These constraints might shrink the set of regulatory mechanisms, rendering our adaptive sandbox infeasible---what then? We think this is an important and exciting avenue for further work.

\setstretch{0.8}
\setlength{\bibsep}{0pt}
\bibliography{references}

\appendix
\crefalias{section}{appendix}

\normalsize 
\setstretch{1.15}

\begin{center}
    \large{\textbf{APPENDIX TO ROBUST TECHNOLOGY REGULATION}}\end{center}
\textbf{Organization.} The appendix is organized as follows. \cref{app:proofs} develops some mathematical preliminaries and proves results in \cref{sec:robustness} (\cref{lem:compstat,prop:dualrobust_characterization}, \cref{thrm:robustness,thrm:dominance,thrm:time-consistency}). \cref{appendix:proofs_worstcase} proves results in \cref{sec:importance} (\cref{thrm:worstcase}, \cref{prop:Bayesian,prop:worstcasecompstat}). 

\section{Proofs of results in Section \ref{sec:robustness}} \label{app:proofs}

\subsection{Mathematical preliminaries and notation} Filtrations can be delicate so it is worthwhile to flesh out some details. Our probability space is $\big( \Omega, (\mathcal{F}, \mathcal{G}), \mathbb{P} \big)$ where $\mathcal{F} := (\mathcal{F}_l)_{l \in \mathcal{L}}$ and $\mathcal{G} := (\mathcal{G}_l)_{l \in \mathcal{L}}$ filter $\Omega$. We will keep the space $\Omega$ abstract but emphasize that it is sufficiently rich to capture both randomness in what is learnt about the state (`first-order'), randomness in what is learnt about what will be learnt (`second-order'), as well as higher-order learning. $\mathcal{G} = (\mathcal{G}_l)_l$ is the natural filtration generated by some right-continuous signal process $(X^{\theta}_l)_l$. $\mathcal{F} = (\mathcal{F}_l)_l$ is any right-continuous filtration such that $\mathcal{F}_l \supseteq \mathcal{G}_l$ for each $l \in \mathcal{L}$. As in the main text, we denote all such filtrations with $\mathbb{F}$. 

\subsection{Comparative statics for stopping}

We start with the following lemma in which we compare the principal and agent's preferences over stopping levels (which need not be optimal). We will use it to show \cref{lem:compstat}, as well as some of our subsequent proofs that will rank the agent's and principal's preferences over stopping levels. 

\begin{lemma}\label{lem:compare_two_stoppingtimes}
    For any two $\mathcal{F}$-stopping levels $\ell^-$, $\ell^+$ such that $\ell^- \leq \ell^+$ a.s.: 
    \[
    \bigg\{ \, V(\mu_{\ell^-},\ell^{-}) \underset{(<)}{\leq} \Ex \big[V(\mu_{\ell^+},\ell^{+}) \big| \mathcal{F}_{\el^-} \big]   \, \bigg\} \subseteq 
    \bigg\{\, U(\mu_{\ell^-},\ell^{-}) \underset{(<)}{\leq} \Ex \big[U(\mu_{\ell^+},\ell^{+}) \big| \mathcal{F}_{\el^-} \big] \, \bigg\}.
    \]
\end{lemma}

\begin{proof}[Proof of \cref{lem:compare_two_stoppingtimes}] We proceed in several steps. 

\underline{Step 1: Lower-bound value of continuing to $\ell^+$ under $U$ and $\mathcal{F}_{\ell^-}$.} \\
    Consider that 
    \begin{align}
    \Ex\Big[U(\mu_{\ell^+},\ell^+) \Big|\mathcal{F}_{\ell^-} \Big] &= \Ex\Big[\mu_{\ell^+} \cdot U(1,\ell^+) \Big|\mathcal{F}_{\ell^-} \Big] + \Ex\Big[(1-\mu_{\ell^+}) \cdot  U(0,\ell^+) \Big|\mathcal{F}_{\ell^-}\Big] \nonumber\\
    &= \Ex\Big[\mu_{\ell^+} \cdot g \circ V(1,\ell^+) \Big|\mathcal{F}_{\ell^-}\Big]  + \Ex\Big[(1-\mu_{\ell^+}) \cdot g \circ V(0,\ell^+) \Big|\mathcal{F}_{\ell^-}\Big] \nonumber  \\
    &\geq \mu_{\ell^-} \cdot g \bigg( \underbrace{\Ex\bigg[ \frac{\mu_{\ell^+}}{\mu_{\ell^-}} \cdot V(1,\ell^+) \Big\lvert \mathcal{F}_{\ell^-} \bigg]}_{\eqqcolon \overline{V}_1}\bigg) \nonumber \\
    &\quad + 
    (1-\mu_{\ell^-}) \cdot g \bigg(\underbrace{\Ex\bigg[ \frac{1-\mu_{\ell^+}}{1-\mu_{\ell^-}} \cdot V(0,\ell^+) \Big\lvert \mathcal{F}_{\ell^-} \bigg]}_{\eqqcolon \overline{V}_0}\bigg), \label{eqn: lower bound of U}
    \end{align}
        where the last inequality is from a version of Jensen's inequality for random measures e.g., for the inequality in the first term we have: 
    \begin{align*}
        &\Ex\Big[\mu_{\ell^+} \cdot g \circ V(1,\ell^+) \Big| \mathcal{F}_{\ell^-}\Big] \\
        &= \int_{\omega \in \Omega} \mu_{l^+} \cdot g\big( V(1,\el^+)\big) d{\mathbb{P}}(\cdot|\mathcal{F}_{\ell^-})  \\ 
        &= \int\mu_{\el^+} d{\mathbb{P}}(\cdot|\mathcal{F}_{\ell^-}) \cdot \bigg(\int_{\omega \in \Omega}  g\big( V(1,\ell^+)\big) \cdot \cfrac{\mu_{\ell^+}}{\int \mu_{\ell^+} d{\mathbb{P}}(\cdot|\mathcal{F}_{\ell^-})} d{\mathbb{P}}(\cdot|\mathcal{F}_{\ell^-})\bigg)\tag{Random transform of measure}
        \\
        &\geq \mu_{\ell^-} g\Big( {\Ex}\Big[\frac{\mu_{\ell^+}}{\mu_{\ell^-}} V(1,\ell^+) \Big| \mathcal{F}_{\ell^+}\Big] \Big) \tag{Jensen + Optional Stopping + $\ell^+ > \ell^-$} 
    \end{align*}
    and an analogous argument applies to the second term. Since $\ell^+ \geq \ell^-$ and $V(1,\cdot)$ is increasing, we have $V(1,\ell^+) \geq V(1,\ell^-)$ while $V(0,\cdot)$ is decreasing so $V(0,\ell^+) \leq V(0,\ell^-)$. Thus,
    \begin{align*}
        \overline V_1 := {\Ex}\bigg[ \frac{\mu_{\ell^+}}{\mu_{\ell^-}}V(1,\ell^+) \Big\lvert \mathcal{F}_{\ell^-} \bigg] 
        &\geq {\Ex}\bigg[ \frac{\mu_{\ell^+}}{\mu_{\ell^-}}V(1,\ell^-) \Big\lvert \mathcal{F}_{\ell^-}  \bigg] \tag{$V(1,\cdot)$ increasing}\\
        &=\frac{V(1,\ell^-)}{\mu_{\ell^-}} {\Ex}[\mu_{\ell^+} | \mathcal{F}_{\ell^-}] \\
            &= V(1,\ell^-). \tag{Optional Stopping \& $\{\ell^+ > \ell^-\}$}
    \end{align*}
    By a similar argument, $\overline{V}_0 \leq V(0, \ell^-)$, and since $V(1,l) \geq V(0,l)$ we have $\overline{V}_1 \geq \overline{V}_0$. \\~\\
\underline{Step 2: Upper-bound value of stopping at $\ell^-$ under $U$.}  \\
Define the following $\mathcal{F}_{\el^-}$-measurable event
\begin{align*}
    A \coloneqq \bigg\{ \omega \in \Omega: V(\mu_{\ell^-},\ell^{-}) \leq \Ex \big[V(\mu_{\ell^+},\ell^{+}) \big| \mathcal{F}_{\el^-} \big] \bigg\}
\end{align*}
which is the event that at $\mathcal{F}_{\ell^-}$ with preference $V$ it is optimal to stop.  

The inequality constraint within the event $A$ can be equivalently written in terms of $\overline{V}_0$ and $\overline{V}_1$ as follows:
\begin{align*} 
    \mu_{\ell^-} \cdot \overline{V}_1 + (1-\mu_{\ell^-})\cdot \overline{V}_0 \geq \mu_{\ell^-}\cdot V(1,\ell^-) + (1-\mu_{\ell^-}) \cdot V(0,\ell^-).
\end{align*}
Since we have already argued from Step 1 that $\overline{V}_1 \geq V(1,\ell^-)$ and $\overline{V}_0 \leq V(0,\ell^-)$, the above inequality implies that under the event $A$ 
\[
\Big(\mu_{\ell^-} \cdot \overline{V}_1, (1-\mu_{\ell^-})\cdot \overline{V}_0\Big) \quad \text{majorizes}  \quad \Big(\mu_{\ell^-} \cdot V(1,\ell^-),(1-\mu_{\ell^-}) \cdot V(0,\ell^-)\Big).
\]
The weighted Karamata inequality then implies, under the event $A$
\begin{align} 
    \mu_{\ell^-} \cdot g(\overline{V}_1) + (1-\mu_{\ell^-})\cdot g(\overline{V}_0) & \geq \mu_{\ell^-}\cdot g \circ V(1,\ell^-) + (1-\mu_{\ell^-})\cdot g \circ V (0,\ell^-) \nonumber \\
    &= U(\mu_{\ell^-},\ell^-). \label{eqn: upper bound of U}
\end{align} 
\\~\\
\underline{Step 3: Putting things together} \\
Equations \eqref{eqn: lower bound of U} and \eqref{eqn: upper bound of U} imply that under the event $A$
\begin{align*}
    \Ex\Big[U(\mu_{\ell^+},\ell^+) \Big|\mathcal{F}_{\ell^-} \Big] &\geq \mu_{\ell^-} \cdot g(\overline{V}_1) + (1-\mu_{\ell^-})\cdot g(\overline{V}_0) \tag{\cref{eqn: lower bound of U}}
    \\
    &\geq U(\mu_{\ell^-},\ell^-) \tag{\cref{eqn: upper bound of U}}
\end{align*}
so at $\mathcal{F}_{\ell^-}$ with preference $U$ it is also optimal to stop, as required. The proof for the strict inequality is exactly the same. 
\end{proof}

We now show \cref{lem:compstat}. 

\begin{proof}[Proof of \cref{lem:compstat}]
Fix the learning process $\mathcal{F} \in \mathbb{F}$, $\mathcal{F}$-stopping levels $\ell_1 \leq \ell_2$. For the indirect utilities $U,V: \Delta(\Theta) \times \mathcal{L} \to \mathbb{R}$ let $\ell_U$ and $\ell_V$ be any selection of optimal stopping levels that solve 
\[
\begin{array}[t]{l}
     \sup_{s \in [\ell_1,\ell_2]} \Ex\big[U(\mu_s,s)| \mathcal{F}_{\ell_1}\big] \\
    \text{s.t. $s$ is $\mathcal{F}$-adapted}
    \end{array} \quad \text{and} \quad 
    \begin{array}[t]{l}
     \sup_{s \in [\ell_1 ,\ell_2]} \Ex\big[V(\mu_s,s)| \mathcal{F}_{\ell_1}\big] \\
    \text{s.t. $s$ is $\mathcal{F}$-adapted}
    \end{array} 
\]
respectively. We will show that $\el^{\vee} := \el_U \vee \el_V$ and $\el^{\wedge} \coloneqq \el_U \wedge \el_V$ also solve the first and second problems respectively. 

    Define the event $\widetilde{\Omega} \coloneqq\{\omega \in \Omega: \ell_V > \ell_U\}$.  If $\Pr(\widetilde{\Omega}) = 0$ there is nothing to show since $\ell_V \leq \ell_U$ a.s. so we suppose $\Pr(\widetilde{\Omega}) > 0$.

    For any $\mathcal{F}_{\ell_U}$-measurable event $A \subset \widetilde{\Omega}$, consider the stopping time \[\ell'_A = \begin{cases}
        \ell_U &\text{if $\omega \in A$} \\ \ell_V &\text{otherwise.}
    \end{cases} \]
    This is indeed a stopping time because $\ell_U < \ell_V$ under the event $A \subset \widetilde{\Omega}$. The optimality of $\ell_V$ implies
    \begin{align}
    &\Ex\Big[V(\mu_{\ell_V},\ell_V) | \mathcal{F}_{\ell_1}\Big] 
    \geq {\Ex}\Big[V(\mu_{\ell'_A }, \ell'_A) | \mathcal{F}_{\ell_1} \Big] \nonumber \\
    &\Longrightarrow \quad {\Ex} \Big[ V(\mu_{\ell_V},\ell_V)1\{\omega \in A\}  \Big]\geq {\Ex} \Big[ V(\mu_{\ell_U},\ell_U)1\{\omega \in A\} \Big]  \nonumber \\
        &\Longrightarrow \quad  \Ex\Big[ V(\mu_{\ell_V},\ell_V)  \Big| \mathcal{F}_{\ell_U}\Big] \geq V(\mu_{\ell_U},\ell_U) \quad \text{a.s. under the event $\widetilde{\Omega}$}    \label{eqn: V(l_V) > l_U}
    \end{align}
    From \cref{lem:compare_two_stoppingtimes}, since $U = g \circ V$ we have the same under preference $U$: 
    \begin{align}
        \Ex\Big[ U(\mu_{\ell_V},\ell_V)  \Big| \mathcal{F}_{\ell_U}\Big] \geq U(\mu_{\ell_U},\ell_U)  \quad \text{a.s. under the event $\widetilde{\Omega}$.}
        \label{eqn: U(l_V) > l_U}
    \end{align}

    To finish, we will show that (i) the stopping time $\ell^{\vee} := \el_U \vee \el_V$ yields an identical expected value as $\ell_{V}$ for preference $V$ under $\mathcal{F}_{\ell_1}$ which, recall, is the coarsest filtration over our space $[\ell_1,\ell_2]$; and (ii) the stopping time $\ell^{\wedge} := \el_U \wedge \el_V$ yields the same expected value as $\ell_{U}$ for preference $U$ under $\mathcal{F}_{\ell_1}$.

    \cref{eqn: U(l_V) > l_U} implies
    \begin{align*}
        \Ex\Big[U(\mu_{\ell_V},\ell_V) 1\{\omega \in \widetilde{\Omega}\} \Big| \mathcal{F}_{\ell_1}\Big] &= \Ex\Big[\Ex\big[U(\mu_{\ell_V},\ell_V) \cdot 1\{\omega \in \widetilde{\Omega}\} \, \big| \, \mathcal{F}_{\ell_U} \big] \, \Big|  \, \mathcal{F}_{\ell_1} \Big] \tag{Iterated expectation} \\
        &= \Ex\Big[1\{\omega \in \widetilde{\Omega}\} \cdot \Ex\big[U(\mu_{\ell_V},\ell_V) \, \big|  \,\mathcal{F}_{\ell_U} \big] \, \Big|\, \mathcal{F}_{\ell_1} \Big] \\
        &\geq \Ex\Big[ U(\mu_{\el_U},\el_U) \cdot 1\{\omega \in \widetilde{\Omega}\} \, \Big| \, \mathcal{F}_{\ell_1}  \Big] \tag{\cref{eqn: U(l_V) > l_U}}.
    \end{align*}
    Thus,
    \begin{align*}
        0 &\leq \Ex\Big[U(\mu_{\ell_U},\ell_U) \Big| \mathcal{F}_{\ell_1}\Big] - \Ex\Big[U(\mu_{\ell^\vee},\ell^\vee) \Big|\mathcal{F}_{\ell_1}\Big] \\
        &= \Ex\Big[U(\mu_{\ell_U},\ell_U) 1\{\omega \in \widetilde{\Omega}\} \Big| \mathcal{F}_{\ell_1}\Big] - \Ex\Big[U(\mu_{\ell_V},\ell_V) 1\{\omega \in \widetilde{\Omega}\} \Big|\mathcal{F}_{\ell_1}\Big] \\
        &\leq 0,
    \end{align*}
    which implies all the above inequalities, including \cref{eqn: U(l_V) > l_U}, must be equalities. Thus,
    \begin{align*}
        \Ex\Big[U(\mu_{\ell_U},\ell_U) \Big| \mathcal{F}_{\ell_1}\Big] = \Ex\Big[U(\mu_{\ell^\vee},\ell^\vee) \Big|\mathcal{F}_{\ell_1}\Big],  
    \end{align*}
    implying $\ell^{\vee}$ solves the optimal stopping problem $\sup_{s \in [\el_1,\el_2]} \Ex[V(\mu_s,s) | \mathcal{F}_{\el_1}]$. 
    
    It remains to show $\ell^{\wedge}$ yields the same expected value as $\ell_U$ for preference $U$ under $\mathcal{F}_{\ell_1}$. Since \cref{eqn: U(l_V) > l_U} must be an equality:
    \begin{align*}
        \Ex\Big[ U(\mu_{\ell_V},\ell_V)  \Big| \mathcal{F}_{\ell_U}\Big] = U(\mu_{\ell_U},\ell_U) \quad \text{a.s. under the event $\widetilde{\Omega}$.}
    \end{align*}
    Now observe we have 
    \[
    \Ex\Big[ V(\mu_{\ell_V},\ell_V)  \Big| \mathcal{F}_{\ell_U}\Big]  \underset{\cref{lem:compare_two_stoppingtimes}}{\overset{\text{\cref{eqn: V(l_V) > l_U}}}{\gtreqless}}  V(\mu_{\ell_U},\ell_U) \quad \text{a.s. under the event $\widetilde{\Omega}$}.
    \]
    so this must be an equality. Hence we have 
    \begin{align*}
        \Ex\Big[ V(\mu_{\ell_V},\ell_V)  \Big| \mathcal{F}_{\ell_U}\Big] = V(\mu_{\ell_U},\ell_U) \quad \text{a.s. under the event $\widetilde{\Omega}$.}
    \end{align*}
    This implies (once again using the fact that $\mathcal{F}_{\ell_1} \subseteq \mathcal{F}_{\ell_U}$ and the law of iterated expectation): 
    \begin{align*}
     \Ex\Big[V(\mu_{\ell_U},\ell_U) 1\{\omega \in \widetilde{\Omega}\} \Big| \mathcal{F}_{\ell_1}\Big] &= \Ex\Big[V(\mu_{\ell_V},\ell_V) 1\{\omega \in \widetilde{\Omega}\} \Big|\mathcal{F}_{\ell_1}\Big] \\
     \implies 
        \Ex\Big[V(\mu_{\ell_V},\ell_V) \Big| \mathcal{F}_{\ell_1}\Big] &= \
        \Ex\Big[V(\mu_{\ell_V},\ell_V) 1(\omega \in \widetilde{\Omega}) + V(\mu_{\ell_V},\ell_V) 1(\omega \notin \widetilde{\Omega})  \Big| \mathcal{F}_{\ell_1}\Big] 
        \\
        &=\Ex\Big[V(\mu_{\ell_U},\ell_U) 1(\omega \in \widetilde{\Omega}) + V(\mu_{\ell_V},\ell_V) 1(\underbrace{\omega \notin \widetilde{\Omega}}_{\ell_V < \ell_U}
        )  \Big| \mathcal{F}_{\ell_1}\Big]  \\
        &=
        \Ex\Big[V(\mu_{\ell^\wedge},\ell^\wedge) \Big|\mathcal{F}_{\ell_1}\Big],  
    \end{align*}
    implying $\ell^{\wedge}$ solves the optimal stopping problem $\sup_{s \in [\el_1,\el_2]} \Ex[V(\mu_s,s) | \mathcal{F}_{\el_1}]$, as required.
\end{proof}

Inspecting the proof yields the same result for any set of states on which payoffs can be totally ordered by replacing the binary state with a vector and invoking the same Schur convexity argument. In Online Appendix \ref{app:experimentation discussion} we show that this nests and generalizes the main results of \cite{keller2019note,chancelier2009risk}.

\subsection{Proof of \cref{thrm:robustness}} 
\begin{proof}
Recall that in the main text we defined the adaptive sandbox mechanism as a $\overline{\mathbb{R}}$-valued stochastic process $\phi^{*}$ which is adapted to the principal's filtration $\mathcal{G}$. 

\underline{\smash{Step 1. $\phi^*$ achieves the payoff guarantee under the learning process $\mathcal{F} = \mathcal{G}$.}} \\
Observe 
\begin{align*}
    \eqref{eqn:ADV} &:= \sup_{\phi \in \Phi}\inf_{\mathcal{F} \in \mathbb{F}} \Ex\Big[V\big(\mu_{\el^*},\el^*(\phi,\mathcal{G},U)\big)\Big] 
    \\
    &\leq \sup_{\phi}\Ex\Big[V\big(\mu_{\el^*},\el^*(\phi,\mathcal{G},U)\big)\Big]  
    \\
    &\leq \sup_s \Ex\Big[ V\big(\mu_s,s\big) \Big] \quad \text{s.t. $s$ is $\mathcal{G}$-adapted} 
    \tag{$\el^*$ is $\mathcal{G}$-adapted}
    \\
    &= \Ex\Big[V\big(\mu_{\overline{\el}}, \overline{\el}\big)\Big] \tag{defn. of $\overline{\el}$}
\end{align*}

\underline{Step 2. No additional info is the worst-case under $\phi^*$.} Take any $(\mathcal{F},U) \in \mathbb{F} \times \mathbb{U}$. Notice that at filtration $\mathcal{F}_{\el^*}$ with $\el^* < \overline{\el}$ we have 
\begin{align*} 
 \Ex\big[U(\mu_{\el^*},\el^*) \big| \mathcal{F}_{\el^*}\big] &\geq
 \begin{array}[t]{l}
     \sup_{s \geq \el^*} \Ex\big[U(\mu_s,s) - \phi^*_s \big| \mathcal{F}_{\el^*}\big] \\
    \text{s.t. $s$ is $\mathcal{F}$-adapted}
    \end{array} \tag{$l^*$ solves \eqref{prob:OSP}}
    \\
    &= 
    \begin{array}[t]{l}
     \sup_{\overline{\el} \geq s \geq \el^*} \Ex\big[U(\mu_s,s)\big| \mathcal{F}_{\el^*}\big] \\
    \text{s.t. $s$ is $\mathcal{F}$-adapted}
    \end{array} \tag{defn. of adaptive sandbox}
\end{align*}
and from \cref{lem:compstat} this implies that at $\mathcal{F}_{\el^*}$
\begin{align*}
 \Ex\big[V(\mu_{\el^*},\el^*) \big| \mathcal{F}_{\el^*}\big] = 
  \begin{array}[t]{l}
     \sup_{\overline{\el} \geq s \geq \el^*} \Ex\big[V(\mu_s,s)\big| \mathcal{F}_{\el^*}\big] \\
    \text{s.t. $s$ is $\mathcal{F}$-adapted}
    \end{array} 
\end{align*}
because, for any stopping time $s$ that solves the optimal stopping problem in the RHS of the above equality, $\el^* = \el^* \wedge s$ must also solve the same stopping problem. 

Now denoting $\gamma\in \Delta(\mathcal{L})$ as the probability distribution i.e., some Borel measure (of $\el^*(\phi^*,\mathcal{F},U)$), we thus have: 
\begin{align*}
     \Ex\Big[V\big(\mu_{\el^*},\el^*(\phi^*,\mathcal{F},U)\big)  \Big| \big\{ \el^* < \overline{\el}\big\} \Big]   &= 
     \int_{\el^* < \overline{\el}} 
     \begin{array}[t]{l}
     \sup_{\overline{\el} \geq s > \el^*} \Ex\big[V(\mu_s,s)\big| \mathcal{F}_{\el^*}\big] \\
    \text{s.t. $s$ is $\mathcal{F}$-adapted}
    \end{array}
    d\gamma(\el^*) \\
    &\geq 
     \int_{\el^* < \overline{\el}} 
     \begin{array}[t]{l}
     \sup_{\overline{\el} \geq s > \el^*} \Ex\big[V(\mu_\el,\el)\big| \mathcal{F}_{\el^*}\big] \\
    \text{s.t. $s$ is $\mathcal{G}$-adapted}
    \end{array}
    d\gamma(\el^*) \\
    &\geq 
    \Ex\big[V(\mu_{\overline{\el}},\overline{\el})  \big| \big\{ \el^* < \overline{\el}\big\} \big]. 
\end{align*}
From the law of total expectation, 
\begin{align*}
    \Ex\Big[V(\mu_{\el^*},\el^*(\phi, \mathcal{F},U))\Big] &= \mathbb{P}\big(\big\{\el^* = \overline{\el}\big\}\big) \cdot  \Ex\Big[V\big(\mu_{l^*},\el^*(\phi^*,\mathcal{F},U)\big)  \Big|  \big\{\el^* = \overline{\el}\big\} \Big] \\
    &\quad + \mathbb{P}\big(\big\{\el^* < \overline{\el}\big\}\big) \cdot  \Ex\Big[V\big(\mu_{\el^*},\el^*(\phi^*,\mathcal{F},U)\big)  \Big| \big\{ \el^* < \overline{\el}\big\} \Big] 
\end{align*}
hence 
\[
\Ex\Big[V(\mu_{\el^*},\el^*(\phi, \mathcal{F},U))\Big] \geq \Ex\Big[V\big(\mu_{\overline{\el}},\overline{\el}\big)\Big]. 
\]

To conclude, observe 
\begin{align*}
    \inf_{\mathcal{F} \in \mathbb{F}} \Ex\Big[V\Big(\mu_{\el^*},\el^*(\phi^*,\mathcal{F},U)\Big)\Big] &\geq \Ex\Big[V(\mu_{\overline{\el}},\overline{\el})\Big]  \tag{Step 2} \\
    &\geq \eqref{eqn:ADV} \tag{Step 1}
\end{align*}
i.e., $\phi^*$ solves \eqref{eqn:ADV}. Since $\phi^*$ does not depend on $U$, it also solves \eqref{eqn:DR}. 
\end{proof}

\begin{figure}[H]
\centering 
\vspace{-0.5em}
{\includegraphics[width=0.55\textwidth]{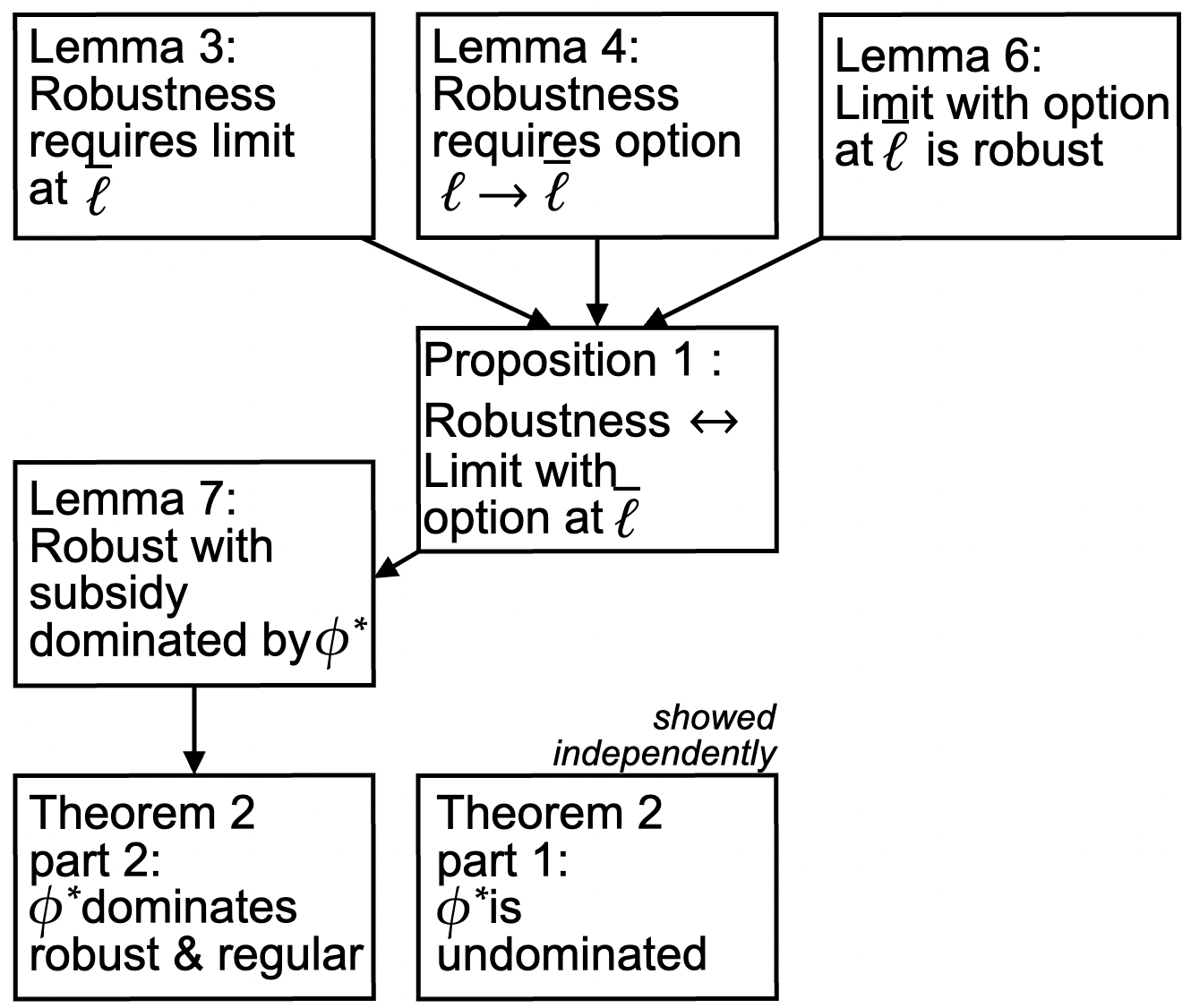}}
\caption{Outline of proof steps for \cref{thrm:dominance}}\label{fig:T2_proof_outline}
\end{figure}

\textbf{Outline of proof of \cref{prop:dualrobust_characterization} and \cref{thrm:dominance}.} We proceed in several steps. \cref{lem: no floor mechanism} argues that `floor mechanisms' (those that impose an infinite subsidy) are either not dually-robust, or are equivalent to a quota. Next, \cref{lemma:dual_hard_deadline} shows a dually robust mechanism must have a hard limit at $\overline{\el}$. \cref{lemma:zero_marginal} shows a dually robust mechanism must have zero marginal transfer until the hard limit. These two parts together imply the forward direction of \cref{prop:dualrobust_characterization}. Finally, \cref{lemma:converse_dual} shows the converse of \cref{prop:dualrobust_characterization}. Then, \cref{prop:dualrobust_characterization} is used to show $phi^*$ dominates robust and regular mechanisms; that $\phi^*$ is undominated is showed separately. \cref{fig:T2_proof_outline} illustrates. 

\subsection{Proof of \cref{prop:dualrobust_characterization}} We prove the complete characterization of dually-robust mechanisms given in the main text. Denote the set of dually-robust adaptive mechanisms with  
\[
    \Phi^* \coloneqq \Big\{\phi \in \Phi: \phi \text{ solves \eqref{eqn:DR}}\Big\}.
\]

We first show that it if a dually-robust mechanism delivers an infinite subsidy with positive probability, it must do so only at $\bar \ell$: 

\begin{lemma} \label{lem: no floor mechanism} If $\phi \in \Phi^*$ and there exists a $\mathcal{G}-$stopping level $s$ such that such that $\phi(s) = -\infty$ with positive probability, then $s = \overline{\el}$ almost surely. 
\end{lemma}
\begin{proof} [Proof of \cref{lem: no floor mechanism}]
     Consider the agent's filtration $(\mathcal{G}_l)_l = (\mathcal{F}_l)_l$ and the agent's utility function $U=V$. Under the mechanism $\phi$, the latest stopping level $s_0$ is an optimal stopping level for the agent because it yields the positively infinite utility. However, since the principal's optimal stopping is unique and $s_0$ is adapted to the filtration $(\mathcal{G}_l)_l,$ we must have $s_0 = \overline{\el}$ almost surely.

     Thus, under $\phi \in \Phi^*$, $\overline{\el}$ is always the agent's optimal stopping no matter what his filtration is. This implies the principal's value under $\phi$ always equals $\Ex[V(\theta,\overline{\el})]$ regardless of the agent's filtration. 
     
     To show that $\phi$ is dominated by the sandbox mechanism, it suffices to find the agent's filtration that yields strictly better the principal's outcome than that under no further information and the sandbox mechanism $\phi^*$. Consider the agent's filtration $(\mathcal{G}_l)_l$ such that $\mathcal{G}_l = \mathcal{F}_l \vee \sigma(\theta)$, i.e., the agent fully learn the state, and the agent's utility function $U=V.$ If $\theta = 0$, the agent always stop at level $0$ under the sandbox mechanism. On the other hand, if $\theta = 1$, the agent always stops as late as possible, which is $\overline{\el}$, under the sandbox mechanism. Thus, the principal's expected utility is
    \begin{align*}
        \underbrace{(1-\mu_0)\Ex\big[V(0,0)\big]}_{\theta = 0} + \Ex\big[V(1,\overline{\el}) 1\{\theta  = 1\}\big]  &> \Ex\big[V(0,\overline{\el})1\{\theta = 0\}\big] + \Ex\big[V(1,\overline{\el})1\{\theta = 1\}\big] \\
        &= \Ex\big[V(\theta,\overline{\el})\big]
    \end{align*}
    which is the robustly optimal payoff for the principal, as desired.
\end{proof}

\cref{lem: no floor mechanism} implies that the only floor mechanism that is dually-robust is one that is equivalent to the following quota mechanism $\phi' \in \Phi^*$ such that $\phi'(s) = 0$ a.s. for all $s \neq \bar \ell$, and $\phi(\bar \ell) = -\infty$ a.s. Hence, we will focus on mechanisms for which $\phi(s) > -\infty$ a.s. for every $\mathcal{G}$-stopping level $s$. 

We now show every robustly optimal mechanism must have hard limit at $\overline{\el}$. We begin with a useful lemma.

\begin{lemma}[Necessity of limit]\label{lemma:dual_hard_deadline} 
Fix a dually-robust mechanism $\phi \in \Phi^{*}$. Let $s$ be a stopping level with respect to the filtration $\mathcal{G}$.  Then, $\Ex[\phi_s | \mathcal{G}_{\overline{\el}}] = + \infty $ under the event $\{\omega: s>\overline{\el}\}$ almost surely. 
    

\begin{proof}
     Suppose towards a contradiction that $\phi$ is dually-robust but there exists some $\mathcal{G}$-stopping level $s$ such that 
        \[\Pr\Big( \big\{\Ex[\phi_s\mid \mathcal{G}_{\overline{\el}}]< +\infty\big\} \cap \big\{s > \overline{\el} \big\}\Big)>0.\]
        Since $s > \overline{\el}$, we must have \[\Ex\big[V(0,\overline{\el}) - V(0,s) \big| \mathcal{G}_{\overline{\el}}\big] > 0\]
        almost surely under the event $\{\omega: s > \overline{\el}\}$. Since $\phi(\overline{\el}) > -\infty$ and $\Pr(\theta = 1| \mathcal{G}_{\overline{\el}}) > 0$ almost surely, there exist $\overline{M},\underline{M} > 0$ and $\underline{\mu},\delta > 0$ such that
        \begin{align*}
            \Pr\left( \Big\{\underbrace{\Ex[\phi_s\mid \mathcal{G}_{\overline{\el}}]< \overline{M} \Big\} \cap \Big\{\phi(\overline{\el}) > -\underline{M} \Big\} \cap \Big\{\mu_{\overline{\el}} \geq \underline{\mu}\Big\} \cap \Big\{\Ex[V(0,\overline{\el}) - V(0,s) | \mathcal{G}_{\overline{\el}}] > \delta \Big\} \cap \Big\{ s > \overline{\el}}_{\eqqcolon \Omega^*} \Big\}
            \right)>0.
        \end{align*}        
        Note that $\Omega^*$ is measurable with respect to $\mathcal{G}_{\el^*}.$
        
        We will construct the agent's preferences $U$ such that
        \begin{align*}
            \Ex\big[U(\mu_{s},s) - \phi_s \big| \mathcal{G}_{\overline{\el}}\big] > \Ex\big[ U(\mu_{\overline{\el}},\overline{\el}) - \phi_{\overline{\el}}\big| \mathcal{G}_{\overline{\el}}\big]
        \end{align*}
        under the event $\Omega^*$. Choose 
        \[U(1,l) = \alpha \beta \cdot V(1,l) \quad \text{and} \quad 
        U(0,l) = \beta \cdot V(0,l)
        \]
        where $\alpha$ and $\beta$ are constants to be determined later and $\alpha > 1$. 
        
        Now observe we can lower-bound the difference (sans mechanism) in the agent's payoff between stopping levels $s$ and $\overline \ell$: 
        \begin{align*}
            \Ex\big[& U(\mu_{s},s) - U(\mu_{\overline{\el}},\overline{\el}) \big| \mathcal{G}_{\overline{\el}}\big]  \\
            &= \Ex\big[U(\mu_{\overline{\el}},s) - U(\mu_{\overline{\el}},\overline{\el}) \big| \mathcal{G}_{\overline{\el}}\big] \tag{Optional stopping theorem} \\
            &= \mu_{\overline{\el}} \cdot \Ex\big[U(1,s) - U(1,\overline{\el})\big| \mathcal{G}_{\overline{\el}}\big] + (1-\mu_{\overline{\el}}) \cdot  \Ex\big[U(0,s) - U(0,\overline{\el}) \big| \mathcal{G}_{\overline{\el}} \big] \\
            &\geq \underline{\mu}\cdot \alpha\cdot \beta\cdot \Ex\big[V(1,s) - V(1,\overline{\el})\big| \mathcal{G}_{\overline{\el}}\big] + (1-\underline{\mu})\cdot \beta \cdot \Ex\big[V(0,s) - V(0,\overline{\el}) \big| \mathcal{G}_{\overline{\el}} \big] \\
            &= \beta\cdot \Ex\left[\underline{\mu}\alpha(V(1,s) - V(1,\overline{\el}) ) + (1-\underline{\mu}) \cdot \big(V(0,s) - V(0,\overline{\el})\big) \big| \mathcal{G}_{\overline{\el}}\right]
        \end{align*}
        Since $V$ is continuously differentiable, we have, for every $l_0 \ne l_1 \in \mathcal{L},$
        \begin{align*}
            \bigg| \frac{V(0,l_0) - V(0,l_1)}{V(1,l_0) - V(1,l_1)}\bigg| \leq \sup_{l} \frac{\partial V(0,l)/\partial l}{\partial V(1,l)/\partial l} \eqqcolon M.
        \end{align*}
        Note that $M$ is also well-defined when technology space is discrete because there are finite values of $l_0 \ne l_1$. Now choose
        \[\alpha = \max\Bigg\{\frac{2M(1-\underline{\mu})}{\underline{\mu}},1 \Bigg\} \quad \text{and} \quad\beta = \frac{\overline{M} + \underline{M}}{(1-\underline{\mu}) \delta}.\]
        Then, our above lower-bound simplifies further to: 
        \begin{align*}
            \Ex\big[U(\mu_{s},s) - U(\mu_{\overline{\el}},\overline{\el}) \big| \mathcal{G}_{\overline{\el}}\big] 
            &\geq \beta (1-\underline{\mu}) \cdot  \underbrace{\Ex\big[V(0,\overline{\el}) - V(0,s) \big| \mathcal{G}_{\overline{\el}}\big]}_{\geq \delta} \\
            &\geq \Ex\big[\phi_s \big| \mathcal{G}_{\overline{\el}}\big] -\phi_{\overline{\el}},
        \end{align*}
        as desired. 
        
        Now consider the agent's learning process $\mathcal{F} = \mathcal{G}$ (no extra information) and the preference $U$ we constructed above. Under the adaptive sandbox $\phi^*$, the agent's optimal stopping level is $\overline{\el}$ from our comparative static \cref{lem:compstat}. Under the mechanism $\phi$, however, $\overline{\el}$ is no longer an agent's optimal stopping because stopping at level $s > \overline \ell$ is strictly better than $\overline{\el}$ under the event $\Omega^*$.\footnote{Choosing to stop at time $s$ under the event $\Omega^*$ and stop at $\overline{\el}$ otherwise is still a valid stopping level because $\Omega^*$ is $\mathcal{G}_{\overline{\el}}$-measurable.} But since $\overline \ell$ was defined as the latest optimal stopping level that solved the principal's direct experimentation problem 
        \[
        \sup_{\ell} \Ex[V(\mu_\ell,\ell)] \quad \text{s.t. $\ell$ is $\mathcal{G}$-adapted}
        \]
        and $\mathbb{P}(\Omega^*) > 0$, by total probability the principal's payoff is strictly lower than its payoff guarantee so $\phi$ cannot be dually-robust. 
 \end{proof}
\end{lemma}

\cref{lemma:zero_marginal} shows that the any dually-robust mechanism $\phi$ must be such that for any $\mathcal{G}$-adapted stopping level $\ell$  for which $\ell >\bar \ell$ with positive probability, $\mathbb{P}(\phi_{\ell} = +\infty) > 0$. We now argue that this implies the same for any $\mathcal{F}$-adapted level. To see this, choose 
\[
\hat \ell := \inf \Big\{l > \bar \ell: \phi_{l} < +\infty \Big\} \wedge L
\]
and notice that this is $\mathcal{G}$-adapted. Now for any $\mathcal{F}$-adapted stopping level $\ell$, we claim
\[
\mathbb{P}(\phi_{\ell} = +\infty) \geq \mathbb{P}(\phi_{\hat \ell} = +\infty)
\]
because for each path of $\phi(\omega)$, if there exists some level $s> \bar \ell$ such that $\phi(\omega)_s < +\infty$ then $\phi(\omega)_{\hat \ell} < +\infty$ and if not, then $\phi(\omega)$ must be equal to infinity everywhere so  $\phi(\omega)_{\ell} = \phi(\omega)_{\hat \ell} = +\infty$. But since $\hat \ell$ is $\mathcal{G}$-adapted, $\mathbb{P}(\phi_{\hat \ell} = +\infty) > 0$ hence $\mathbb{P}(\phi_{\ell} = +\infty) > 0$. Hence, for any learning process $\mathcal{F}$, the agent never stops after $\bar \ell$ as formalized by the following corollary: 

\begin{corollary}
    For every $\phi \in \Phi^*$ and the agent's preference and filtration $(U,\mathcal{F})$, we must have $\ell^*(\phi, \mathcal{F}, U) \leq \overline{\el}$ almost surely.
\end{corollary}

Now we prove the transfer before the hard limit must be weakly higher that that at the hard limit for every dually robust mechanism.

\begin{lemma}[Necessity of option] \label{lemma:zero_marginal}
    For every $\phi \in \Phi^*$, we must have $\phi_\ell \geq \phi_{\overline{\el}}$ for any $\mathcal{G}$-stopping level $\el\leq \overline{\el}$ almost surely. 
\end{lemma}
\begin{proof}[Proof of \cref{lemma:zero_marginal}] We proceed in four steps. 

\underline{Step 1: Construct agent's learning process $\mathcal{F}^{\epsilon}$.}
Suppose towards a contradiction there exists a $\mathcal{G}$-stopping level $\ell_0$ such that the event
\[A^* \coloneqq \{ \omega \in \Omega: \phi_{\ell_0} < \phi_{\overline{\el}}\}\]
in which the agent faces a strictly positive subsidy between $\ell_0$ and $\overline \ell$ happens with positive probability. Define $A^*_1 := A \cap \{\omega: \theta = 1 \}$. This event must happen with strictly positive probability since $\Pr(\theta = 1 | \mathcal{G}_L) \in (0,1)$ almost surely. 

Let $\nu$ be Bernoulli random variable such that $\Pr(\nu = 1) = \epsilon$ which is is independent of the principal's learning $\mathcal{G}_L$. Define the random variable  
\[Z^\epsilon \coloneqq 1\{\omega \in A^*_1, \nu = 1\}.\]
Now construct the agent's filtration $(\mathcal{F}^\epsilon_l)_l$ as follows:
\begin{align*}
    \mathcal{F}^\epsilon_l \coloneqq \sigma\left( \underbrace{\Big\{ E \cap \{\omega: l < \ell_0\} : E \in \mathcal{G}_l \Big\}}_{\text{Same info before $\ell_0$}} \bigcup \underbrace{\Big\{ E \cap \{\omega: l \geq \ell_0, Z^\epsilon =z\}  : E \in \mathcal{G}_l, z \in \{0,1\}\Big\}}_{\text{Observe $Z^{\epsilon}$ after $\ell_0$}}  \right).
\end{align*}
In words, our construction does the following:
\begin{itemize}
    \item Before level $\el_0$, the agent learns the same information as the principal. 
    \item At and after level $\ell_0$, the agent observes the value of $Z^\epsilon$: 
    \begin{enumerate}
        \item Under event $(A^{*}_1)^c$, the agent still receives the same information as the principal does.
        \item Under event $A^*_1$, the agent learns the current transfer is strictly less than the transfer at the limit $\overline \ell$ but the technology state is good ($\theta = 1$) with probability $\epsilon$.
    \end{enumerate}
\end{itemize}
In particular, for every stopping level $\el$, if $\el < \el_0$, then $\mathcal{F}^{\epsilon}_\el = \mathcal{G}_\el$. On the other hand, if $\el \geq \el_0$, then $\mathcal{F}^{\el^*}_\el = \mathcal{G}_\el \vee \sigma(Z^\epsilon)$. The intuition why this filtration makes the principal worse off is the agent is disincentivized to continue pushing the technology under event $A^*_1$ even though the technology state is good because she knows that she will be taxed more in the future.

\underline{Step 2: Under $\mathcal{F}^{\epsilon}$, the agent stops after $\ell_0$.} Let $s^\epsilon_U \coloneqq \ell^*(\phi, \mathcal{F}^\epsilon_l,U)$ be the largest stopping level that solves the agent's optimal stopping problem under preference $U$ and filtration $\mathcal{F}^\epsilon$. We want to show that  $s^\epsilon_U \geq l_0$ almost surely. To do so, we will construct a few auxillary stopping levels and `thread' $s^{\epsilon}_U$ through them. 

Take \emph{any} stopping level $\tilde s$ solves the agent's optimal stopping problem under $U$ and filtration $\mathcal{F}^\epsilon_l$. Define 
    \begin{align*}
    s' \coloneqq \begin{cases}
        \overline{\el} &\text{if $\tilde s \geq \ell_0$}, \\ \tilde s &\text{if $\tilde s < \ell_0$},
    \end{cases}
    \quad \quad s'_1 \coloneqq \begin{cases}
        \tilde s &\text{if $\tilde s \geq \ell_0,$} \\
        \overline{\el} &\text{if $\tilde s < \ell_0$}.
    \end{cases}
    \end{align*}
    Note that $s'_1$ is adapted $\mathcal{F}^\epsilon$, and $s'$ is adapted to $\mathcal{G}$. Consider that
    \begin{align*}
        0 &\leq  \Ex[U^\phi(\theta,\tilde s)] - \Ex[U^\phi(\theta,s'_1)] = \Ex[U^\phi(\theta,s')] - \Ex[U^\phi(\theta,\overline{\el})], 
    \end{align*}
    where the first inequality follows from the optimality of $\tilde s$. This implies $\Ex[U^\phi(\theta,\overline{\el})] \leq \Ex [U^\phi(\theta,s')]$. 
    
    Next we will once again make the familiar observation that if $\phi \in \Phi^*$ then the agent's latest optimal stopping stopping level $\ell^*(\phi,\mathcal{G},U)$ under no extra information $(\mathcal{G} = \mathcal{F}$) must be equal to $\overline{\el}$ almost surely. To see this, notice that dual robustness of $\phi$ demands $\Ex[V(\mu_{\ell^*}, \ell^*(\phi,\mathcal{G},U))] \geq \Ex[V(\mu_{\overline{\el}} , \overline{\el})].$ However, but since $\ell^*(\phi,\mathcal{G},U)$ is $\mathcal{G}$-adapted so we must have $\ell^*(\phi,\mathcal{G},U) = \overline{\el}$ almost surely. That is, the agent must optimally stop at the principal's optimal stopping level under every robustly optimal mechanism with no extra information. 
    
    Since $s'$ is also adapted to $\mathcal{G}$, we must have $\Ex[U^\phi(\theta,\overline{\el})] = \Ex [U^\phi(\theta,s')]$, implying the above inequality must be an equality and so 
    \[
    \Ex[U^\phi(\theta,\tilde s)] = \Ex[U^\phi(\theta,s'_1)]
    \]
    i.e., $s'_1$ is also an optimal stopping level under $U$ and filtration $\mathcal{F}^\epsilon_l$. This implies $s^\epsilon_U \geq s'_1$ almost surely since $s^\epsilon_U$ was defined as the largest optimal stopping level. But since $s'_1 \geq \el_0$, we must have $s^\epsilon_U \geq \el_0$ almost surely, as desired.

\underline{Step 3: Decompose $s^{\epsilon}_U$.} Since we have showed that $s^\epsilon_U \geq \ell_0$ a.s., it must solve the following optimal stopping problem.
\begin{align*}
    \sup_{s \geq \el_0 } \Ex[U(\theta,s) - \phi_s] \quad \text{s.t. $s$ is adapted to the filtration $\mathcal{F}^\epsilon_l$}.
\end{align*}
We can decompose $s^{\epsilon}_U$ as follows: 
\begin{align*}
    s^\epsilon_U \coloneqq \begin{cases}
        s^1 &\text{if $Z^{\epsilon} = 1$} \\ s^\epsilon &\text{if $Z^{\epsilon} = 0$}
    \end{cases}
\end{align*}
almost surely, where $s^1$ is a $\mathcal{F}^{\epsilon}$-stopping level that solves 
\begin{align*}
    &\sup_{s \geq \el_0} \Ex\big[U^\phi(\theta,s) \cdot  1\{\omega \in A^*_1\}\big] \quad \text{s.t. $s$ is adapted to $\mathcal{G}_l$} \tag{OSP-1}\label{prob:OST-1} 
\end{align*}
and $s^{\epsilon}$ is a $\mathcal{F}^{\epsilon}$-stopping level that solves 
\[
\sup_{s \geq \el_0} \Ex\big[U^\phi(\theta,s) \cdot 1\{Z^\epsilon = 0\}\big]
\quad \text{s.t. $s$ is adapted to $\mathcal{G}_l$.} 
\tag{OSP-$\epsilon$}\label{prob:OST-0}
\]  
To see this, observe that
\begin{align*}
    \Ex[U^\phi(\theta, s)] 
    &= \Ex[U^\phi(\theta,s)| Z^{\epsilon} = 1] \Pr(Z^\epsilon = 1)+ \Ex[U^\phi(\theta,s) 1\{ Z^{\epsilon} = 0\}] \\
    &= \Ex[U^\phi(\theta,s) \cdot 1\{\omega \in A^*_1\}] \epsilon + \Ex[U^\phi(\theta,s) \cdot 1\{Z^{\epsilon} = 0\}].
\end{align*}
This implies, $s^\epsilon|Z^\epsilon = 1$ and $s^\epsilon|Z^{\epsilon} = 0$ must solve \eqref{prob:OST-1} and \eqref{prob:OST-0}, respectively. 

\underline{Step 4: The principal's payoff is strictly below dual-robust payoff guarantee.} 

Notice that the principal's payoff when the agent's preference is $U$ and her filtration is $\mathcal{F}^{\epsilon}$ can be written as:
\begin{align*}
    \Ex[V(\theta, s^\epsilon_U)] &= \Ex[V(\theta,s)1\{\omega \in A^*_1\}] \epsilon + \Ex[V(\theta,s) 1\{Z^{\epsilon} = 0\}] \\
    &\leq \underbrace{\Ex[V(\theta,s^1)1\{\omega \in A^*_1\}] \epsilon + \Phi(\epsilon)}_{\eqqcolon \Psi(\epsilon)},
\end{align*}
where
\begin{align*}
    \Phi(\epsilon) \coloneqq \sup_{s \geq \el_0} \Ex\big[V(\theta, s)\cdot 1\{Z^{\epsilon} = 0\}\big] \quad \text{s.t. $s$ is $\mathcal{G}$-adapted}.
\end{align*}

Our goal is to show that $\Psi'(0+) < 0 $ for some choice of the agent's utility function. We start by investigating the term $\Ex[V(\theta,s^1)1\{\omega \in A^*_1\}]$. We choose the agent's utility function as $U = \epsilon V$ such that
\begin{align*}
    \epsilon < \frac{\Ex\big[(\phi_{\overline{\el}} - \phi_{\el_0})1 \{\omega \in A^*_1\} \big]}{\Ex \big[(V(1,\overline{\el}) - V(1,\el_0) )1\{\omega \in A^*_1\}\big]}.
\end{align*}
This is well-defined because $\Pr(A^*_1) > 0$ so that both of the numerator and denominator are strictly positive. Consider that 
\begin{align*}
    \Ex[(U(1,\el_0) - \phi(\el_0))1\{\omega \in A^*_1\}] &= \Ex[(\epsilon V(1,\el_0) - \phi_{\el_0}) 1\{\omega \in A^*_1\}] \\
    &> \Ex[(\epsilon V(1,\overline{\el}) - \phi_{\overline{\el}})1\{\omega \in A^*_1\}].
\end{align*}

This implies choosing $\overline{\el}$ is suboptimal under the event $A^*_1$, so $s^1 \ne \overline{\el}$ positive probability. \cref{lemma:dual_hard_deadline} implies $s^1 \leq \overline{\el}$, where the inequality is strict with positive probability. Since $V(1, \cdot)$ is strictly increasing, we must have
\begin{align*}
    \Ex[V(\theta, s^1) 1\{ \omega \in A^*_1\}] &=  \Ex[V(1, s^1) 1\{ \omega \in A^*_1\}] \\
    &< \Ex [V(1,\overline{\el})1\{\omega \in A^*_1\}] \\
    &= \Ex [V(\theta,\overline{\el})1\{\omega \in A^*_1\}].
\end{align*}

Next, we compute $\Phi'(0+)$. Consider that $\Ex[V(\theta,s) 1\{Z^\epsilon = 0\}] = \Ex[X^\epsilon_s],$ where $X^\epsilon_l \coloneqq \Ex[V(\theta,t)1\{Z^\epsilon_0 = 0\} | \mathcal{G}_l].$ This means $X^\epsilon_s$ is $\mathcal{G}$-adapted. Thus, the Snell envelope is well-defined, and the maximum is attainable. For every stopping level $s$ adapted to $\mathcal{G}_l$ and a nonnegative real number $\epsilon > 0$, define $f(\epsilon,s) \coloneqq \Ex[V(\theta,s) 1\{Z^\epsilon = 0\}]$. Thus, $\Phi(\epsilon) = \sup_{s \geq \el_0} \Ex[f(\epsilon,s)]$ subject to $s$ is adapted to $\mathcal{G}_l$.

We know that $\Phi(0)$ is uniquely attained by $s = \overline{\el}$ because $\Pr(Z^0 = 0) = 0 $. Moreover, we have the following lemma whose proof is technical and relegated to Online Appendix \ref{app:technical}.
\begin{lemma}\label{lem:technical_dom}
    For any decreasing sequence $(\epsilon_n)_n$ such that $\epsilon_n \downarrow 0$, define $s^{\epsilon_n}$ as the latest stopping time that solves (OSP-$\epsilon_n$) for every $n$. Then, $s^{\epsilon_n} \uparrow \overline{\el}$ as $n \to \infty$. 
\end{lemma}

Now consider that 
\begin{align*}
    f(\epsilon,s) &= \Ex[V(\theta,s)] - \Ex[V(\theta, s) \cdot Z^\epsilon] \\
    &= \Ex[V(\theta,s)] - \epsilon V[(\theta,s)1\{\omega \in A^*_1\}].
\end{align*}
Thus, $\frac{\partial f}{\partial \epsilon} = -\Ex[V(\theta,s)1 \{\omega \in A^*_1\}]$, which is constant in $\epsilon $. From Theorem 3 of \cite{milgrom2002envelope} since $\Phi$ is right-differentiable at $0$ and
\begin{align*}
    \Phi'(0+) &= \lim_{\epsilon \to 0^+} \frac{\partial f}{\partial \epsilon}(s^\epsilon, 0 )\\
    &= - \lim_{n \to \infty} \Ex\Big[V(\theta,s^{\epsilon_n}) \cdot 1\{\omega \in A^*_1\}\Big].
\end{align*}
Since $s^{\epsilon_n} \uparrow \overline{\el}$ as $n \to \infty$ we have
\begin{align*}
    \Phi'(0+) &= - \lim_{n \to \infty} \Ex[V(\theta,s^{\epsilon_n})1\{\omega \in A^*_1\}] =-\Ex[V(\theta,\overline{\el})1 \{\omega \in A^*_1\}],
\end{align*}
by the dominated convergence theorem. This implies
\begin{align*}
    \Psi'(0+) &= \Ex[V(\theta,s^1) 1\{\omega \in A^*_1\}] -\Ex[V(\theta,\overline{\el})1 \{\omega \in A^*_1\}] < 0.
\end{align*}
Thus, there exists $\epsilon >0$ such that $\Psi(\epsilon) < \Psi (0)$, which means $\phi$ is not robustly optimal, as desired.
\end{proof}
\cref{lemma:dual_hard_deadline} and \cref{lemma:zero_marginal} imply the forward direction of \cref{prop:dualrobust_characterization} i.e., necessacity.  

The next lemma establishes sufficiency.
\begin{lemma} \label{lemma:converse_dual}
    If $\phi$ satisfies two conditions stated in \cref{prop:dualrobust_characterization}, then $\phi$ is dually-robust.
\end{lemma}
\begin{proof} [Proof of \cref{lemma:converse_dual}]
    Given any agent's filtration $(\mathcal{F}_l)_l$ and agent's utility function $U$, let $\ell^*_U$ be the agent's corresponding optimal stopping level facing $\phi$. This implies $\ell^*_U \leq \overline{\el}$. Moreover, the optimality of $\el^*_U$ implies
    \begin{align*}
        &U^\phi(\mu_{\el^*_U},\el^*_U) \geq \Ex\big[U^\phi(\mu_{\overline{\el}},\overline{\el})\big| \mathcal{F}_{\el^*_U}\big] = \Ex\big[U^\phi(\mu_{\el^*_U},\overline{\el})\big| \mathcal{F}_{\el^*_U}\big] \text{ a.s.}\\
        &\Longrightarrow U(\mu_{\el^*_U},\el^*_U) - \Ex\big[U(\mu_{\el^*_U},\overline{\el}) \big| \mathcal{F}_{\el^*_U}\big] \geq \phi_{\el^*_U} - \Ex\big[\phi_{\overline{\el}} \big| \mathcal{F}_{\el^*_U}\big] \geq 0 \text{ a.s.},
    \end{align*}
    where the last inequality is because $\phi$ offers the option of continuing under $\bar \ell$ and enjoying a weakly positive subsidy. 
    
    This implies $U(\mu_{\el^*_U},\el^*_U) \geq \Ex\big[U(\mu_{\el^*_U},\overline{\el}) \big| \mathcal{F}_{\el^*_U}\big]$ almost surely. Since $\overline{\el} \geq \el^*_U$ almost surely and $U = g\circ V$ for some convex function $g$, \cref{lem:compare_two_stoppingtimes} implies\footnote{Note here that $\ell^*_U$ need not be an optimal stopping level for the agent with preference $U$ facing the empty mechanism so \cref{lem:compstat} does not apply. But \cref{lem:compare_two_stoppingtimes} does since it compares preferences over two stopping times.} 
    \[V(\mu_{\el^*_U},\el^*_U) \geq \Ex\big[V(\mu_{\el^*_U},\overline{\el}) \big| \mathcal{F}_{\el^*_U}\big] \, \text{a.s.}\] 
    Thus,
    \begin{align*}
        \Ex[V(\mu_{\el^*_U},\el^*_U)] &\geq \Ex\big[\Ex[V(\mu_{\el^*_U},\overline{\el}) | \mathcal{F}_{\el^*_U}]\big] \\
        &= \Ex\big[\Ex[V(\mu_{\overline{\el}},\overline{\el}) | \mathcal{F}_{\el^*_U}]\big] \\
        &= \Ex[V(\mu_{\overline{\el}},\overline{\el})] \\
        &= \Ex[V(\mu_0,\overline{\el})]
    \end{align*}
    where the first and last equalities are implied by the optional stopping theorem, as desired.
\end{proof}
\subsection{Proof  of \cref{thrm:dominance}}
 We start with an important lemma stating that any mechanism whose transfer is decreasing until the hard limit is dominated by the sandbox mechanism.
\begin{lemma} \label{lem:decreasing is dominated}
    Every mechanism $\phi \in \Phi^*$ such that $\phi(\el_1) \geq \phi(\el_2)$ for every $\mathcal{G}$-stopping level $\el_1 \leq  \el_2 \leq \overline{\el}$ almost surely is dominated by $\phi^*$.\footnote{This is also true under an adversarial selection of optimal stopping level. See Online Appendix \ref{appendix:adversarialstoppingselection}}
\end{lemma}
\begin{proof}
Fix any $U \in \mathcal{U}$ and any agent's filtration $\mathcal{F}$. We will construct the simplest coupling of the agent's stopping behavior under $\phi^*$ and under $\phi$: let the realized learning process be the same (that is, the processes are indistinguishable). Suppose the agent's optimal stopping level under $\phi^{*}$ is $\el_{*,U}$. Then, $\el_{*,U}$ must solves the restricted stopping problem:
\begin{align*}
    \sup_{s \leq \overline{\el}} \Ex [U(\theta,s)] \quad \text{where $s$ is $(\mathcal{F}_l)_l$-adapted stopping level.}
\end{align*}
We will show that the agent's latest optimal stopping level $\el^*_{\phi}$ under $(\phi,\mathcal{F})$ must satisfy $\el^*_{\phi} \geq \el_{*,U}$ almost surely.
Define $\ell_1 = \el^*_{\phi} \vee \el_{*,U}$ and $\el_0 = \el^*_{\phi} \wedge \el_{*,U}$. The optimality of $\el^*_{\phi}$ implies
\begin{align*}
    0 &\leq \Ex[U^\phi(\theta, \el^*_{\phi})] - \Ex[U^\phi(\theta,\el_1)] \\
    &= \Ex[(\phi_{\el_{*,U}} - \phi_{\el^*_{\phi}})1\{\el_{*,U} \geq \el^*_\phi\}] + \Ex[(U(\theta,\el^*_{\phi}) - U(\theta,\el_{*,U})) 1\{\el_{*,U} \geq \el_\phi\}]  \\
    &\leq \Ex[(U(\theta,\el^*_{\phi}) - U(\theta,\el_{*,U}) 1\{\el_{*,U} \geq \el^*_\phi\}] \tag{$\phi$ is decreasing up until $\overline{\el}$}
\end{align*}
The optimality of $\el_{*,U}$ implies
\begin{align*}
    0 \leq \Ex[U(\theta,\el_{*,U})] - \Ex[U(\theta,\el_0)] = \Ex[(  U(\theta,\el_{*,U}) - U(\theta,\el^*_{\phi})) 1\{\el_{*,U} \geq \el^*_\phi\}] 
\end{align*}
Thus, two above inequalities must be equalities, so $\el_1$ must be the agent's optimal stopping level under $(\phi,\mathcal{F})$ as well. The definition of $\el^*_\phi$ implies $\el^*_\phi \geq \el_1,$ meaning $\el^*_\phi \geq \el_{*,U},$ as desired.

Now applying \cref{lem:compstat} to the restricted (random) domain $[0,\overline{\el}]$ we have the solution to the fictional problem in which the principal is in control of the stopping level from time $\el_{*,U}$ onward 
\[
\sup_{\el_{*,U} \leq s \leq \overline{\el}} \Ex \Big[ V(\mu_{s}, s) \Big| \mathcal{F}_{\el_{*,U}} \Big] \quad \text{s.t. $s$ is $(\mathcal{F}_l)_{l}$-adapted}
\]
is simply $s = \el_{*,U}$ i.e., to stop immediately.

Our argument thus far has been path-wise. Now letting $F_{\phi^*} \in \Delta([0,L])$ be the distribution of the agent's stopping level under $\phi^{*}$, we have: 
\begin{align*}
    \Ex\Big[V(\bm{\mu}, \phi^{*}, U)\Big] &= \int \Ex\Big[V(\mu_s,s) \Big| \el_{*,U} = s\Big] dF_{\phi^{*}} \tag{Total expectation}
    \\
    & = 
    \int \bigg\{ \sup_{\el_{*,U} \leq \el \leq \overline{\el}} 
    \Ex\Big[V(\mu_{\el},\el) \Big| \el_{*,U} = s\Big]\bigg\}dF_{\phi^{*}}  \tag{Stopping at $\el_{*,U}$ is optimal for principal}
    \\
    &\geq 
    \int_{s \leq \overline{\el}} \Ex\Big[V(\mu_{\el^*_{\phi}},\el^*_{\phi}) \Big| \el_{*,U} = s\Big] dF_{\phi^*} \tag{$\el^*_\phi \geq \el_{*,U}$ almost surely} \\
    &= \Ex\Big[V(\bm{\mu}, \phi, U)\Big]  
\end{align*}
where the first equality is from the law of total probability, the first inequality is because we showed the solution to the principal's restricted stopping problem is to stop at $\el_{*,U}$ and furthermore conditioning on just $\{\el_{*,U} = s\}$ is a coarser event, and the last inequality is because $\el^*_{\phi} \geq \el_{*,U}$ almost surely. This establishes that $\phi^*$ performs weakly better under any learning process and any agent preference. 
\end{proof}

We are now ready to prove \cref{thrm:dominance}. 
\begin{proof}[Proof of \cref{thrm:dominance}]  We show each part of \cref{thrm:dominance} in turn.

\textbf{Step 1: $\phi^*$ dominates other dually-robust and regular mechanisms.} 

Here we focus on the case that the principal's learning process $\mathcal{G}$ and preferences $V$ (both primitives of our environment) are such that
$\overline{\el}<L$ almost surely. The other case is more technical and is handled in Online Appendix \ref{app:technical}. 

Consider any regular and robustly optimal mechanism $\phi$. For every $\mathcal{G}$-stopping level $\el_1 \leq \el_2 \leq \overline{\el}$, we will show that $\phi_{\el_1} > \phi_{\el_2}.$ We know that $\phi_{\el_1},\phi_{\el_2} \geq \phi_{\overline{\el}} $ almost surely by \cref{prop:dualrobust_characterization}. Since $\overline{\el}< L$, we must have $\phi_L = \infty$. This implies $\phi_{\el_1},\phi_{\el_2} \geq \phi_{\overline{\el}} < \phi_{L}$. Since $\phi$ is regular, $\phi_{\el_1} \geq \phi_{\el_2}$ a.s. i.e., $\phi$ is decreasing almost surely up until $\overline{\el}$. Then, \cref{lem:decreasing is dominated} implies $\phi$ is dominated by $\phi^*$. This delivers the second half of \cref{thrm:dominance}.

\textbf{Step 2: $\phi^*$ is undominated.} Suppose towards a contradiction that a mechanism $\phi$ dominates the sandbox mechanism. Then, the principal's worst case under $\phi$ must be weakly greater than that under $\phi^*$. However, since $\phi^* \in \Phi^*$, we must have $\phi \in \Phi^*$. We will show that $\phi_{\el^*} = \phi_{\overline{\el}}$ almost surely for every $\mathcal{G}$-adapted stopping level $\el^*$ such that $\el^* \leq \overline{\el}$. 
    
Define the following random variable: 
\begin{align*}
    \varphi^{\el^*} \coloneqq \begin{cases}
        1 & \text{if }\phi_{\el^* } > \phi_{\overline{\el}} \\ 0 & \text{if }\phi_{\el^*} = \phi_{\overline{\el}}\\ -1 & \text{if }\phi_{\el^*} < \phi_{\overline{\el}}
    \end{cases}
\end{align*}
which encodes information about whether the tax between $\ell^*$ and $\bar \ell$ is positive or negative. 

We choose the agent's filtration $\mathcal{F}^{\el^*}_l$ as follows:
\begin{align*}
    \mathcal{F}^{\el^*}_l \coloneqq \sigma \Big(\Big\{ &A \cap \{\omega: l < \el^*\} : A \in \mathcal{G}_l \Big\} \\
    &\cup \Big\{A \cap \{\omega: l \geq \ell^*, \theta = \theta_0,\varphi^{\ell^*} = \varphi \} : A\in \mathcal{G}_l, \theta_0 \in \Theta,\varphi \in \{-1,0,1\}\Big\} \Big).
\end{align*}
In words, we are constructing a filtration under which, at the fixed stopping level $\el^* \leq \overline{\el}$, the agent (i) learns the true state $\theta$; and (ii) whether the transfer at the hard limit is higher or lower than the current transfer i.e., she learns about what the principal will learn in the future. In particular, for every stopping level $\el$, if $\el < \el^*$, then $\mathcal{F}^{\el^*}_\el = \mathcal{G}_\el$. On the other hand, if $\el \geq \el^*$, then $\mathcal{F}^{\el^*}_\el = \mathcal{G}_\el \vee \sigma(\theta,\varphi^{\ell^*})$.

We now construct the agent's preferences by setting $U = \epsilon \cdot V$ for some small $\epsilon > 0$ that will be chosen later. Under our adaptive sandbox mechanism, since the agent's utility is a linear transformation of the principal's, the principal's expected utility under mechanism $\phi$ is bounded above by the value of the following optimal stopping problem:
\begin{align*}
    \sup_{s \leq \overline{\el}} \Ex\big[V(\theta, s)\big] \quad \text{s.t. $s$ is adapted to $(\mathcal{F}^{\el^*}_l)_l$} 
\end{align*}

Now suppose that $s^*$ is optimal stopping under the sandbox mechanism and the agent's filtration $(\mathcal{F}_l^{\el^*})_l$. We know that $s^* \leq \overline{\el}$ since continuing past the limit $\bar \ell$ is always dominated by stopping. This implies $s^*$ must solve the above optimal stopping problem. We claim
\begin{align*}
    s^* = \begin{cases}
        \el^* & \text{if $\theta = 0$} \\
        \overline{\el} & \text{otherwise}
    \end{cases}
    \quad \text{almost surely.}
\end{align*}
To see this, we will show it is always suboptimal to stop before $\el^*$ with positive probability. Define
\begin{align*}
s' := \begin{cases} \overline{\el} &\text{if $s^* < \ell^*$} \\ s^*&\text{if $s^* \geq \ell^*$}\end{cases} \quad  s'_1 := \begin{cases} s^* &\text{if $s^* < \ell^*$} \\ \overline{\el}&\text{if $s^* \geq \ell^*$}\end{cases}
\end{align*}
Note that both $s'$ and $s'_1$ are adapted to the agent's filtration $\mathcal{F}^{\el^*}$, and that $s'_1$ is in addition adapted to $\mathcal{G}$. 
Then,
\begin{align*}
    \Ex[V(\theta,s')] - \Ex[V(\theta ,s^*)] &= \Ex[(V(\theta,\overline{\el}) - V(\theta,s^*))1\{s^* < \el^*\}]  \\
    &= \Ex[V(\theta ,\overline{\el})] - \Ex[V(\theta, s'_1)]  \\
    &> 0,
\end{align*}
where the last inequality follows from the fact that $\overline{\el}$ is the unique optimal stopping of the principal under the filtration $\mathcal{G}$, and $s'_1$ is adapted to the filtration $\mathcal{G}$. This contradicts the optimality of $s',$ as desired. Thus, $s^* \geq \el^*$. If $\theta = 0$, the principal wants to stop immediately at $\el^*$. If $\theta = 1$, the principal wishes to continue until $\overline{\el}$. Thus, $s^*$ does, in fact, solve the principal's stopping problem and the principal's payoff under the adaptive sandbox $\phi^*$ against the agent's learning process $(\mathcal{F}^{\ell^*}_l)_l$ and preference $U = \epsilon \cdot V$ hits the upper-bound. 

Next, we show that $s^*$ cannot be the agent's optimal stopping level under $\phi$ and the filtration $\mathcal{F}^{\el^*}$. Consider the following alternative stopping level
\begin{align*}
    s = \begin{cases}
        \el^* &\text{if $\theta = 1$ and $\varphi^{\el^*} = -1$}\\
        \overline{\el} & \text{if $\theta = 0$ and $\varphi^{\el^*} = 1$} \\
        s^* & \text{otherwise}.
    \end{cases}
\end{align*}
In words, the agent stops at $\ell^*$ if the state is good and the agent has learnt that $\phi_{\ell^*} > \phi_{\bar \ell}$ i.e., she will be taxed if she continues until $\bar \ell$, and continues until $\bar \ell$ if the state is bad and the agent has learnt that she will be subsidized. This is clearly a $\mathcal{F}^{\ell^*}$-stopping level. We show it is optimal for the agent. 

The difference of the agent's utility under stopping levels $s$ and $s^*$ is 
\begin{align*}
    \Ex[U^\phi(\theta,s)]& - \Ex[U^\phi(\theta,s^*)]  \\
    &= \underbrace{\Ex\big[(U^\phi(1,\el^*) - U^\phi(1,\overline{\el})) 1\{\theta = 1, \varphi^{\el^*} = -1\}\big]}_{ \eqqcolon A} \\
    &+ \underbrace{\Ex\big[(U^\phi(0,\overline{\el}) - U^\phi(0,\el^*)) 1\{\theta = 0, \varphi^{\el^*} = 1\}\big] }_{\eqqcolon B}.
\end{align*}
Now recalling that $U = \epsilon \cdot V$, we can rewrite $A$ and $B$ as: 
\begin{align*}
    A &= \epsilon \cdot \underbrace{\Ex[(V(1,\el^*) - V(1,\overline{\el})) 1\{\theta = 1, \varphi^{\el^*} = -1\}]}_{\leq 0} + \underbrace{\Ex [(\phi_{\overline{\el}} - \phi_{\el^*} ) 1\{\theta = 1, \varphi^{\el^*} = -1\}]}_{\geq 0}  \\
    B &= \epsilon \underbrace{\Ex[(V(0,\overline{\el}) - V(1,\el^*)) 1\{\theta = 0, \varphi^{\el^*} = 1\}]}_{\leq 0} + \underbrace{\Ex [(\phi_{\el^*} - \phi_{\overline{\el}} ) 1\{\theta = 0, \varphi^{\el^*} = 1\}]}_{\geq0}  
\end{align*}
We can choose sufficiently small but positive $\epsilon > 0$ so that both $A$ and $B$ are weakly positive and strictly positive whenever $\Pr(\theta=1,\varphi^{\el^*} = -1) > 0$ and $\Pr(\theta=0,\varphi^{\el^*} = 1) > 0$, respectively. Thus, if either $\Pr(\theta=1,\varphi^{\el^*} = -1) > 0$ or $\Pr(\theta=0,\varphi^{\el^*} = 1) > 0$, then the agent strictly prefers stopping level $s$ to $s^*$, which contradicts the optimality of $s^*$. 

Under $\phi$, we must thus have 
\[
\mathbb{P}\Big( \big\{\theta=1,\varphi^{\el^*} = -1\big\} \cup \big\{ \theta=0,\varphi^{\el^*} = 1\big\}
\Big) = 0 \implies \Pr\Big(\varphi^{\el^*} \in \{-1,1\}\Big) = 0. 
\]
Hence, $\phi_{\el^*} = \phi_{\overline{\el}}$ almost surely, as desired. But since the stopping level $\ell^* \leq \bar \ell$ was arbitrary, this implies $\phi_{\ell} = \phi_{\overline{\el}} = \phi_0$ for any $\ell \leq \bar \ell$. Since $\phi_0$ is $\mathcal{G}_0$-measurable, this implies that $\phi$ is equivalent to the adaptive sandbox mechanism $\phi^*$, as desired.
\end{proof}

\subsection{Proof of \cref{thrm:time-consistency} (Time-consistency)} 
Consider any $\mathcal{G}$-stopping level $\ell_0 \in (0,L)$. We first show the sufficiency direction that $\phi^*$ is time-consistent starting following from $\ell_0$. Then, we show the necessity direction that among dually-robust mechanisms, only $\phi^*$ is time-consistent. 

\underline{\smash{$\phi^*$ is time-consistent upon stopping.}}

We wish to show that $\phi^*$ is conditionally undominated upon the agent stopping at $\el_0$. Suppose, towards a contradiction, that there exists some other other mechanism $\phi'$ that conditionally dominates $\phi^*.$ This implies for every $(\mathcal{F},U) \in \mathbb{F} \times \mathbb{U}$ and $\mathcal{G}_{\ell_0}$-measurable set $A_{\ell_0},$
\begin{align*}
    \Ex\big[V(\mu_{\ell^*},\ell^*) \cdot 1\{\omega \in A_{\ell_0}, \ell^* = \ell_0\}\big] \leq \Ex\big [V(\mu_{\ell_{\phi'}}, \ell_{\phi'}) \cdot 1\{\omega \in A_{\ell_0},\ell^*= \ell_0\}\big],
\end{align*}
with strict inequality for some $(\mathcal{F},U) \in \mathbb{F} \times \mathbb{U}$ and some $\mathcal{G}_{\ell_0}$-measurable set $A_{\ell_0}$.  

For each $\mathcal{G}_{\ell_0}$-measurable set $A_{\ell_0}$, we can partition it as follows:  
\[A_{\ell_0} = \underbrace{\Big\{A_{\ell_0} \cap \{\omega: \ell_0 < \overline{\el}\} \Big\}}_{A^<_{\ell_0}} \cup \underbrace{\Big\{A_{\ell_0} \cap \{\omega: \ell_0 \geq \overline{\el}\}\Big\}}_{A^\geq_{\ell_0}}.\]
Note that the event $\{\ell_0 = \ell^*\} \cap A^<_{\ell_0}$ is the event that the agent stops strictly before the sandbox limit while $\{\ell_0 = \ell^*\} \cap A^{\geq}_ {\ell_0}$ is the event that the agent stops right at the sandbox limit. We proceed in two cases.

\underline{Case A: Under $A^\geq_{\ell_0}$ the $\phi^*$ is undominated.} 
Consider any filtration $(\mathcal{F}_l)_l$ such that the agent and the principal learn identically before level $\el_0,$ i.e., $\mathcal{F}_{\el_0} = \mathcal{G}_{\el_0}$, and pick any agent preference $U \in \mathbb{U}$. Now let $\el^*_V$ be the principal's optimal stopping level \emph{as if} they observe the agent's information $(\mathcal{F}_l)_l$. An identical argument to Step 2 of the proof of \cref{thrm:dominance} implies that if $\ell_0 \leq \overline{\el}$, then $\ell_0 \leq \ell^*_V$, i.e., it is suboptimal for the principal to stop before $\ell_0$ under the event $\{\ell_0 \leq \overline{\el}\}$ because $\mathcal{F}_{\ell_0} = \mathcal{G}_{\ell_0}$. By \cref{lem:compstat}, we must also have, if $\ell_0 \leq \overline{\el}$, then $\ell_0 \leq \ell^*_V \leq \ell^*$. Thus, $\{\el^* = \el_0\} = \{\overline{\el} = \el_0\}$ because $\el^* \leq \overline{\el}$. In words, if the agent stops at $\ell_0$, then $\ell_0$ must coincide with the limit.  

This implies 
\[A^\geq_{\ell_0} \cap \{\ell_0 = \ell^*\} = A_{\ell_0} \cap \{\ell_0 = \overline{\el}\} \eqqcolon A^=_{\ell_0},\]
which is $\mathcal{G}_{\ell_0}$-measurable, i.e., the principal does not learn any additional information from the observation that the agent has stoped at $\ell_0$.
Since $\phi'$ conditionally dominates $\phi^*$ upon stopping at $\el_0$, for every $\mathcal{G}_{\el_0}$-measurable set $A'_{\el_0} \subset A^=_{\el_0},$
\begin{align*}
    &\Ex[V(\mu_{\ell^*}, \ell^*)1\{\omega \in A'_{\ell_0}, \el^* = \el_0\}] \leq \Ex [V(\mu_{\ell_{\phi'}},\ell_{\phi'})1\{\omega \in  A'_{\ell_0}, \el^* = \el_0\}] \\
    \Longrightarrow &\Ex[V(\mu_{\ell^*}, \ell^*)1\{\omega \in A'_{\ell_0}\}] \leq \Ex [V(\mu_{\ell_{\phi'}},\ell_{\phi'})1\{\omega \in  A'_{\ell_0}\}] \\
     \Longrightarrow  & \Ex[V(\mu_{\el^*}, \el^*) | \mathcal{G}_{\el_0}] \leq \Ex[V(\mu_{\el_{\phi'}}, \el_{\phi'}) | \mathcal{G}_{\el_0}] \quad \text{a.s. under the event $A^=_{\el_0}$}
\end{align*}
for every $(\mathcal{F},U) \in \mathbb{F} \times \mathbb{U}$ such that $\mathcal{F}_{\el_0} = \mathcal{G}_{\el_0}$. In other words, $\phi'$ dominates $\phi^*$ in terms of expected payoff under the event $A^=_{\el_0}$ with no restriction of learning process and agent's utility function beyond level $\el_0$. Since $\phi^*$ remains an adaptive sandbox after level $\el_0$, we can apply the same construction from the proof that $\phi^*$ is undominated from Step 2 of \cref{thrm:dominance} otherwise by the same argument from Step 2 of \cref{thrm:dominance}, we can construct some $(\mathcal{F}, U) \in \mathbb{F} \times \mathbb{U}$ such that $\mathcal{F}_{\el_0} = \mathcal{G}_{\el_0}$ under which $\phi^*$ yields strictly higher payoff than $\phi'$, a contradiction. Hence, choosing the initial level $\el_0$ instead of $0$ to show that $\phi'$ must be equivalent to $\phi^*$ under the event $A^=_{\ell_0}$. 
Thus,
\begin{align}
    \Ex[V(\mu_{\ell^*}, \ell^*)\underbrace{1\{\omega \in A^\geq_{\ell_0}, \ell^*= \ell_0\}}_{=  A^=_{\ell_0} }] = \Ex [V(\mu_{\ell_{\phi'}},\ell_{\phi'})1\{\omega \in A^\geq_{\ell_0},\ell^* = \ell_0\}]. \label{eqn: eqvn when stop correct}
\end{align}

\underline{Case B: Under $A^<_{\ell_0}$ $\phi^*$ is undominated.} Next, we analyze $\phi'$ under the event $A^<_{\el_0}.$ The following lemma implies $\phi$ must also have a hard limit at $\overline{\el}$. 
\begin{lemma} \label{lemma: hard limit - conditionally undominated}
    For every $\mathcal{G}$-stopping level $\ell_1 > \bar \ell$, we must have $\phi'_{\ell_1} = \infty$ almost surely under the event $\{\phi'_{\el_0} > -\infty\} \cap A^<_{\el_0}.$ Moreover, $\phi'_{\ell_1} > -\infty$ almost surely under the event $\{\phi'_{\el_0} = -\infty\} \cap A^<_{\el_0}$.
\end{lemma}
We relegate the proof to Online Appendix \ref{app:technical}. The proof is similar to \cref{lemma:dual_hard_deadline}, but we need choose $(U,\mathcal{F})$ more carefully so that the event that the agent stops at $\el_0$ happens with strictly positive probability. This implies stopping after $\overline{\el}$ under the event $A^<_{l_0} \cap \{\phi'_{\el_0} > -\infty\}$ with positive probability is always suboptimal. Moreover, under the event $A^<_{\el_0} \cap\{\phi'_{\el_0} = -\infty\},$ stopping after $\overline{\el}$ is suboptimal as well because it gives finite utility, whereas stopping at $\el_0$ gives the positive infinite utility. Then, we yield the following corollary.
\begin{corollary}
    $\ell_{\phi'} \leq \overline{\el}$ almost surely under the event $A^<_{\el_0}$.
\end{corollary}

Since $\phi'$ conditionally dominates $\phi^*$, under any $(\mathcal{F},U)$ we obtain: 
    \begin{align}
        &\Ex\big[V(\mu_{\ell^*},\ell^*)\cdot 1\{\omega \in A^<_{\ell_0}, \ell^* = \ell_0\}\big] \leq \Ex\big[V(\mu_{\ell_\phi},\ell_\phi)\cdot 1\{\omega \in A^<_{\ell_0}, \ell^* = \ell_0\}\big] \nonumber \\
        \overset{\text{\cref{lem:compare_two_stoppingtimes}}} {\Longrightarrow} &\Ex\big[U(\mu_{\ell^*},\ell^*) \cdot 1\{\omega \in A^<_{\ell_0}, \ell^* = \ell_0\}\big] \leq \Ex\big[U(\mu_{\ell_{\phi'}},\ell_{\phi'}) \cdot 1\{\omega \in A^<_{\ell_0}, \ell^* = \ell_0\}\big] \label{eqn: u condtional on stopping}
    \end{align}
    since $A^<_{\ell_0}$ is measurable with respect to $\mathcal{G}_{\ell_0}$ and $\ell_{\phi'} \geq \ell^*,$ and the last inequality becomes strict when $\Pr (\omega \in A^<_{\ell_0}, \ell^* = \ell_0 < \ell_{\phi'}) > 0.$ 
    
    However, the agent optimally stops at $\ell^*$ under the adaptive sandbox mechanism instead of stopping at $\ell_{\phi'}$ under the event $A^<_{\ell_0} \cap \{ \ell^* = \ell_0\}$, which is $\mathcal{F}_{\ell_0}$-measurable.  We must have
    \begin{align*}
        \Ex\big[U^{\phi^*}(\mu_{\ell^*},\ell^*)\cdot 1\{\omega \in A^<_{\ell_0}, \ell^* = \ell_0\}\big] &\geq \Ex\big[U^{\phi^*}(\mu_{\ell_{\phi'}},\ell_{\phi'})\cdot 1\{\omega \in A^<_{\ell_0}, \ell^* = \ell_0\}\big] \\
        \Longrightarrow \Ex\big[U(\mu_{\ell^*},\ell^*)\cdot 1\{\omega \in A^<_{\ell_0}, \ell^* = \ell_0\}\big] &\geq \Ex\big[U(\mu_{\ell_{\phi'}},\ell_{\phi'}) \cdot 1\{\omega \in A^<_{\ell_0}, \ell^* = \ell_0\}\big]
    \end{align*}
    because $\ell_{\phi'} \leq \overline{\el}.$ These together imply
    \[\Ex\big[U(\mu_{\ell^*},\ell^*) \cdot 1\{\omega \in A^<_{\ell_0}, \ell^* = \ell_0\}] = \Ex\big[U(\mu_{\ell_{\phi'}},\ell_{\phi'})\cdot  1\{\omega \in A^<_{\ell_0}, \ell^* = \ell_0\}\big].\]
    Then, \cref{eqn: u condtional on stopping} must become an equality. Hence, 
    \[\Pr\Big(\big\{\omega \in A^<_{\ell_0}\big\} \cap \big\{ \el^* = \el_0 < \el_{\phi'}\big\} \Big) = 0.\]
    Together with \cref{eqn: eqvn when stop correct}, we obtain
    \[\Ex\big[V(\mu_{\ell^*},\ell^*)\cdot 1\{\omega \in A_{\ell_0}, \ell^* = \ell_0\}\big] = \Ex\big[V(\mu_{\ell_{\phi'}},\ell_{\phi'})\cdot 1\{\omega \in A_{\ell_0}, \ell^* = \ell_0\}\big],\]
    for every $\mathcal{G}_{\el_0}$-measurable set $ A_{\ell_0}$ and every $(\mathcal{F},U) \in \mathbb{F} \times \mathbb{U}$.
    From Cases A and B, we have that $\phi'$ and $\phi^*$ yield identical payoffs for every $\mathcal{G}_{\el_0}$-measurable set $ A_{\ell_0}$ and every $(\mathcal{F},U) \in \mathbb{F} \times \mathbb{U}$, contradicting the hypothesis that $\phi'$ conditionally dominates $\phi^*$, as desired. This yields the first part of \cref{thrm:time-consistency}.

\underline{$\phi^*$ is time-consistent upon continuing.}
We wish to show that $\phi^*$ is conditionally undominated on this history $A_{\ell_0}$. Suppose, towards a contradiction, that there exists some other other mechanism $\phi'$ that conditionally dominates $\phi^*.$ This implies, for every $(\mathcal{F},U) \in \mathbb{F} \times \mathbb{U}$ and $\mathcal{G}_{\ell_0}$-measurable set,
\begin{align}
    \Ex[V(\mu_{\ell^*},\ell^*)1\{\omega \in A_{\ell_0}, \ell^* > \ell_0\}] \leq \Ex [V(\mu_{\ell_{\phi'}}, \ell_{\phi'})1\{\omega \in A_{\ell_0},\ell^* > \ell_0\}], \label{eqn: dominate upon continuing}
\end{align}
and the inequality is strict for some $(\mathcal{F},U) \in \mathbb{F} \times \mathbb{U}$ and a $\mathcal{G}_{\ell_0}$-measurable set $ A_{\ell_0}.$ We proceed similarly as we consider $\phi^*$ under the event $A^\geq_{\ell_0}$.

 We consider any filtration $(\mathcal{F}_l)_l$ that the agent and the principal learn equally before level $\el_0,$ and any preference $U \in \mathbb{U}.$ We showed earlier that, if $\el_0 \leq \overline{\el}$, then $\el_0 \leq \el^*$. Because of that sandbox limit, $\ell^* = \overline{\el}$ under the event $\{\overline{\el} \leq \ell_0\}$. Thus, $\{\el^* > \el_0\} = \{\overline{\el} > \el_0\}$ and the latter is $\mathcal{G}_{\el_0}$-measurable. Hence, the principal does not learn any extra information by observing that the agent continues beyond $\ell^*$. From \cref{eqn: dominate upon continuing}, we obtain
\begin{align*}
    \Ex[V(\mu_{\ell^*}, \ell^*)| \mathcal {G}_{\el_0}] \leq \Ex [V(\mu_{\ell_{\phi'}},\ell_{\phi'})| \mathcal{G}_{\el_0}] \quad \text{under the event $\{\el^*>\el_0\}$}.
\end{align*} 
Since $\phi^*$ remains an adaptive sandbox after level $\el_0$, we can apply the same argument from the proof that $\phi^*$ is undominated from \cref{thrm:dominance} to show that $\phi'$ must coincide with $\phi^*$ when $\ell \geq \ell_0$ under the event $\{\el^*>\el_0\}$. Thus,
\[ \Ex[V(\mu_{\ell^*}, \ell^*)1\{\omega \in A_{\ell_0}, \ell^*> \ell_0\}] = \Ex [V(\mu_{\ell_{\phi'}},\ell_{\phi'})1\{\omega \in A_{\ell_0},\ell^* > \ell_0\}] \]
 for every $\mathcal{G}_{\ell_0}$-measurable set $A_{\ell_0}$, but this contradicts that the inequality must be strict somewhere, as desired.

\underline{\smash{All other dually-robust mechanisms are not time-consistent.}} We now show the converse of \cref{thrm:time-consistency}. Consider any time-consistent mechanism $\phi$ under the finite technology space $\mathcal{L} := \{l_0,l_1,\dots,l_n\}$. We will show via induction backwards in levels that for every $l \in \mathcal{L}$, $\phi_{l \wedge \overline{\el}}$ must be a constant. 

    \underline{Base step: $\phi_{l_{n-1} \wedge \overline{\el}} = \phi_{l_n \wedge \overline{\el}}$ almost surely.} Consider the event the agent continues at $l_{n-1}$. We pick the agent's filtration so that $\mathcal{F}_l = \mathcal{G}_l$ for every $l < n-1$. Since $\phi$ is a dually-robust mechanism, by \cref{prop:dualrobust_characterization} we must have $\phi_{l \wedge \overline{\el}} \geq \phi_{\overline{\el}}$. Under the event $\{\overline{\el} > l_{n-1}\}$, we must have $\phi_{l_{n-1}} \geq \phi_{l_n}$. Thus, condition on the agent continuing until $l_{n-1}$ under the event $\{\overline{\el} > l_{n-1}\}$, the transfer is weakly decreasing starting from $l_{n-1}$ until $\overline{\el}$ almost surely i.e., $\phi_{\bar \ell} \leq \phi_{l_{n-1}}$.

    If it is strictly decreasing, then from \cref{lem:decreasing is dominated}, $\phi$ is dominated by the sandbox mechanism $\phi'$ constructed such that $\phi'_{\bar \ell} = \phi'_{l_{n-1}}$ under the event $\{\overline{\el} > l_{n-1}\}$ as long as $\phi_{l_{n-1} \wedge \overline{\el}} > \phi_{l_n \wedge \overline{\el}}$ with positive probability. Thus, for $\phi$ to be conditionally undominated at $l_{n-1}$, $\phi_{l_{n-1} \wedge \overline{\el}} = \phi_{l_n \wedge \overline{\el}}$ almost surely. 

    \underline{Inductive step: $\phi_{l_k \wedge \overline{\el}} = \phi_{l_{k+1} \wedge \overline{\el}}$ almost surely if $\phi_{l_{k+1} \wedge \overline{\el}} = \phi_{l_{k+2} \wedge \overline{\el}} = \dots = \phi_{l_n \wedge \overline{\el}}$.} The proof is similar to the base step. We pick the agent's filtration $\mathcal{F}_l = \mathcal{G}_l$ for every $l < l_k.$ Since $\phi$ is a dually-robust mechanism, we must have $\phi_{l \wedge \overline{\el}} \geq \phi_{\overline{\el}}$. Under the event $\{\overline{\el} > l_{k}\}$, we must have $\phi_{l_k} \geq \phi_{\overline{\el}} = \phi_{l_{k'} \wedge \overline{\el}}$ for every $k'>k$ by the induction hypothesis. Thus, condition on the agent continuing until $l_{k}$ under the event $\{\overline{\el} > l_k\}$, transfer is weakly decreasing starting from $l_{k}$ until $\overline{\el}$ almost surely. However, from \cref{lem:decreasing is dominated}, $\phi$ is dominated by the sandbox mechanism $\phi'$ constructed such that $\phi'_{\ell} = \phi_{l_{k}}$ for all $l_k \leq \el \leq \bar \ell$ under the event $\{\overline{\el} > l_{k}\}$ as long as $\phi_{l_{k} \wedge \overline{\el}} > \phi_{l_{k+1} \wedge \overline{\el}}$ with positive probability. 
    
    Thus, for $\phi$ to be conditionally undominated at $l_{k}$, $\phi_{l_{k} \wedge \overline{\el}} = \phi_{l_{k+1} \wedge \overline{\el}}$ almost surely. 

    The argument yields that $\phi_{l_k \wedge \overline{\el}} = \phi_{l_0}$ for every $k \in \{0,\dots,n\}$, i.e., $\phi_l$ must be almost surely constant up until $\overline{\el}$. But then $\phi$ must be equivalent to our adaptive sandbox $\phi^*$ as desired.

\section{Proofs of results in Section \ref{sec:importance}} \label{appendix:proofs_worstcase}

\subsection{Proof of \cref{thrm:worstcase}} 
\begin{figure}[H]
\vspace{-1em}
\centering
{\includegraphics[width=1.0\textwidth]{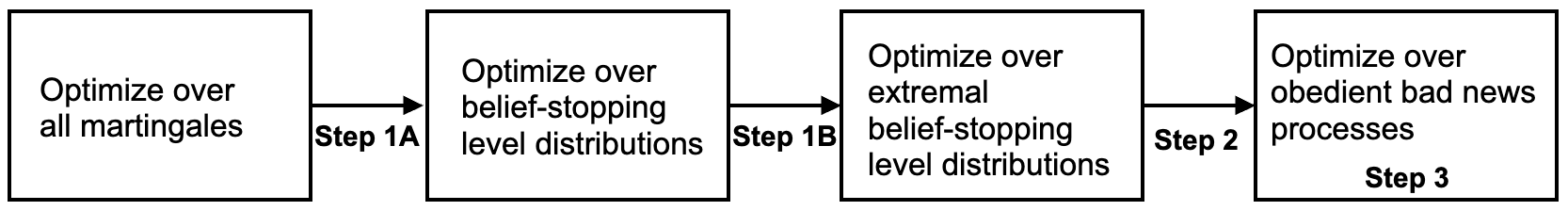}}
    \caption{Outline of proof of \cref{thrm:worstcase}}
    \label{fig:T3_proof_outline}
\end{figure}

The proof of \cref{thrm:worstcase} is quite involved. We will prove the result via the following steps:
\begin{enumerate}
    \item There exists a `direct and extremal' worst-case learning process
    \item A worst-case learning process is an obedient bad news process
    \item The obedient bad news process has arrival of bad news supported on $[l^*_{U^{\phi}}(\mu_0),L)$ such that continuation beliefs are $\gamma(l) = \mu^*_{U^{\phi}}(l)$ for each level in the support. 
    \item Argue that as $L \to \infty$ worst-case payoffs when both $\mathcal{F}$ and $U$ are chosen adversarially are arbitarily poor. 
\end{enumerate}
Steps 1-3 are outlined in \cref{fig:T3_proof_outline}.

\begin{proof}[Proof of Theorem \ref{thrm:worstcase}] We proceed along the steps outlined above.\\
\underline{\smash{Step 1. There is a direct and extremal worst-case process}} \\
Nature's problem is 
\begin{align*}
    &\inf_{\mathcal{F} \in \mathbb{F}} \Ex\Big[V(\mu_{\el^*}, \el^*(\phi, \bm{\mu},U)\Big] \tag{N} \label{prob:natureprimal}
    \\
    &\quad \text{s.t. $\el^*(\phi, \bm{\mu}, U)$ is a $(\mathcal{F}_l)_l$-adapted optimal stopping level for the agent}     
\end{align*}
From the revelation principle \citep{myerson1986multistage,forges1986approach} developed for dynamic information design \citep*{koh2022attention,koh2024persuasion}, it is without loss to directly consider the joint distribution over beliefs and stopping levels $f \in \Delta(\Delta(\Theta) \times \mathcal{L})$ i.e., it is without loss to rule out learning processes with continuous sample paths i.e., those driven by Brownian noise such as the drift-diffusion model. In particular, if $f$ fulfills a series of conditions laid out below, we can construct a jump process corresponding to the learning process which induces $f$ via agent optimal stopping; conversely, if it does not fulfill these conditions, then we cannot find any learning process which induces $f$. 

Thus, Nature's problem can be recast as that of choosing a joint distribution over agent's beliefs and stopping levels as follows: 
    \begin{align*}
        \inf_{f \in \Delta(\Delta(\Theta) \times \mathcal{L})} &\int V(\mu,l) f(d\mu,dl) \tag{N$^*$} \label{prob:nature}
        \\
        \text{s.t.} \quad  &\int \mu f(d\mu,dl) = \mu_0 \tag*{[Martingale]}
        \\
        &
        \int_{l>s} U^\phi(\mu,l) f(d\mu,dl)\geq  \int_{l > s} U^\phi(\mu,s ) f(d\mu,dl)\quad \forall s \in \mathcal{L} \tag*{[OC-C]}
        \\
        & \mu \leq \mu^*_{U^{\phi}}(l) \quad \forall (\mu,l) \in \text{supp } f \tag*{[OC-S]}
    \end{align*}
    
    We claim $\eqref{prob:nature} = \eqref{prob:natureprimal}$ which follows from applying Lemma 1 of \cite*{koh2024persuasion}. We briefly sketch what each of these conditions correspond to, and the intuition for why these conditions are tight. [Martingale] is a standard requirement on beliefs which any learning process $\bm{\mu}$, projected onto its induced stopping level, must continue to fulfill. [OC-C] is the requirement that the agent must find it optimal to continue pushing the technology given the joint distribution $f$.\footnote{Since $U$ is linear in belief, this is equivalent to the condition on expected payoff from pushing the technology until the level prescribed by $f$ vs stopping at s under the current belief: 
    \[
    \cfrac{\int_{l>s} U^\phi(\mu,l) f(d\mu,dl)}{\int_{l > s} f(d\mu,dl)} \geq  U^\phi\bigg(\cfrac{\int_{l>s} \mu f(d\mu,dl)}{\int_{l > s} f(d\mu,dl)},s \bigg) \quad \forall s \in \mathcal{L}.
    \]
    } [OC-S] is the requirement that the agent finds it optimal to stop. This is necessary because at belief and level pair $(\mu,l) \in \text{supp} f$, the agent cannot find it optimal to continue even in the absence of further information i.e., we cannot have $\mu > \mu^*_{U^{\phi}}(l)$. 
    On the other hand, it is also sufficient since nature can choose the learning process so that agent indeed learns nothing following $(\mu,l)$. 

    We can further reduce the set of candidate joint distributions over beliefs and stopping levels. Define the set of direct and extremal belief-level pairs as follows: 
    \[
    \mathcal{E} := \Bigg\{f \in \Delta(\Delta(\Theta) \times \mathcal{L}): \text{supp} f \subseteq 
    \underbrace{\Big\{\{0, \mu^*_{U^{\phi}}(l)\} \times [0,L) \Big\}
    }_{\text{extremal beliefs before $L$}} 
    \bigcup 
    \underbrace{\Big\{\{0,1\} \times \{L\}\Big\}}_{\text{extremal beliefs at $L$}} 
    \Bigg\}.
    \]
    \begin{lemma} \label{lemma:extreme_dist}
        There exists $f \in \mathcal{E}$ such that $f$ solves \eqref{prob:nature}. 
    \end{lemma}
    This follows from noticing that nature's objective is a linear functional, $V,U$ are linear in belief, [OC-C], [Martingale], and [OC-S] are linear constraints. Then applying Bauer's maximum principle, is without loss to extremize each belief $\mu$ at time $l$ to extreme beliefs on the boundaries which, for level $l$, are exactly $\{0,\mu^*_{U^{\phi}}(l)\}$.

\underline{\smash{Step 2. Every worst-case learning process is an obedient bad news process}} \\
Say that $f \in \Delta(\Delta(\Theta) \times \mathcal{L})$ is an \emph{obedient bad news process} if 
\[
    \text{supp} f \subseteq \Big\{ \{0\} \times [0,L] \Big\} \bigcup \Big\{\{1\} \times L \Big\}.
\]
since $f$ must fulfil [OC-C], this means that conditional on bad news not arriving, the agent pushes the technology. 

\begin{lemma}[Worst-case is obedient bad news]\label{lemma:badnews}
    If $\phi$ is such that $U^{\phi} = g\circ V$ for some increasing convex function $g$ then there exists an obedient bad news process $f$ that solves \eqref{prob:nature}. 
\end{lemma}
\begin{proof}
    Suppose that $f$ solves \eqref{prob:nature}. We will modify it so that it is an obedient bad news process. From \cref{lemma:extreme_dist}, it is without loss to suppose that $f \in \mathcal{E}$. 

    \begin{figure}[H]
    \begin{minipage}[t]{0.5\linewidth}
    We will define a family of joint distributions parametrized by $l_0 \in \text{int } \mathcal{L}$. Write $\widehat{f}_{l_0} \in \Delta(\Delta(\Theta) \times [0,L])$ such that 
    (i) the support of stopping beliefs are on $0$ over the interval $[l_0,L)$, and on $0$ and $1$ at $L$ i.e., 
    \[\text{supp} \widehat{f}_{l_0} = \Big \{\{0\} \times [l_0, L]\Big\} \bigcup
    \Big\{ \{1\} \times L\Big\} 
    \]
    which is depicted in \cref{fig:hatf}.
    \end{minipage}%
    \hfill%
    \begin{minipage}[t]{0.5\textwidth}\vspace{0pt}
    \vspace{-1em}
    \centering
    {\includegraphics[width=0.95\textwidth]{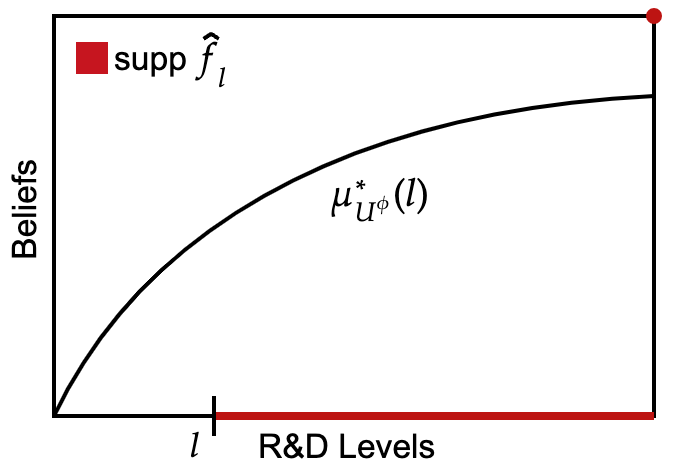}}
        \caption{Support of $\hat f_{l_0}$}\label{fig:hatf}
    \end{minipage}
    \end{figure}
    Moreover, the distribution $\widehat f_{l_0}$ is constructed such that (ii) the agent is indifferent between continuing and stopping for every $s \in [l_0,L]$: 
    \begin{align*}
        \widehat{f}_{l_0}\Big(\{(1,L)\}\Big) &= \mu^*_{U^{\phi}}(l_0)
        \quad 
        \quad 
        \widehat{f}_{l_0}\Big(\Big\{ \{0\} \times [s,L] \Big\}\Big) = \frac{\mu^*_{U^{\phi}}(l_0)}{\mu^*_{U^{\phi}}(s)} -  \mu^*_{U^{\phi}}(l_0).
    \end{align*}
    and we use the convention that $\mu^*_{U^{\phi}}(L) = 1$ i.e., there is an atom at the belief-level pair $(0,L)$. This is well-defined because $\mu^*_{U^{\phi}}(s)$ is weakly increasing in $s$. 

    \begin{lemma} \label{lemma:fhat_binds_OCC}
    Under $\widehat f_{l_0}$, [OC-C] is tight for all times $s \geq l_0$.
    \end{lemma}

    \begin{proof}
        Direct calculation yields: 
        \begin{align*}
            \int_{l > s} U^{\phi}(\mu,l) \widehat{f}_{l_0}(d\mu,dl)
            &= U^{\phi}(1,L)\cdot \mu^*_{U^{\phi}}(l_0)  - \Big[ U^{\phi}(0,l) \cdot \Big(\frac{\mu^*_{U^{\phi}}(l_0)}{\mu^*_{U^{\phi}}(l)} -  \mu^*_{U^{\phi}}(l_0) \Big)\Big]_{l = s}^{l = L} \\
            &\quad\quad + \int_{l > s} \cfrac{\partial U^{\phi}(0,l)}{\partial l} \Big( \frac{\mu^*_{U^{\phi}}(l_0)}{\mu^*_{U^{\phi}}(l)} -  \mu^*_{U^{\phi}}(l_0) \Big) dl  
            \\
            &=  U^{\phi}(1,L)\cdot \mu^*_{U^{\phi}}(l_0)  + U(0,s) \cdot  \Big( \frac{\mu^*_{U^{\phi}}(l_0)}{\mu^*_{U^{\phi}}(s)} -  \mu^*_{U^{\phi}}(l_0) \Big) \\
            &\quad\quad + \mu^*_{U^{\phi}}(l_0) \int_{l > s} \underbrace{\cfrac{\partial U^{\phi}(0,l)}{\partial l} \Big( \frac{1}{\mu^*_{U^{\phi}}(l)} -  1 \Big)}_{= - \partial U^{\phi}(1,l)/\partial l \, (\diamondsuit)} dl \\
            &= U^{\phi}(1,s) \mu^*_{U^{\phi}}(l_0) + U^{\phi}(0,s) 
            \underbrace{\Big( \frac{\mu^*_{U^{\phi}}(l_0)}{\mu^*_{U^{\phi}}(s)} -  \mu^*_{U^{\phi}}(l_0) \Big)}_{= \widehat{f}_{l_0}(\{ \{0\} \times [s,L] \})}
            \\
            &= \int_{l > s} U^{\phi}(\mu,s) f(d\mu,dl)
        \end{align*}
        where the first equality is via integration by parts and $(\diamondsuit)$ is because $\mu^*_{U^{\phi}}$ must almost everywhere (a.e.) satisfy,  
        \[
        \cfrac{\partial U^{\phi} \Big( \mu^*_{U^{\phi}}(l), l \Big)}{\partial l} = 0 \implies 
        (1- \mu^*_{U^{\phi}}(l)) \cfrac{\partial U^{\phi}(0, l)}{\partial l} + 
        \mu^*_{U^{\phi}}(l) \cfrac{\partial U^{\phi}(1, l)}{\partial l} = 0.
        \]
        noting that $U^{\phi}$ is differentiable in its second argument a.e. because $V$ is the weighted sum of monotone continuous functions which imply that they are differentable a.e., and $U^{\phi}$ inherits this because by assumption $U^{\phi} = g \circ V$ where $g$ is convex. 
    \end{proof}

    Now define $f^* \in \Delta(\Delta(\Theta) \times [0,L])$ as the `weighted projection' of our original distribution $f$ onto the beliefs $\{0,1\}$ so that $\text{supp} f^* \subseteq \{(1,L)\} \cup \{(0,l): l \in[0,L]\}$. 
    \begin{align*}
        f^*\Big(\{(1,L)\}\Big) &= f\Big(\{(1,L)\}\Big) + \int_{l \in [0,L)} \mu^*_{U^{\phi}}(l) 1\Big\{\mu = \mu^*_{U^{\phi}}(l)\Big\} f(d\mu,dl)
    \end{align*}
    and for each $s \in \mathcal{L}$, 
    \begin{align*}
        f^*\Big(\Big\{ \{0\} \times [s,L] \Big\}\Big) = &f\Big(\Big\{ \{0\} \times [s,L] \Big\}\Big)  \\
        &+ \int_{l \in [0,L)}\widehat{f}_{l}\Big(\Big\{ \{0\} \times [s,L] \Big\}\Big) 1\Big\{\mu = \mu^*_{U^{\phi}}(l)\Big\} f(d\mu,dl).
    \end{align*}
    where we recall that $f \in \mathcal{E}$ is the distribution which is hypothesized to solve \eqref{prob:nature}. We first verify that $f^*$ satisfies the constraints in \eqref{prob:nature}; this is a little tedious but the calculations are straightforward:  
    \begin{enumerate}[leftmargin = *]
        \item[1.] [$f^* \in \Delta(\Delta(\Theta) \times \mathcal{L})$] Notice that $f^*$ is a valid probability distribution because:
        \begin{align*}
            \int df^* 
            &= f^*\Big(\{(1,L)\}\Big) +  f^*\Big(\Big\{ \{0\} \times [0,L] \Big\}\Big) \\
            &= f\Big(\{(1,L)\}\Big) + f\Big(\Big\{ \{0\} \times [0,L] \Big\}\Big) + \int_{l \in [0,L)} 1\Big\{\mu = \mu^*_{U^{\phi}}(l)\Big\} f(d\mu,dl) \\
            &= \int df =  1
        \end{align*}
        where the second-last equality is because $f \in \mathcal{E}$ and 
        the last is because $f$ is by assumption a valid probability distribution. 
        
        \item[2.] [Martingale]: Directly integrating:
        \begin{align*}
            \int \mu f^*(d\mu,dl) &= \int \mu f(d\mu,dl) - \int \mu 1\Big\{\mu = \mu^*_{U^{\phi}}(l)\Big\} f(d\mu,dl) \\
            &\quad + \int_{l \in [0,L)}  \mu^*_{U^{\phi}}(l)1\Big\{\mu = \mu^*_{U^{\phi}}(l)\Big\}f(d\mu,dl) \\
            &= \int \mu f(d\mu,dl) \\
            &= \mu_0.
        \end{align*}
        \item[3.] [OC-C]: for every $s \in [0,L)$, we can write the difference in [OC-C] as : 
    \begin{align*}
        &\int_{l>s}  U(\mu,l) f^*(d\mu,dl) - \int_{l>s}  U(\mu,s ) f^*(d\mu,dl) \\
        &= \int_{l>s} \Big(U(\mu,l) - U(\mu,s) \Big) f^*(d\mu,dl)  
        \\
        &= \int_{l>s} \Big(U(\mu,l) - U(\mu,s) \Big) f(d\mu,dl) \tag{Definition of $f^*$}
        \\
        &\quad - \int_{l>s} \Big[U(\mu^*_{U^{\phi}}(l),l) - U(\mu^*_{U^{\phi}}(l),s) \Big] \cdot 1\Big\{\mu = \mu^*_{U^{\phi}}(l)\Big\} f(d\mu,dl)\\
        & \quad + \int_{l} \bigg[\mu^*_{U^{\phi}}(l) \big(U(1,L) - U(1,s)\big) + \int_{l'\in[s,L)} (U(0,l') - U(0,s)) \widehat{f}_l(0,dl') \bigg] \\
        &\quad \quad \quad \cdot 1\Big\{\mu = \mu^*_{U^{\phi}}(l)\Big\} f(d\mu,dl)
    \end{align*}
    Note that from \cref{lemma:fhat_binds_OCC}, since the agent is indifferent between continuing and stopping at $s \vee l$ under $\widehat{f}_l$ we have 
    \begin{align*}
        0 &= \int_{l' > s \vee l}\big(U(\mu,l) - U(\mu,s) \big) \widehat{f}_l(d\mu,dl') \\
        &= \mu^*_{U^{\phi}}(l) \big(U(1,L) - U(1,s \vee l)\big) + \int_{l'\in[s \vee l,L)} (U(0,l') - U(0,s \vee  l)) \widehat{f}_l(0,dl').
    \end{align*}
    Hence 
    \begin{itemize}[leftmargin =*]
        \item 
    If $s \geq l$, we have
    \begin{align*}
        \mu^*_{U^{\phi}}(l) \big(U(1,L) - U(1,s)\big) + \int_{l'\in[s,L)} (U(0,l') - U(0,s)) \widehat{f}_l(0,dl') = 0.
    \end{align*}
        \item  If $s < l$, we have
    \begin{align*}
        &\mu^*_{U^{\phi}}(l) \big(U(1,L) - U(1,s)\big) + \int_{l'\in[s,L)} (U(0,l') - U(0,s)) \widehat{f}_l(0,dl') \\
        &= \mu^*_{U^{\phi}}(l) (U(1,l) - U(1,s)) +  (U(0,l) - U(0,s)) \widehat{f}_l\big( \{(0,l): l \in (l,L]\}\big) \\
        &= U(\mu^*_{U^{\phi}}(l),l) - U(\mu^*_{U^{\phi}}(l),s).
    \end{align*}
    \end{itemize}
   and these together imply 
    \begin{align*}
        &\int_{l>s} U(\mu,l) f^*(d\mu,dl) - U\bigg(\int_{l>s} \mu f^*(d\mu,dl),s \bigg) =\int_{l>s} \big(U(\mu,l) - U(\mu,s) \big) f(d\mu,dl) \geq 0,
    \end{align*}
    as desired. 
    \item[4.] [OC-S]: fulfilled by construction. 
    \end{enumerate}
 
 We have verified that $f^*$ satisfies the conditions in \eqref{prob:nature}. We now show that $f^*$ achieves a lower value than $f$:
    \begin{align*}
        &\int V(\mu,l) f^*(d\mu,dl) \\
        &= \int V(\mu,l)f(d\mu,dl) - \int V(\mu^*_{U^{\phi}}(l),l) 1\{\mu = \mu^*_{U^{\phi}}(l)\} f(d\mu,dl) \\
        &\quad + \int\bigg( \mu^*_{U^{\phi}}(l)V(1,L) + \int V(0,l') \widehat{f}_l(0,dl') \bigg) 1\{\mu = \mu^*_{U^{\phi}}(l)\} f(d\mu,dl).
    \end{align*}
    From \cref{lemma:fhat_binds_OCC}, $\widehat{f}_l$, the agent is indifferent between continuing and stopping at $l$. From our comparative static \cref{lem:compstat}, the principal must weakly prefer stopping at $l$ under $\widehat{f}_l$ i.e., 
    \begin{align*}
        V(\mu^*_{U^{\phi}}(l),l) \geq \mu^*_{U^{\phi}}(l)V(1,L) + \int V(0,l') \widehat{f}_l(0,dl'),
    \end{align*}
    for every $l$. Therefore, 
    \[\int V(\mu,l) f^*(d\mu,dl) \leq \int V(\mu,l)f(d\mu,dl),\]
    as desired.
\end{proof}

\underline{{Step 3. The arrival of bad news is supported on $[l^*_{U^{\phi}}(\mu_0),L)$}} \\
The previous step established that a worst-case learning process is obedient bad news; we now establish its specific form is that claimed in \cref{thrm:worstcase}. An advantage of any obedient bad news process $f$ is that the joint distribution over stopping beliefs and stopping levels can be summarized by a single cumulative distribution function $F: \mathcal{L} \to [0,1)$ where $F(s)$ gives the unconditional probability that bad news arrives before level $s \in \mathcal{L}$:
\[
F(s) := f\big(\{0\} \times [0,s] \big) 
\]
and we require $F(L) = 1-\mu_0$. This choice of $F$ induces the principal's payoff
    \begin{align*}
        \mu_0 V(1,L) + \int_0^{L} V(0,s) dF(s),
    \end{align*}
    subject to the agent's continuation constraint [OC-C] at each level $s$ which can be rewritten in terms of $F$ as: 
    \begin{align*}
        &\mu_0 U^\phi(1,L) + \int_{l > s} U^\phi(0,l) dF(l) \geq \mu_0 U^\phi(1,s) + \int_{l > s} U^\phi(0,s) dF(l) \\
        \iff & \mu_0 \Big(U^\phi(1,L) - U^\phi(1,s)\Big)  \geq  \int_{l > s} \Big(U^\phi(0,s)-U^\phi(0,l)\Big) dF(l).
    \end{align*}
    We can thus restate problem \eqref{prob:nature} in terms of optimizing over $F$:  
    \begin{align*}
        &\inf_{F:[0,L] \to [0,1)}  \int_0^{L} V(0,l) dF(l) \tag{N$^{**}$} \label{prob:nature_CDF}
        \\
        &\text{s.t. } \text{$F$ is nondecreasing, $F(L) = 1-\mu_0$, and} \\
        & \underbrace{\mu_0 \Big(U^\phi(1,L) - U^\phi(1,s)\Big)}_{\eqqcolon H(s)}  \geq  \int_{l > s} \Big(U^\phi(0,s)-U^\phi(0,l)\Big) dF(l) \quad \text{for all } s \in \mathcal{L}.
    \end{align*}
    where the first constraint on the nondecreasingness of $F$ and the value of $G(L)$ is equivalent to [Martingale] and the condition that the joint distribution $f$ is a probability distribution; the second constraint is equivalent to [OC-C]; [OC-S] is automatically fulfilled. From the previous step, we know it is without loss to look for an obedient bad news process to solve \eqref{prob:nature} and since there is a bijection between obedient bad news processes $f$ and cumulative distribution functions $F$, $\eqref{prob:nature} = \eqref{prob:nature_CDF}$. 
    
    We now solve \eqref{prob:nature_CDF} which is a fairly standard infinite-dimensional program: we proceed via weak duality by constructing an appropriate set of multipliers. Define $\Lambda^*: \mathcal{L} \to \mathbb{R}_{\geq 0}$ as follows: 
    \begin{align*}
        \Lambda^*(s) := \begin{cases}
            \cfrac{\partial V/ \partial l (0,s)}{\partial U^\phi/ \partial l (0,s)} &\text{if $s> l^*_{U^\phi}(\mu_0)$,} \\
            \cfrac{V \Big(0,l^*_{U^\phi}(\mu_0)\Big)}{U^\phi \Big(0,l^*_{U^\phi}(\mu_0)\Big)} &\text{if $0 < s \leq l^*_{U^\phi}(\mu_0)$,} \\
            0 &\text{if $s=0$,}
        \end{cases}
    \end{align*}
    Note that $\Lambda$ is increasing, so $\frac{d\Lambda^*}{ds} \geq 0$ wherever it is differentiable. Define the Lagrangian 
    \[
    \Psi(F,\Lambda) := \int_0^{L} V(0,l) dF(l) + \int_0^{L} \bigg(\int_{l >s}\Big(U^\phi(0,s)-U^\phi(0,l)\Big) dF(l)  - H(s) \bigg) d\Lambda(s) 
    \]
    which, from standard manipulation, can be rewritten
    \begin{align*}
        \Psi(F,\Lambda) &= \int_0^{L} V(0,l) dF(l) + \int_0^{L} \int_{l >s}\Big(U^\phi(0,s)-U^\phi(0,l)\Big) dF(l) d\Lambda(s)   - \int_0^{L} H(s) d\Lambda(s)  
        \\
        &=\int_0^{L}
        \bigg( 
        V(0,l)  + \int_{0}^s \Big(U^\phi(0,s)-U^\phi(0,l)\Big)  d\Lambda(s) \bigg)  dF(l)  - \int_0^{L} H(s) d\Lambda(s)  \\
        &= \int_0^{L}
        \bigg( 
        \int_{0}^s \cfrac{\partial V(0,l)}{\partial l}   - \cfrac{\partial U^{\phi}(0,l)}{\partial l} \Lambda(l)  dl \Big)dF(s)  - \int_0^{L} H(s) d\Lambda(s) 
    \end{align*}

    Now from our construction of $\Lambda^*$, we have 
    \[
    \frac{\partial V(0,s)}{\partial s} - \frac{\partial U^\phi(0,s)}{\partial s} \cdot \Lambda^*(s) = 0 \quad \text{for every $s > l^*_{U^\phi}(\mu_0)$.}
    \]
    For $0 \leq s \leq l^*_{U^\phi}(\mu_0)$, notice that 
    \begin{align*}
        \int_0^s 
        \bigg(\frac{\partial V(0,l)}{\partial l} &- \frac{\partial U^\phi(0,l)}{\partial l} \cdot \Lambda^*(l) \bigg) dl \\
        &= V(0,s) - U^\phi(0, s) \cdot \frac{V(0,l^*_{U^\phi}(\mu_0))}{U^\phi(0,l^*_{U^\phi}(\mu_0))} > 0
    \end{align*}
    because $ \frac{\partial V/ \partial l (0,s)}{\partial U^\phi/ \partial l (0,s)}$ is increasing in $s$ since $U^{\phi} = g \circ V$ for some increasing convex function $g$. 
    
    Therefore, we have the lower-bound (changing the order of integration) 
    \begin{align*}
        \inf_F \Psi(F,\Lambda^*) &= \inf_F \Bigg[  \underbrace{\int_0^{L} \int_l^L 
        \bigg(\frac{\partial V(0,s)}{\partial s} - \frac{\partial U^\phi(0,s)}{\partial s} \cdot \Lambda^*(s) \bigg) ds   dF(l)}_{\geq 0} -  \int_0^{L} g(l)  d\Lambda^*(l) \Bigg] \\
        &\geq -  \int_0^{L} g(l)  d\Lambda^*(l).
    \end{align*}
    But this can be attained by choosing $F^*$ corresponding to $\widehat f_{l^*_{U^{\phi}}(\mu_0)}$ as follows: 
    \begin{align*}
        F^*(s) = \begin{cases}
            0 & s \leq l^*_{U^\phi}(\mu_0) \\
            \widehat{f}_{l^*_{U^{\phi}}(\mu_0)}\Big(\{0\} \times [0,s] \Big) & s \in (l^*_{U^\phi}(\mu_0),L)\\
            1-\mu_0 & s = L.
        \end{cases}
    \end{align*}
    and observe that (i) by construction $F^*$ is nondecreasing; (ii) $F^*(l^*_{U^{\phi}}(\mu_0)) = 0$; (iii) from \cref{lemma:fhat_binds_OCC}, [OC-C] binds from  for each $s \in (l^*_{U^\phi}, L)$. We have found a feasible $F^*$ which attains the lower bound. But this also solves the primal since 
    \[
    \Psi(F^*,\Lambda^*) =  \inf_F \Psi(F,\Lambda^*) \leq \sup_{\Lambda} \inf_F \Psi(F,\Lambda)  \leq \inf_F \sup_{\Lambda} \Psi(F,\Lambda) 
    \]
    where the first equality is because $F^*$ attains $\inf_F \Psi(F,\Lambda^*)$, and the last inequality is from weak duality.

It remains to  verify that continuation beliefs take the form claimed in \cref{thrm:worstcase}:
\[
\gamma^{F*}(l) = \begin{cases}
    \mu_0 \quad &\text{if $l \leq l^*_{U^{\phi}}(\mu_0)$} \\
    \mu^*_{U^{\phi}}(l) \quad &\text{if $l > l^*_{U^{\phi}}(\mu_0)$}.
\end{cases}
\]
The first case is immediate; for the second case $l > l^*_{U^{\phi}}(\mu_0)$, continuation beliefs can be rewritten directly as: 
\begin{align*}
    \gamma^{F^*}(l) &= \cfrac{\mu_0}{\mu_0 + \Big(1 - F^*(l)\Big)}      = \cfrac{\mu_0}{\mu_0 + \frac{\mu^*_{U^{\phi}}(l^*_{U^{\phi}}(\mu_0))}{\mu^*_{U^{\phi}}(l)} -  \mu^*_{U^{\phi}}(l^*_{U^{\phi}}(\mu_0))} 
    = \mu^*_{U^{\phi}}(l) 
\end{align*}
where the last equality is because $\mu^*_{U^{\phi}}(l^*_{U^{\phi}}(\mu_0)) = \mu_0$ by definition.

\underline{\smash{Step 4: Worst-case payoff arbitarily poor.}}

Consider that
\begin{align*}
    \lim_{L \to \infty} V(\mu_0,L) = \lim_{L \to \infty} V(0, L)\bigg( \mu_0 \cdot \frac{V(1,L)}{V(0,L)} + (1-\mu_0) \bigg) = \lim_{L \to \infty} V(0,L) (1-\mu_0).
\end{align*}
Since $V(0, L)< 0 $ and
\[\lim_{L \to \infty} |V(0,L)| = \underbrace{\bigg\lvert \frac{V(0,L)}{V(1,L)}}_{\to \infty} \bigg\rvert \cdot \underbrace{|V(1,L)|}_{>V(1,1) > 0} = \infty. \]
Thus, $\lim_{L \to \infty} V(\mu_0,L) = -\infty.$ Fix some small $\epsilon>0$. For each sufficiently large $L$, 
there exists an agent's utility function $U_{L} \in \mathbb{U}$ such that $l^*_{U^\phi_{L}}(\mu_0) \in [L-\epsilon , L].$ Under the no information learning process, the agent with the utility function $U_{L}$ stops within $[L-\epsilon , L]$ with probability $1$. Thus, 
\[\inf_{\substack{\mathcal{F} \in \overline{\mathbb{F}} \\ U \in \mathbb{U}}} \Ex\Big[V( \mu_{\ell^*}, \ell^*(\phi, \mathcal{F}, U)) \Big] \leq \lim_{L \to \infty} V(\mu_0,L-\epsilon) - \Ex[V(\mu_0,\overline{\el})] \to -\infty
\]
which implies $\lim_{L \to +\infty} \inf_{\substack{\mathcal{F} \in \overline{\mathbb{F}} \\ U \in \mathbb{U}}} \Ex\big[V( \mu_{\ell^*}, \ell^*(\phi, \mathcal{F}, U)) \big] = -\infty$ as desired.
\end{proof}

\subsection{Proof of \cref{prop:worstcasecompstat}}
\begin{proof}
    We know from \cref{thrm:worstcase} that the principal's worst information structure induces a distribution of stopping belief and time as follows: 
    \begin{align*}
        \widehat{f}_{l^*_{U^\phi}(\mu_0)}\Big(\{(1,L)\}\Big) &= \mu_0
        \quad 
        \quad 
        \widehat{f}_{l^*_{U^\phi}(\mu_0)}\Big(\{(0,l) : l \geq l^*_{U^\phi}(\mu_0)\}\Big) = \frac{\mu_0}{  \mu^*_{U^{\phi}}(l)} - \mu_0.
    \end{align*}
    Thus, the principal's worst-case payoff under mechanism $\phi$ can be rewritten as: 
    \begin{align*}
        \mu_0 V(1,L) + \int_{l^*_{U^\phi}(\mu_0)}^{L} V(0,l) \widehat{f}_{l^*_{U^\phi}}(0,dl).
    \end{align*}
    Since $U^{\phi'} (\theta,l) = g \circ U^{\phi}(\theta,l)$, Lemma 1 implies $l^*_{U^\phi}(\mu_0) \leq l^*_{U^{\phi'}}(\mu_0)$ and $\mu^*_{U^\phi}(l) \geq  \mu^*_{U^{\phi'}}(l)$. Thus, the distribution of $\widehat{f}_{l^*_{U^\phi}}(0,\cdot)$ is strictly dominated by that of $\widehat{f}_{l^*_{U^{\phi'}}}(0,\cdot)$ in the first-order stochastic dominance sense. Since $V(0,l)$ is decreasing in $l$, we must have
    \begin{align*}
          \int_{l^*_{U^{\phi}}(\mu_0)}^{L} V(0,l) \widehat{f}_{l^*_{U^{\phi}}}(0,dl) \geq  \int_{l^*_{U^{\phi'}}(\mu_0)}^{L} V(0,l) \widehat{f}_{l^*_{U^{\phi'}}}(0,dl),
    \end{align*}
    implying the principal's payoff under mechanism $\phi$ is better than that under mechanism $\phi'$, as desired.
\end{proof}

\clearpage 

\titleformat{\section}
		{\normalsize\bfseries\center \MakeUppercase }     
         {\thesection}
        {0.5em}
        {}
        []
\renewcommand{\thesection}{\Roman{section}}

\begin{center}
    \large{\textbf{ONLINE APPENDIX TO \\ `ROBUST TECHNOLOGY REGULATION'}}\\
    \small{ANDREW KOH \quad SIVAKORN SANGUANMOO} \\ 
    \footnotesize{\url{ajkoh@mit.edu} \quad \url{sanguanm@mit.edu}}
\end{center}

\setcounter{footnote}{0}

\setcounter{page}{1}
\setcounter{section}{0}

\section{Remaining technical proofs}\label{app:technical}
We now show the remaining technical proofs that were omitted from the main text. 

\subsection{Proof of \cref{lem:technical_dom}}
We begin with the following lemma stating that optimal stopping levels that solve \eqref{prob:OST-0} in the proof of \cref{lemma:zero_marginal} must form a kind of `lattice'.
\begin{lemma}\label{lem:OST forms lattice}
 Let $\epsilon_1>\epsilon_2$ be positive real numbers. Let $\eta^{\epsilon_1}$ and $\eta^{\epsilon_2}$ solve (OSP-$\epsilon_1$) and (OSP-$\epsilon_2$), respectively.  Then, $\eta_* \coloneqq \eta^{\epsilon_1} \wedge \eta^{\epsilon_2} $ and $\eta^* \coloneqq \eta^{\epsilon_1} \vee \eta^{\epsilon_2}$ also solve (OSP-$\epsilon_1$) and (OSP-$\epsilon_2$), respectively.
\end{lemma}
\begin{proof}[Proof of \cref{lem:OST forms lattice}]
    Consider that
    \begin{align*}
        &\Ex[V(\theta,\eta^{\epsilon_1})1\{Z^{\epsilon_1} = 0\}] = \Phi(\epsilon_1) \geq \Ex[V(\theta,\eta_*)1\{Z^{\epsilon_1} = 0\}]\\
        &\implies   \Ex[(V(\theta,\eta^{\epsilon_1}) - V(\theta, \eta^{\epsilon_2}))1\{\eta^{\epsilon_1} > \eta^{\epsilon_2}\}] \\
        &\qquad \geq \epsilon_1 \underbrace{\Ex[V(1,\eta^{\epsilon_1}) - V(1, \eta^{\epsilon_2}))1\{\eta^{\epsilon_1} > \eta^{\epsilon_2}, \omega \in A^*_1\}]}_{\geq 0}.
    \end{align*}
    Similarly,
    \begin{align*}
        &\Ex[V(\theta,\eta^{\epsilon_2})1\{Z^{\epsilon_2} = 0\}] = \Phi(\epsilon_2) \geq \Ex[V(\theta,\eta^*)1\{Z^{\epsilon_2} = 0\}]\\
        &\implies  \Ex[(V(\theta,\eta^{\epsilon_1}) - V(\theta, \eta^{\epsilon_2}))1\{\eta^{\epsilon_1} > \eta^{\epsilon_2}\}] \\
        &\qquad \leq \epsilon_2 \underbrace{\Ex[V(1,\eta^{\epsilon_1}) - V(1, \eta^{\epsilon_2}))1\{\eta^{\epsilon_1} > \eta^{\epsilon_2}, \omega \in A^*_1\}]}_{\geq 0}.
    \end{align*}
    Since $\epsilon_1 > \epsilon_2$, both inequalities must be equalities so
    \begin{align*}
        \Phi(\epsilon_1) = \Ex[V(\theta,\eta_*)1\{Z^{\epsilon_1} = 0\}] \quad \text{and} \quad  \Phi(\epsilon_2) = \Ex[V(\theta,\eta^*)1\{Z^{\epsilon_2} = 0\}]
    \end{align*}
    as desired.
\end{proof}
\begin{proof} [Proof of \cref{lem:technical_dom}]
Suppose $s^{\epsilon_n}$ is the latest optimal stopping level that solves (OSP-$\epsilon_n$). First, we show that $s^{\epsilon_n}$ is increasing in $n$ almost surely. From \cref{lem:OST forms lattice}, $s^{\epsilon_n} \vee s^{\epsilon_{n+1}}$ must solve (OSP-$\epsilon_{n+1}$). The definition of $s^{\epsilon_{n+1}}$ implies $s^{\epsilon_n} \vee s^{\epsilon_{n+1}} \leq s^{\epsilon_{n+1}}$ almost surely. Thus, $s^{\epsilon_n} \leq s^{\epsilon_{n+1}}$ almost surely, as desired. Thus, $s^{\epsilon_n} \uparrow s'$ for some $s'$ almost surely and $s'$ is still a stopping level because $(\mathcal{G}_l)_l$ is a right-continuous filtration. 

Next, we verify $\Phi$ must be a continuous function. Consider that, for every positive real numbers $\epsilon_1$ and $\epsilon_2$,
\begin{align*}
    \Phi(\epsilon_2) - \Phi(\epsilon_1) &= \Ex[V(\theta,s^{\epsilon_2})1\{Z^{\epsilon_2} = 0\}] - \Ex[V(\theta,s^{\epsilon_1})1\{Z^{\epsilon_1} = 0\}] \\
    \\&\leq \Ex[V(\theta,s^{\epsilon_2})1\{Z^{\epsilon_2} = 0\}] - \Ex[V(\theta,s^{\epsilon_2})1\{Z^{\epsilon_1} = 0\}] \\
    &= \Ex[V(\theta,s^{\epsilon_2})1\{Z^{\epsilon_1} = 1\}] - \Ex[V(\theta,s^{\epsilon_2})1\{Z^{\epsilon_2} = 1\}] \\
    &= (\epsilon_1 - \epsilon _2) \Ex[V(\theta , s^{\epsilon_2})1\{\omega \in A^*_1\}] \\
    &\leq M|\epsilon_1- \epsilon_2|, 
\end{align*}
where $M \coloneqq \sup_{\theta , l} |V(\theta,l)|$, which is well-define because the level space is compact.
Similarly, $\Phi(\epsilon_1) - \Phi(\epsilon_2) \leq M|\epsilon_1- \epsilon_2|$. This implies $|\Phi(\epsilon_1) - \Phi(\epsilon_2)| \leq M|\epsilon_1 - \epsilon_2|,$ so $\Phi$ is continuous.

Finally, we show that $s' = \overline{\el}$ almost surely. Consider that
\begin{align*}
    \Ex[V(\theta,s')] &= \lim_{n \to \infty} \Ex[V(\theta , s^{\epsilon_n})] \tag{Dominated convergence theorem} \\
    &= \lim_{n \to \infty} \Ex[V(\theta , s^{\epsilon_n}) 1\{Z^{\epsilon_n} = 0\}] + \lim_{n \to \infty} \Ex[V(\theta , s^{\epsilon_n}) 1\{Z^{\epsilon_n} = 1\}] \\
    &= \lim_{n \to \infty} \Ex[V(\theta , s^{\epsilon_n}) 1\{Z^{\epsilon_n} = 0\}] \tag{$\Pr(Z^{\epsilon_n} = 1) \to 0$ as $n \to \infty$}\\
    &=\lim_{n \to \infty} \Phi(\epsilon_n) \\
    &= \Phi(0) \tag{$\Phi$ is continuous} \\
    &= \Ex[V(\theta, \overline{\el})].
\end{align*}
Since $\overline{\el}$ is the principal's unique optimal stopping level, we must have $s' = \overline{\el}$ almost surely. Thus,  $s^{\epsilon_n} \uparrow \overline{\el}$, as desired.
\end{proof}
\subsection{Proof of \cref{thrm:dominance} when $\bar \ell < L$ w.p.p.} \label{OA: remaining proof of dominance}
We assumed in the main text that $\overline{\el} < L$ almost surely. We relax this assumption and show \cref{thrm:dominance}. We start with the following lemma stating that the hard limit is strictly before $L$ with positive probability given any interim history.
\begin{lemma} \label{lem: stop before L with postive prob}
For every stopping level $\ell<L,$ we must have $\Pr(\overline{\el} < L | \mathcal{G}_\ell) > 0$ almost surely.    
\end{lemma}
\begin{proof}[Proof of \cref{lem: stop before L with postive prob}]
Suppose towards a contradiction there exists a stopping level $\ell<L$ such that the event
\[ \mathbb{P}\Big(\big\{ \omega \in \Omega: \Pr(\overline{\el} = L | \mathcal{G}_\el) = 0 \big\}\Big) > 0.\] Our assumption on the law of the principal's signal process $\Pr^\theta$ implies the event
\[ A = \Big\{ \omega \in \Omega: \Pr(\overline{\el} = L | \mathcal{G}_\el) = 0, \mu^P_{\el+\epsilon} < \delta \Big\},\]
where $\mu^P_l \coloneqq \Pr(\theta = 1| \mathcal{G}_\el)$ also happens with positive probability, where we choose $\delta>0$ so that $V(\delta,L) < V(\delta,l)$ for every $l < L$.\footnote{We simply choose small $\delta$ so that $\frac{\delta}{1-\delta} < \sup_{l \in [0,L]} |\frac{\partial V(1,l)/\partial l}{\partial V(0,l)/\partial l}|$.} Note that the event $A$ is $\mathcal{G}_{\el+\epsilon}$-measurable. We define a stopping level $\el'$ as follows
\begin{align*}
    \ell' \coloneqq \begin{cases}
        \overline{\el} &\text{if $\omega \notin A$}\\
        \ell+\epsilon &\text{if $\omega \in A$}
    \end{cases}
\end{align*}
It is easy to see that $\ell'$ is a valid stopping level because $A$ is $\mathcal{G}_{\el+\epsilon}$-measurable. Under event $A$, we must also have $\overline{\el} = L $ almost surely. Thus,
\begin{align*}
    \Ex[V(\mu_{\overline{\el}},\overline{\el}) - V(\mu_{\ell'} , \ell')] &= \Ex[(V(\mu_{\overline{\el}},\overline{\el}) - V(\mu_{\ell'} , \ell')) 1 \{\omega \in A\}]  \\
    &= \Ex[(V(\mu_{\ell^*}, L) - V(\mu_{\ell + \epsilon} , \ell+\epsilon))1 \{\omega \in A\}]\\
    &= \Ex[(V(\mu_{\ell+\epsilon}, L) - V(\mu_{\ell + \epsilon} , \ell+\epsilon))1 \{\omega \in A\}] \tag{optional stopping theorem}\\
    & < 0,
\end{align*}
where the inequality follows from $\mu_{\ell+\epsilon} < \delta$ and $V(\delta,L) < V(\delta , l)$ for every $l < L$, and $\Pr(A) > 0.$ This contradicts the optimality of $\overline{\el},$ as desired.
\end{proof}

 We are now ready to show the proof of \cref{thrm:dominance} for the case that $\overline{\ell} = L$ with positive probability.
\begin{proof} [Remaining proof of \cref{thrm:dominance}]
    Consider any stopping times $\el_1 \leq \el_2 \leq \overline{\el}$. We will show that $\phi_{\el_1} \geq \phi_{\el_2}$ almost surely. Suppose toward a contradiction that the event $A \coloneqq \{ \omega : \phi_{\el_1} < \phi_{\el_2}\}$ happens with positive probability. From \cref{lemma:zero_marginal}, we have, under event $A$, $\phi_{\overline{\el}} \leq \phi_{\el_1} < \phi_{\el_2}$. 
    Since $A$ is $\mathcal{G}_{\el_2}$-measurable, $\Pr(A \cap \{\overline{\el} < L\}) > 0$. Under event $A \cap \{\overline{\el} < L\}$, we have $\phi_{\el_1} < \phi_{\el_2} \geq \phi_{\overline{\el}} < \infty = \phi_L$, which contradicts single-peakedness, as desired. Thus, $\phi$ is decreasing until the hard limit, and then the remaining proof follows the same as the proof in the main text.
\end{proof}
\subsection{Proof of \cref{lemma: hard limit - conditionally undominated}} \label{oa: proof of hard limit - time consistence}

\begin{proof} [Proof of \cref{lemma: hard limit - conditionally undominated}]
 Suppose, towards a contradiction, there exists a stopping level $\ell_1 > \overline{\el}$ such that the event
    \[ A_0 = \Big\{\omega : \big\{\ell_1 - \epsilon > s \geq  \overline{\el} > \ell_0 \big\} \bigcap \big\{ \phi'_{\ell_1} - \phi'_{\el_0}< M \text{ or } \phi'_{\el_1} = -\infty \big\}\Big\} \cap A^<_{\el_0}
    \]
    happens with positive probability for some $\epsilon > 0, s \in \mathcal{L}$, and some $M < +\infty$. Note that the event is $\mathcal{G}_L$-measurable.
    
    We construct an auxiliary random variable. Let $S$ be a Bernoulli random variable (taking values in $\{0,1\}$) such that 
    \[ \Pr\big(S=0 \big| \mathcal{G}_L,\theta = 0\big)  = 1 \quad \text{and} \quad \Pr\big(S=0\big| \mathcal{G}_L,\theta = 1\big) = \max\bigg\{  \frac{\delta(1-\mu_L)}{\mu_L (1-\delta)},1\bigg\},\]
    where $\mu_L = \Pr(\theta = 1 | \mathcal{G}_L) $ and $\delta$ is sufficiently small so that $V(\delta,l_0) > V(\delta,l_1)$ for every $l_1 > l_0$. This condition is equivalent to
    \begin{align*}
    \frac{\delta}{1-\delta} < \inf_{l_1 > l_0} \frac{V(1,l_1) - V(1,l_0)}{V(0,l_0) - V(0,l_1)},
    \end{align*}
    where the RHS is weakly greater than $\inf_{l \in [0,L]}\Big\lvert \frac{\partial V(1,l)/\partial l}{\partial V(0,l)/\partial l} \Big\rvert$, which is strictly positive. The belief about the state given by all principal's information and $S$ is
    \begin{align*}
    \Pr(\theta = 1 | \mathcal{G}_L, S = 1) &= 1 \\
    \Pr(\theta = 1| \mathcal{G}_L, S = 0) &= \frac{\Pr(S=0| \mathcal{G}_L, \theta = 1)\mu_L}{\Pr(S=0| \mathcal{G}_L, \theta = 1) \mu_L + \Pr(S=0| \mathcal{G}_L, \theta = 0) (1-\mu_L)} = \min\{\delta, \mu_L\}.
    \end{align*}
    Since $A_0$ is  $\mathcal{G}_L$-measurable, the event $A^* \coloneqq A_0 \cap \{\omega: S= 0, \mu_{L} > \underline{\delta} \}$ happens with positive probability for some $\underline{\delta} > 0$.

Consider the following agent's filtration:
    \begin{align*}
    \mathcal{F}^\epsilon_l \coloneqq \sigma\Big( \Big\{ X \cap \{\omega: l < \ell_0\} : X \in \mathcal{G}_l \Big\} \cup \Big\{ X \cap \{\omega: l \geq \ell_0, S =s\}  : X \in \mathcal{G}_L, s \in \{0,1\}\Big\}  \Big).
\end{align*}
In words, at $\ell_0$, the agent learns everything the principal will learn in the future as well as information from the extra signal $S$. In particular, for every stopping level $\el$, if $\el< \el_0$, then $\mathcal{F}_\el = \mathcal{G}_\el$. On the other hand, if $\el \geq \el_0$, then $\mathcal{F}_\el = \mathcal{G}_L \vee \sigma(S).$

This implies, for every $l$, under the event that $\{ \ell_0 \leq l\}$,
\begin{align*}
\Pr(\theta = 1| \mathcal{F}_l) = \Pr(\theta = 1| \mathcal{G}_L, S) = \begin{cases} 1 &\text{if $S=1$} \\ \min\{\delta, \mu_L\} &\text{if $S=0$.} \end{cases} \eqqcolon \mu^U_{l_0}
\end{align*}
Suppose $\el^* \coloneqq \el^*(\phi^*, \mathcal{F},U)$ is the agent's stopping time under the adaptive sandbox mechanism $\phi^*$, facing the learning process $\mathcal{F}$, with preference $U$. We can use the same argument from Step 2 in the proof of \cref{thrm:dominance} to show that $\el^* \geq \overline{\el} \wedge \el_0$ almost surely. Under the event $\{S=1, \overline{\el} > \el_0\}$, the agent learns the state is perfectly good at level $\el_0$, so he will continue developing until the hard limit $\overline{\el}$ under the sandbox mechanism. Therefore, $\Pr(S = 0| \mathcal{G}_{\el_0}, \el^* = \el_0) = 1$ under the event that $\{\overline{\el} > \el_0\}$, i.e., the principal believes that $S=0$ with probability one after observing that the agent stopped at $\el_0$ strictly before the sandbox limit. Thus, under the event $\{\el^*= \el_0\}$, we must have $\mu^U_{\el_0} \leq \delta$.

We now construct the agent's preference $U \in \mathbb{U}$ as follows. We set 
\begin{align*}
    U(\theta,l) = \begin{cases}
        V(1,s) + \beta(V(1,l) - V(1,s)), &\text{if $l \geq s$ and $\theta = 1$} \\
        V(\theta , l), &\text{otherwise},
    \end{cases}
\end{align*}
for sufficiently large $\beta>1$ which will be chosen later. 
In words, the agent and the principal have aligned preferences until the level $s$, and then the agent becomes highly optimistic about the technology state. It is easy to see that $U \in \mathbb{U}.$\footnote{This is because $U = g \circ V$, where $g: \mathbb{R} \to \mathbb{R}$ is a convex function such that \begin{align*}
        g(x) = \begin{cases}
        x  &\text{if $x < V(1,s)$},\\
        V(1,s) + \beta(x - V(1,s)) &\text{if $x \geq V(1,s)$},
        \end{cases}
    \end{align*}}
    
 First, we will show that $A^* \subset \{\omega: \el^* = \el_0\}$, i.e., under the event $A^*$, the agent must stop at $\el_0$ under the adaptive sandbox mechanism. Due to the hard limit, the optimal stopping $\el^*$ must be before the hard limit $\overline{\el}$. Since $\overline{\el} < s$ under the event $A^*$, we must have $\el^* < s$. Under event $A^*$, we must have
\begin{align*}
&\Ex[U(\mu^U_{\el^*},\el^*)] \geq U(\mu^U_{\el_0},\el_0)\\
\Longrightarrow & \mu^U_{\el_0}(V(1,\el^*) - V(1,\el_0)) + (1-\mu^U_{\el_0})(V(0,\el^*) - V(0,\el_0)) \geq 0 \tag{$U = V$ whenever $l < s$} \\
\Longrightarrow &\delta(V(1,\el^*) - V(1,\el_0)) + (1-\delta)(V(0,\el^*) - V(0,\el_0))  \geq 0 \tag{Since $S=0$, $\mu^U_{\el_0} \leq \delta$} \\
\Longrightarrow & V(\delta, \el^*) \geq V(\delta,\el_0).
\end{align*}
By the construction of $\delta$, we must have $\el^* = \el_0$, as desired.

Suppose $\el_{\phi'} \coloneqq \el^*(\phi, \mathcal{F},U)$ is the agent's stopping time under the mechanism $\phi'$, facing the learning process $\mathcal{F}$, with preference $U$ (after the agent has been developing until the level $\el_0$). Thus, $\el_{\phi'}$ must solves the following stopping problem
\begin{align*}
&\sup_{\el \geq \el_0} \Ex[U^{\phi'}(\theta,\el)] \quad \text{s.t. $\el$ is adapted to $\mathcal{F}_l$}
\end{align*}
This implies $\el_{\phi'} \in \text{arg max}_{\el \geq \el_0} U^{\phi'}(\mu_{\el_0}, \el)$ almost surely because
\begin{align*}
\Ex[U^{\phi'}(\theta,\el)] = \Ex[\Ex[U^{\phi'}(\theta,\el)|\mathcal{G}_L ,S]] \leq \Ex[\max_{\el \geq \el_0}\Ex[U^{\phi'}(\theta,l) | \mathcal{G}_L,S]] = \Ex[\max_{\el \geq l_0} U^{\phi'}(\mu_{\el_0},\el)],
\end{align*}
where the last inequality follows from the fact that $\Ex[\theta | \mathcal{G}_L,S] = \mu_{\el_0}$ and $U^{\phi'}$ is linear in belief. Consider that  $\el_\phi$ must be strictly greater than $\el_0$ under the event $A^*$ because we have
    \begin{align*}
    U(\mu_{\el_0}, \el_1) - U(\mu_{\el_0} , \el_0) 
    &\geq U(\underline{\delta},\el_1)  - U(\underline{\delta},\el_0) \\
    &= (\underline{\delta} \beta V(1,\el_1) + (1-\underline{\delta}) V(0,\el_1)) - (\beta-1)\underline{\delta}V(1,s) - V(\underline{\delta},\el_0) \tag{$\el_1 > s$}\\
    &\geq (\underline{\delta} \beta V(1,s + \epsilon) + (1-\underline{\delta}) V(0,s+\epsilon)) - (\beta-1)\underline{\delta}V(1,s) - V(\underline{\delta},\el_0) \tag{$\el_1 > s+\epsilon $}\\
    &\geq  \underline{\delta}\beta\underbrace{\big( V(1,s + \epsilon) - V(1,s)  \big)}_{>0} \\
    &\qquad + \big(\underline{\delta} V(1,s) + (1-\underline{\delta}) V(0,s + \epsilon) - V(\underline{\delta},0) \big) \tag{$V(\underline{\delta},0) > V(\underline{\delta},\el_0)$}   \\
    &\geq M+1,
    \end{align*}
    by setting $\beta$ large enough so that the second and third inequality hold.\footnote{For the second inequality, a sufficient condition is
    \begin{align*}
         \beta > \frac{1-\underline{\delta}}{\underline{\delta}} \sup_{l\in [0,L]} \Big\lvert \frac{\partial V(0,l)/\partial l}{\partial V(1,l)/\partial l} \Big\rvert\Longrightarrow \frac{\partial}{\partial l} \big(\underline{\delta} \beta V(1,l) + (1-\underline{\delta}) V(0,l)\big) > 0 \quad\forall l
    \end{align*}
    For the third inequality, the multiplier of $\beta$ is strictly positive, so the inequality must hold when $\beta$ is sufficiently large.} Under the event $A^*$, we know either $\phi'_{\el_1} - \phi'_{\el_0} <M$ or $\phi_{\el_1} = -\infty$. If $\phi_{\el_1} - \phi_{\el_0} < M$  we must have
    \begin{align*}
    U^{\phi'}(\mu_{\el_0},\el_1) - U^{\phi'}(\mu_{\el_0},\el_0) \geq (M+1) - M =1,
    \end{align*}
    so stopping at $\el_0$ is not optimal under the event $A^*$, implying $\el_{\phi'} > \el_0$ under the event $A^*$, as desired. On the other hand, if $\phi'_{\el_1} = -\infty $, then $\el_0$ is not the latest optimal stopping under the event $A^*$, implying $\el_{\phi'} > \el_0$ under the event $A^*$, as desired.
    
    Now we compare the principal's payoff under $\phi^*$ and $\phi'$ under the event $\{\omega: \ell^* = \ell_0\} \cap A^<_{\ell_0}, $. We showed earlier that $\mu_{\el_0} \leq \delta$ under the event $\{\el^* = \el_0\}$. Because $V(\delta,\el_0) > V(\delta , \el)$ for every $\el >\el_0$, we must have $V(\mu_{\el_0},\el_0) \geq V(\mu_{\el_0},\el_{\phi'})$ under the event $\{\el^* = \el_0\}$, and the inequality is strict under the event $A^*$ because $\el_{\phi'} > \el_0$. Since the event $A^*$ happens with positive probability, we must have
    \begin{align*}
    &\Ex[V(\theta,\el_{\phi'})1\{\el^* = \el_0 , \omega \in A^<_{\ell_0}\}] \\
    &= \Ex[V(\mu_{\el_0},\el_{\phi'}) )1\{\el^* = \el_0 , \omega \in A^<_{\ell_0}\}] \\
    &= \Ex[V(\mu_{\el_0},\el_{\phi'}) 1\{\omega \in A^*\}] + \Ex[V(\mu_{\el_0},\el_{\phi'}) 1\{\omega \notin A^*, \el^* = \el_0, \omega \in A^<_{\ell_0}\}  ] \tag{$A^* \subset \{\omega: \el^* = \el_0$\}} \\
    &< \Ex[V(\mu_{\el_0},\el_0) 1\{\omega \in A^*\}] + \Ex[V(\mu_{\el_0},\el_0) 1\{\omega \notin A^*, \el^* = \el_0, \omega \in A^<_{\ell_0}\}  ]  \\
    &= \Ex[V(\mu_{\el_0},\el_0) 1\{\el^* = \el_0, \omega \in A^<_{\ell_0}\}].
    \end{align*}
This implies $\phi'$ performs strictly worse than $\phi^*$ under the event $A^<_{\ell_0}$ given the principal knows the agent stops $\el_0$, which contradicts $\phi'$ conditionally dominates $\phi^*$ upon stopping at $\el_0$, as desired.

\end{proof}


\section{Experimentation under risk-aversion} \label{app:experimentation discussion}

We relate our comparative static on optimal experimentation to extant work on optimal stopping and risk aversion.

\subsection{Generalizing learning processes} We show that \cref{lem:compstat} generalizes the main result of \cite{keller2019note} (henceforth KNW). We start by mapping their exponential bandit setting into our framework. Interpret $l$ as the level (in KNW, time) the agent switches from the risky to the safe arm. Let $\theta = 1$ correspond to the risky arm paying out a transfer of $h > 0$ at rate $\lambda > 0$, and $\theta = 0$ correspond to never paying out. The safe arm pays out a transfer of $s > 0$ at rate $1$.

Then, we have that if the arm is good: 
\begin{align*}
    u(1,l) &= \Ex\Big[\int^l_0 e^{-rs} \cdot 1\{\text{arrive at $l$} \} \cdot u(h) ds \Big]  = \cfrac{\lambda \cdot u(h) - u(s) }{r} \Big(1 - e^{-rl} \Big) \tag{Bandit-1} \label{eqn:bandit_good}
\end{align*}
where the second equality makes use of the fact (sometimes called `Poissonization') that conditional $K$ arrives over the interval $[0,l]$ the arrival locations are uniformly distributed over $[0,l]$. 

But if the arm is bad: 
\begin{align*}
    u(0,l) &= -\Ex\Big[\int^l_0 e^{-rs} \cdot 1\{\text{arrive at $l$} \} \cdot u(s) ds \Big]  = - \cfrac{u(s) }{r} \Big(1 - e^{-rl} \Big) \tag{Bandit-0} \label{eqn:bandit_bad}
\end{align*}

Suppose that for two increasing concave utility functions $u_1, u_2$ normalized to $u_1(0) = u_2(0) = 0$, we have that $u_2 = f \circ u_1$ for some increasing concave function $f$. We call $u_2$ more \emph{flow risk averse} than $u_1$ where the qualifier `flow' is to emphasize that this risk aversion is imposed on flow payoffs rather than on the timing of rewards. 

\begin{corollary}[Generalizing \cite{keller2019note}] \label{cor:generalized_keller} Consider the model with payoffs given by \eqref{eqn:bandit_good} and \eqref{eqn:bandit_bad}. 
\begin{itemize}
    \item[(i)] If $h > s$ then under any learning process, the more flow risk averse agent stops sooner. 
    \item[(ii)] If $h < s$ then under under any learning process, the more flow risk averse agent stops later. 
\end{itemize}
    Proposition 1 of \cite{keller2019note} is recovered as the special case with learning process is given by the arrival of perfect good news with rate $\lambda > 0$. 
\end{corollary}
By decoupling the learning process from the reward process, this is also in the spirit of \citep*{lizzeri2024disentangling}. To see \cref{cor:generalized_keller}, define the following function: 
\[
g(v) = 
\begin{cases}
        \cfrac{\lambda u_2(h) - u_2(s)}{\lambda u_1(h) - u_1(s)} \cdot v \quad &\text{if $x \geq 0$} \\
        \cfrac{u_2(s)}{u_1(s)} \cdot v \quad &\text{if $x <0$} 
\end{cases}
\]
so that $v_2 = g \circ v_1$. Now this function is convex if and only if 
\begin{align*}
    &\cfrac{\lambda u_2(h) - u_2(s)}{u_2(s)} \geq \cfrac{\lambda u_1(h) - u_1(s)}{u_1(s)} \\
    &\iff \cfrac{u_2(h)}{u_2(s)} \geq \cfrac{u_1(h)}{u_1(s)}
\end{align*}
which requires $h < s$. By \cref{lem:compstat}, an agent with preference $v_2$ stops later than the agent with preference $v_1$. But recalling that $v_2$ corresponds to the flow payoff of $u_2$ (more flow risk-averse), so this delivers the corollary. The learning process in KNW is a special case of `perfect good news' arriving at rate $\lambda > 0$ so they can characterize the stopping boundary exactly. 

\subsection{General comparative static over time and risk}
\cref{lem:compstat} is useful for settings beyond exponential discounting. For instance, both \cite*{chancelier2009risk} (henceforth CDP) and \cite*{quah2013discounting} (henceforth QS) deliver conditions under which changing the preferences (flow risk aversion in CDP, impatience in QS) can lead an agent to experiment less. \cref{lem:compstat} can be viewed as partially generalizing these insights by incorporating risk-aversion both within and across time/levels. 

CDP works within an (indexable) multi-armed bandit setting and argues that higher \emph{flow risk aversion} ensures that the Gittins index for each risky arm is accordingly lower than that for an agent with lower flow risk aversion. QS apply their results on a \emph{patience order} over discount rates to experimentation (see Section II.F of QS), showing that a \emph{more impatient} agent explores less in multi-arm bandit environments. 

We now develop the simplest model of experimentation that allows for both time and risk preferences. This partially nests CDP who fix time preference and vary risk preferences, and in the spirit of QS who fix risk preference and vary time preferences. There is a single risky arms with unknown flow payoff given by the independent random variable $X_l \sim F^{\theta} \in \Delta(\mathbb{R}_+)$ where $\theta \in \Theta$ parameterizes the distribution over risky arns and $\theta \geq \theta' \implies F^{\theta} \geq_{FOSD} F^{\theta'}$. The safe arms pays out a flow payoff of $1$ per unit time and to make the problem non-trivial we assume there exists $\underline \theta \in \Theta$ such that $\Ex[X|\underline\theta] < 1$. Payoffs are aggregated across levels (in QS, time) via exponential discounting so we can define: 
\begin{align*}
&v(\theta,l) := \Ex\Big[\int^l_0 \alpha(s) \cdot \big[u(X_s) - u(1)\big] ds \Big| \theta \Big]   \\
&\quad \text{for discount factor $\alpha: \mathcal{L} \to \mathbb{R}_+$ with bounded variation} \\
&\quad \text{and flow utility $u: \mathbb{R} \to \mathbb{R}$.}    
\end{align*}
The \emph{patience order} of QS requires that for two discount factors $\alpha, \alpha'$ their ratio $\alpha(l)/\alpha(l')$ is increasing in $l$. The main result of QS shows that the patience order is quite generally a sufficient condition to order the duration of experimentation (fixing other aspects of the decision problem). Within the above bandit setting, the condition on patience implies the existence of increasing convex functions $(g^{\theta})_{\theta}$ such that $u(\theta,\cdot) = g^{\theta} \circ v(\theta,\cdot)$ for each $\theta \in \Theta$. In fact, we can pick a single $g: \mathbb{R} \to \mathbb{R}$ since just taking partials at $l \in \mathcal{L}$: 
\[
\Big(\cfrac{\partial u(\theta,\cdot)}{\partial l}  \Big / \cfrac{\partial v(\theta,\cdot)}{\partial l}\Big)= \cfrac{\alpha(l) \Ex[u(X_l) | \theta]}{\alpha'(l) \Ex[u(X_l) | \theta]} = \cfrac{\alpha(l)}{\alpha'(l)} 
\]
which via straightforward manipulation means that $g' = \alpha(l)/\alpha'(l)$ is increasing and hence $g$ is convex. 

The \emph{flow risk aversion} from the previous section ranks $u,v$ in the usual way: there exists some increasing convex function $f$ such that $u = f \circ v$. Supposing that the discount rate is identical and suppose, as is assumed in Theorem 1 of CDP that we have that $u(1) \leq v(1)$ while for all $\theta \in \Theta$ we have $\Ex[u(X)|\theta] \geq \Ex[v(X)|\theta]$ then 
\[
g^{\theta}(l) = \cfrac{\Ex[u(X)|\theta] - u(1)}{\Ex[v(X)|\theta] - v(1)} \cdot l
\]
with slope either $[0,1)$ when the state is such that the safe arm is better, and $[1,+\infty)$ if the state is such that the risky arm is better for the agent with flow utility $u = f \circ v$ and with slight modification of the proof of \cref{lem:compstat} we can show that this implies that the agent who is more flow risk-averse stops experimentation sooner.  

Of course, there there are experimentation problems in which both impatience and flow payoffs cannot be ranked along the patience order and flow risk-aversion. But some of them can still be ranked according to our notion and, in those situations, \cref{lem:compstat} applies.

\section{Examples of risk-aversion}\label{app:examples}



The following example illustrates different channels through which differential risk appetites (\cref{assumption:reg}) might arise: 

\begin{example}[Constant relative risk aversion] \label{eg:riskaversion}
We consider the case of AI risk and follow \cite{jones2024ai} by considering CRRA utility over consumption; differently, we will suppose that the damage from a higher technological level in the dangerous state is continuous which allows us to capture a wide set of potential harms.\footnote{\cite{jones2024ai} focuses on existential risk which he models as the event that consumption is zero for all future periods. It is also straightforward to do an analogous exercise for other kinds of technologies (maps from $(l,\theta)$ to consumption) and other preferences (maps from consumption to utility) e.g., CARA.} Consumption is a random variable determined by the state and technology level: 
\[
\underbrace{c(\theta = 1,l) = c_0 \cdot \exp\big(a\cdot l\big)}_{\text{Beneficial $\implies$ grows at rate $a > 0$}}
\quad \text{and} \quad 
\underbrace{\quad c(\theta = 0,l) = c_0 \cdot \exp\big(-b\cdot l\big)}_{\text{Harmful $\implies$ shrinks at rate $b > 0$}}
\]
for some constant $c_0 > 0$. The principal's preferences over consumption (which might represent the preferences of society) is CRRA with risk-aversion parameter $\gamma_P 
 \neq 1$.\footnote{That is, $v(c(\theta,l)) = \bar u + \frac{c^{1-\gamma_P}}{1 - \gamma_P}$ for $\bar u > 0$ chosen to normalize zero technology to zero utility.} The following examples all fulfill \cref{assumption:reg}:\footnote{See Online Appendix \ref{app:examples} for details.}
\begin{enumerate}
    \item \textbf{Intrinsic differences in risk aversion.} The agent's preferences over consumption is also CRRA, but the Principal has a higher risk-aversion coefficient than the agent i.e., $\gamma_P > \gamma_A$ by choosing 
     \[u = g(v) = 
        \frac{\big(v ( 1 - \gamma_P) + 1\big)^{\frac{1 - \gamma_A}{1 - \gamma_P}} - 1}{ 1- \gamma_A}\]
    \item \textbf{Negative externalities.} The principal and agent's preferences are identical if the technology is safe but if the technology is harmful, the agent does not fully internalize those harms e.g., because of limited liability. This might be represented by $U(0,l) = \alpha \cdot V(0,l)$ for $\alpha < 1$ so 
    \[g(v) =\begin{cases}
        v \qquad &\text{for $v \geq 0$} \\
        \alpha \cdot v &\text{for $v < 0$}
    \end{cases}
    \]
    
    \item \textbf{Disproportionate winners.} The principal and agent's preferences are identical if the technology is unsafe but if the technology is beneficial, the agent benefits disproportionately relative to the rest of society. This might be represented by  $U(1,l) = \beta \cdot V(1,l)$ for $\beta > 1$ so 
    \[g(v) =\begin{cases}
        \beta v \qquad &\text{for $v \geq 0$} \\
        v &\text{for $v < 0$}
    \end{cases}
    \]
    \item \textbf{R\&D arms race.} There are $n$ identical firms playing a continuous-time stopping game where learning is public among firms. Let $\bm{l} = (l_i)_{i=1}^n$ be the profile of firms' R\&D. Firm $i$'s payoff is $U(0,\bm{l}) = V(0,\max \bm{l})$ if the technology is dangerous (common disaster), and $U(1,\bm{l}) = \beta_i \cdot V(1,l_i)$ if it is safe (private benefit) where $\beta_1 > \beta_2 > \ldots \beta _n > 0$ represent heterogeneous payoffs across firms. Then in every SPE of this game, all firms choose the same stopping level that a representative firm with $U(0,l) = V(0,l)$ and $U(1,l) = \beta_1 \cdot V(1,l)$.
    $\hfill \diamondsuit$
\end{enumerate}
\end{example} 

\subsection{Microfoundation for R\&D arms race}

We now develop a simple model of R\&D arms races and show the claim made in \cref{eg:riskaversion} about equilibrium play. Let $\mathcal{F} \in \mathbb{F}$ be any learning process, and let $\widetilde{\mathcal{F}}$ be the augmented filtration that contains information about other players' stopping levels i.e., 
\[
\widetilde{\mathcal{F}}_\el = \mathcal{F}_\el \vee \underbrace{\sigma(\el_1,\el_2,\ldots \el_n)}_{\substack{\text{Natural filtration} \\ \text{generated} \\ \text{by $(\el_1,\el_2,\ldots \el_n)$}}}
\]
Strategies are stopping levels that are $\widetilde{\mathcal{F}}-$adapted---letting players condition on the stopping behavior of others players `at the same time', therbey allowing for \emph{instantaneous} responses. This simplifies our exposition, is intuitive, and well-defined (see \cite{simon1989extensive}). 

\begin{lemma}\label{lemma:armsrace}
    For any learning process $\mathcal{F} \in \mathbb{F}$ and any mechanism $\phi \in \Phi$, let $\mathsf{SPE}(\mathcal{F},\phi)$ be the set of subgame-perfect equilibrium: 
    \begin{itemize}
        \item[(i)] For any $\bm{\el}^* \in \mathsf{SPE}(\mathcal{F},\phi)$, $\el_1 = \el_2 = \ldots \el_n =: \el^*$ almost surely. 
        \item[(ii)] $\el^*$ solves 
        \[
        \sup_{\el} \Ex\Big[U_1(\mu_\el,\el) - \phi_\el\Big] \quad \text{s.t. $\el$ is $\mathcal{F}$-adapted} 
        \]
        where $U_1(0,l) = V(0,l)$ and $U_1(1,l) = \beta_1 \cdot V(1,l)$. 
    \end{itemize}
\end{lemma}

\begin{proof} Let $\el^i$ be any selection of firm $i$'s optimal stopping level that solves $\Ex[U_1(\mu_\el,\el) - \phi(\el)]$. 

To show (i), take any firm $i$ that stops at level $\el_i$ strictly before all agents have stopped i.e., $\el_i <  \max \bm{\ell} =: \ell^{max}$. Then observe on this event: 
\begin{align*}
\sup_{\el > \el_i} \Ex\Big[ U_i(\mu_\el,\el) &\Big| \widetilde{\mathcal{F}}_{\el_i} \Big] \quad \text{s.t. $\ell$ is $\widetilde{\mathcal{F}}$-adapted} \\
&=     \sup_{\el > \el_i} \Ex\Big[\mu_\el \cdot \beta_i u(1,\el) + (1-\mu_\el) \cdot u(0,\ell^{max}) \Big| \widetilde{\mathcal{F}}_{\el_i} \Big] \\
&\quad \quad \text{s.t. $\ell$ is $\widetilde{\mathcal{F}}-measurable$}\\
&\geq  \Ex\Big[\mu_{\ell^{max}} \cdot \beta_i u(1,\ell^{max}) + (1-\mu_{\ell^{max}}) \cdot u(0,\ell^{max}) \Big| \widetilde{\mathcal{F}}_{\el_i} \Big] \tag{$\ell^{max}$ is $\widetilde{\mathcal{F}}$-adapted}
\\
&=\Ex\bigg[ \Ex\big[\mu_{\ell^{max}} \cdot \beta_i u(1,\el^{max}) + (1-\mu_{\ell^{max}}) \cdot u(0,\el^{max}) \big| \ell^{max} = s, \widetilde{\mathcal{F}}_{\el_i} \big] \, \bigg| \, \widetilde{\mathcal{F}}_{\el_i} \bigg] \tag{iterated expectations}
\\
&=  \Ex\Big[\mu_{\el_i} \cdot \beta_i u(1,\el^{max}) + (1-\mu_{\el_i}) \cdot u(0,\el^{max}) \Big| \widetilde{\mathcal{F}}_{\el_i} \Big] \tag{martingale} 
\\
& > \Ex\Big[ U_i(\mu_{\el_i},\el) \Big| \widetilde{\mathcal{F}}_{\el_i} \Big] \tag{$u(0,\cdot)$ is strictly decreasing}
\end{align*}
a contradiction. 

We now show (ii). For any $\mathcal{F}$-stopping level $\ell$, if at filtration $\mathcal{F}_{\ell}$ we have 
\[
 \begin{array}[t]{l}
     \sup_{s > \ell} 
        \Ex\Big[U_1(\mu_s,s) - \phi_s \Big| \mathcal{F}_{\ell}\Big]
     \\
    \text{s.t. $s$ is $\mathcal{F}$-adapted}
    \end{array}
    <  U_1(\mu_{\ell},\ell) - \phi_{\ell}
\]
then this must also be true of all other firms. Since the number of firms is finite and firms can respond immediately, backward induction implies that all firms must stop at such histories so $\ell^* = \ell$ a.s. Now suppose, instead, that 
\[
 \begin{array}[t]{l}
     \sup_{s > \ell} 
        \Ex\Big[U_1(\mu_s,s) - \phi_s \Big| \mathcal{F}_{\ell}\Big]
     \\
    \text{s.t. $s$ is $\mathcal{F}$-adapted}
    \end{array}
    >  U_1(\mu_{\ell},\ell) - \phi_{\ell}
\]
then firm $1$ has an individual deviation to do strictly more R\&D since 
\begin{align*}
    \begin{array}[t]{l}
     \sup_{s > \ell} 
        \Ex\Big[U_1(\mu_s,s) - \phi_s \Big| \widetilde{\mathcal{F}}_{\ell}\Big]
     \\
    \text{s.t. $s$ is $\widetilde{\mathcal{F}}$-adapted} 
    \end{array} 
    &= \begin{array}[t]{l}
     \sup_{s > \ell} 
        \Ex\Big[\mu_s \cdot U_1(1,s) + (1- \mu_s) \cdot U_1(0, \ell^{max}) - \phi_s \Big| \widetilde{\mathcal{F}}_{\ell}\Big]
     \\
    \text{s.t. $s$ is $\widetilde{\mathcal{F}}$-adapted} 
    \end{array}  \\
    &=\begin{array}[t]{l}
     \sup_{s > \ell} 
        \Ex\Big[\mu_s \cdot U_1(1,s) + (1- \mu_s) \cdot U_1(0, s) - \phi_s \Big| \widetilde{\mathcal{F}}_{\ell}\Big]
     \\
    \text{s.t. $s$ is $\widetilde{\mathcal{F}}$-adapted} 
    \end{array} \tag{Backward induction from stopping at $s$} \\
    &= \begin{array}[t]{l}
     \sup_{s > \ell} 
        \Ex\Big[U_1(\mu_s,s) - \phi_s \Big| \mathcal{F}_{\ell}\Big]
     \\
    \text{s.t. $s$ is $\mathcal{F}$-adapted}
    \end{array}  \tag{Information about others' stopping levels is trivial}
    \\
    &>  U_1(\mu_{\ell},\ell) - \phi_{\ell} \tag{by hypothesis}
\end{align*}
and the last term on the RHS is, in turn, an upper-bound on firm $1$'s payoff from stopping at $\ell$ so $\ell^* > \ell$ a.s. Hence $\ell^*$ solves the single-agent stopping problem in the statement of \cref{lemma:armsrace}.  
\end{proof} 

\subsection{Quantitative results on worst-case payoffs} (\cref{eg:CRRA_worstcase}) Suppose that consumption is given by 
\[c(\theta,l) = \begin{cases}
    \exp(a\cdot l) \quad &\text{if $\theta = 1$} \\
    \exp(-b\cdot l) \quad &\text{if $\theta = 0$}
\end{cases} \quad \text{for $a,b > 0$}
\]
and that the agent and principal's utility of consumption is CRRA and given by: 
\[
u(\theta,l) = \cfrac{c(\theta,l)^{1-\gamma_A}}{1 - \gamma_A}
\quad \text{and} \quad 
v(\theta,l) = \cfrac{c(\theta,l)^{1-\gamma_P}}{1 - \gamma_P}
\]
where $1 \leq \gamma_A \leq \gamma_P$. Under the empty laissez-faire mechanism i.e., $\phi = 0$ a.s., we have that the lowest belief that can rationalize pushing the technology up to $l$ is: 
\[
\mu^*_{U}(l) = \cfrac{1}{1 + \frac{a}{b}\cdot \exp\Big(l\cdot (1 - \gamma_A) \cdot (a+b) \Big)}
\]
so under weak optimism defined in the main text, we have that 
\[
\mathbb{P}\Big( l^* \leq l + l^*_{U}(\mu_0)\Big) = 1 - \cfrac{\mu_0}{\mu^*_{U}(l)}
\]
and let $F: \mathcal{L} \to [0,1-\mu_0]$ be the (unnormalized) CDF of the technology level $l^*$. Since whenever $l^* < L$ the agent conclusively learns the state is bad, this pins down the equilibrium joint distribution between beliefs and technology levels. Plugging this into the regulator's utility and standard manipulation yields: 
\begin{align*}
    \Ex\big[V(l,\mu)\big] &= \mu_0 \cdot V(1, L) + (1-\mu_0) \cdot \Ex\big[V(\mu,l) \big| \theta = 0\big]  \\
    &=  \mu_0 \cdot V(1, L) +  \int_{0}^{L} V(0,l) dF(l) \\
    &= \mu_0 \cdot V(1, L) + \frac{\mu_0 \cdot a \cdot (\gamma_A - 1) \cdot (a+b)}{b \cdot (1 - \gamma_P) \cdot \lambda} \cdot \Big(\exp(\lambda L) - 1\Big)
\end{align*}
where $\lambda := b(\gamma_P - 1) - (\gamma_A - 1)(a+b)$. Direct calculation yields:  
\[
\lim_{L \to +\infty} \Ex\big[V(l,\mu)\big] = -\infty \iff \frac{\gamma_P - 1}{\gamma_A - 1} > 1 + \frac{a}{b}
\]
and it is immediate to replicate this analysis for linear Pigouvian taxes. 

\section{Extensions and Generalizations}\label{app:ext}
In the \cref{sec:policy} of the main text we outline directions in which our framework and results could be extended and generalized. Here we flesh out some details. 

\subsection{No value of elicitation and randomization} We start by augmenting our space of adaptive mechanisms to allow for additional independent randomization and type reports:

\begin{itemize}[leftmargin = *]
    \item  \textbf{Introducing additional randomization.}  Recall we started on the space $(\Omega, \mathcal{G}, \mathbb{P})$ and since our definition of adaptive mechanisms were $\mathcal{G}$-adapted processes, this did not accommodate additional (independent) randomization. We now enlarge our filtered probability space with an independent Brownian motion. Define an auxillary space $(\Omega', \mathcal{G}',\mathbb{P}')$ that supports an independent Brownian motion $(B_l)_{l \in \mathcal{L}}$. Our augmented space is then 
    \[
    \Big(\Omega \times \Omega', \mathcal{G} \times \mathcal{G}', \mathbb{P} \times \mathbb{P}'\Big) 
    \]
    and define $(\tilde{\mathcal{G}}_l)_l$ as the product filtration of $\mathcal{G}$ and $\mathcal{G}'$.\footnote{That is, $\tilde{\mathcal{G}}_l = \sigma\big(\mathcal{G}_s \otimes \{\emptyset,\Omega'\} \cup \{\emptyset, \Omega' \} \otimes \mathcal{G}_s': s \leq l\big)$ which is the standard way of defining the product filtration via cylinders.} 
    \item  \textbf{Introducing additional elicitation.} Consider a richer space of adaptive mechanisms in which the regulator attempts to elicit reports and condition future transfers on past reports. Let $\bm{r} := (r_s)_{s \leq l}$ denote a path of reports up to level $l$, each taking values in the rich message space $\mathbb{R}$. Let $\mathcal{R}_l$ denote the set of all possible paths of reports up to level $l$ and $\mathcal{R} := \bigcup_{l \in \mathcal{L}} \mathcal{R}_l$. Define 
\[
\widehat{\mathcal{G}}_l := \tilde{\mathcal{G}}_l\vee \sigma(R_l)
\]
as the filtration augmented with the path of reports. 
\end{itemize}

An \emph{augmented adaptive mechanism} is a $(\widehat{\mathcal{G}_l})_l$-adapted stochastic process that satisfies the standard regularity conditions in the main text i.e., uniform integrability where it is finite, and lower-semicontinuity. Let $\widehat{\Phi}$ denote this space of augmented adaptive mechanisms. Given $\widehat{\phi} \in \widehat{\Phi}$ the agent solves a joint stopping and control problem of choosing both an (i) $\mathcal{F}$-adapted stopping time; and (ii) a dynamic reporting strategy which is a progressively measurable $\mathbb{R}$-valued process adapted to $\mathcal{F}$. Clearly the agent does not require additional randomization since (i) she is `best-responding' against $\widehat \phi$; and (ii) is maximizing expected payoffs. We let $\el^*(\widehat{\phi}, \mathcal{F}, U)$ denote the largest stopping time that is part of a solution to the agent's joint problem.

\begin{proposition}\label{prop:novalue} Randomization and elicitation have no value:
\[
\sup_{\widehat{\phi} \in \widehat{\Phi}} \inf_{\mathcal{F} \in \mathbb{F}}\Ex \Big[ V \big(\mu_{\el^*},\el^*(\widehat \phi, \mathcal{F},U)\big) \Big]
= 
\sup_{\phi \in \Phi} \inf_{\mathcal{F} \in \mathbb{F}} \Ex \Big[ V \big(\mu_{\el^*},\el^*(\phi,\bm{\mu},U)\big) \Big] =: \eqref{eqn:ADV}. 
\]
\end{proposition}

\begin{proof}[Proof sketch] We sketch the argument since it follows that of \cref{thrm:robustness} with a few modifications. First notice that Step 1 of the proof of \cref{thrm:robustness} continues to apply: no additional information $(\mathcal{F} = \mathcal{G})$ remains an admissible choice by nature. Facing this mechanism, the adaptive sandbox $\phi^*$ (that does not require randomization) is optimal---loosely speaking, we can find an adaptive mechanism $\widehat \phi \in \widehat{\Phi}$ that is `driven only by variation' in the projection of $\widehat{\mathcal{G}}$ onto $\mathcal{G}$. The point is that at this saddle point, there is no value to (i) elicitation since the agent has no extra information; and (ii) randomization since extra randomization cannot improve solutions to the principal's optimal experimentation problem. Step 2 then proceeds similarly to that of \cref{thrm:robustness}---under any alternate $\mathcal{F} \in \mathbb{F}$ the principal is made weakly better-off, and putting these steps together yields the result. 
\end{proof}

The fact that the regulator cannot do better with elicitation is reminiscent of \cite*{kruse2015optimal,kruse2019inverse}; the fact that the regulator cannot do better with randomization is in contrast to \cite{libgober2021informational}.  

\subsection{Endogenous agent learning} 

Suppose that instead of exogeneously learning about the state as the technology develops, the learning process is optimally chosen by the agent at some cost which is increasing in the flow variation of the belief process. This framework was proposed in the important papers of \cite{zhong2022optimal,hebert2023rational}.\footnote{Our environment is nearby, but non-nested within theirs: our agent's action space is singleton and the payoffs from stopping at different levels varies with the state. By contrast, \cite{zhong2022optimal,hebert2023rational} assumes that the agent always dislikes delay.} We augment it to accommodate outside learning (via $\mathcal{G})$). Some notation: 

Let $H: \Delta(\Theta) \to \mathbb{R}$ be a concave and continuous measure of uncertainty and $C: \mathbb{R} \to \mathbb{R}$ is a weakly increasing cost function with $C(0) = 0$. The \emph{flow information} at level $l$ is given by the random variable  
\[
I_l := \Ex\Bigg[ \lim_{l' \downarrow l} \cfrac{H\big(\Ex[\theta | \mathcal{F}_{l'}]\big) - H\big(\Ex[\theta | \mathcal{F}_l \vee \mathcal{G}_{l'}]\big)}{l'-l} \Bigg| \mathcal{F}_l \Bigg]
\]
Notice here that our variation measure (i) is \emph{infinitesimal} i.e., it depends on how much beliefs move between $l$ and $l'$ (resembling the infinitesimal generator for Feller semigroups); (ii) depends only on movement in \emph{first-order beliefs} about the state, thereby following a long tradition in information economics; and (iii) measures the \emph{excess movement} since change is measured to the agent's `counterfactual' level-$l'$ belief \emph{as if} she did not acquire any excess information interim information, but continued to observe the outside learning process so her first-order belief is given by $\Ex[\theta | \mathcal{F}_l \vee \mathcal{G}_{l'}]$. 

Facing the mechanism $\phi$ and outside learning process $\mathcal{G}$, the agent thus solves the following {stochastic control problem}:
\[
\sup_{\mathcal{F} \in \mathbb{F},\el} \Ex\Big[ U(\mu,\el) - \phi(\el)  - \int^{\el}_0 C(I_s) ds \Big]
\]
where $\el$ is a stopping level adapted to $(\mathcal{F}_l)_l$. Note that the choice of $\mathcal{F} \in \mathbb{F}$ already specifies a complete and contingent plan for how (first-order) beliefs from $\mu_l$ evolve.  Let $\el^*(\phi, (H,C),U)$ denote the agent's optimal solution to the above problem. 

We suppose, instead, that the principal is unsure how hard it is for the agent to acquire additional information about the technology. To reflect this uncertainty, we let $\mathcal{E}$ denote the set of cost functions and uncertainty measure $(C,H)$ pairs that fulfill the above assumptions. The adaptive sandbox $\phi^*$ remains robustly optimal when the ambiguity set is $\mathcal{E}$. 

\begin{proposition}[Robust regulation with endogenous agent information]
    $\phi^{*}$ solves
    \begin{align*}
    \sup_{\phi \in \Phi} \inf_{(C,H) \in \mathcal{E}} &\Ex\big[V\big(\mu_{\el^*},\el^*(\phi, (H,C),U)\big)\big] \\
    &\text{and} \quad    \sup_{\phi \in \Phi} \inf_{\substack{(C,H) \in \mathcal{E} \\ U \in \mathbb{U}}} \Ex\big[V\big(\mu_{\el^*},\el^*(\phi, (H,C),U)\big)\big]     
    \end{align*}
\end{proposition}
\begin{proof}[Proof sketch] Step 1 of the proof of \cref{thrm:robustness} is unchanged, choosing the saddle point in which it is infinitely costly / impossible to acquire excess information over $\mathcal{G}$ so $\mathcal{F} = \mathcal{G}$. The challenge now is to show that away from the `infinite cost' case, the all solutions to the agent's endogenous information acquisition problem must weakly improve the principal's payoff. 

Given $\phi \in \Phi$ and $(H,C) \in \mathcal{E}$, given the agent's solution of endogenous information acquisition and stopping $\mathcal{F}, \el^*$ we will construct an \emph{exogeneous} filtration $\mathcal{F}^{\mathsf{x}}$ that induces an identical joint distribution over stopping levels and first-order beliefs as follows: 
\[
\mathcal{F}^{\mathsf{x}}_l := \begin{cases}
    \mathcal{F}_l \quad &\text{if $l \leq \el^*$}\\
    \mathcal{F}_{\el^*} \vee \mathcal{G}_l \quad &\text{otherwise.}
\end{cases}
\]
Now let $\el^\mathsf{x}$ denote the agent's solution facing mechanism $\phi$ and \emph{exogenous} information $\mathcal{F}^{\mathsf{x}}$. We proceed via cases. 

\underline{Case 1: $\el^{\mathsf{x}} < \el^*$.} Since the agent found it strictly better to continue at $\el^{\mathsf{x}}$ (and acquire information at some cost) we have 
\begin{align*}
0 &< \sup_{\el > \el^{\mathsf{x}}} \Ex\Big[ U(\mu_\el,\el) - \int^{\el^*}_{\el^{\mathsf{X}}} C(I_s) ds - U\big(\Ex[\theta| \mathcal{F}_{\el^{\mathsf{X}}}], \el^{\mathsf{x}}\big) \Big| \mathcal{F}_{\el^*} \Big] \tag{$\el^* > \el^{\mathsf{x}}$}
\\
&\leq \sup_{\el > \el^{\mathsf{x}}} \Ex\Big[ U(\mu_\el,\el)  - U\big(\Ex[\theta| \mathcal{F}_{\el^{\mathsf{x}}}], \el^{\mathsf{x}}\big) \Big| \mathcal{F}_{\el^*} \Big]  \tag{$C \geq 0$}
\\
&= \sup_{\el > \el^{\mathsf{x}}} \Ex\Big[ U(\mu_\el,\el)  - U\big(\Ex[\theta| \mathcal{F}_{\el^{\mathsf{x}}}], \el^{\mathsf{x}}\big) \Big| \mathcal{F}^{\mathsf{x}}_\el \Big] \tag{Construction of $\mathcal{F}^{\mathsf{x}}$}
\\
&\leq 0 
\end{align*}
a contradiction. 

\underline{Case 2: $\el^{\mathsf{x}} > \el^*$.} If the agent finds it strictly better to continue under the exogenous learning process past $\el^*$, then consider instead the deviation by the agent under the endogenous process under which at level $\el^*$, the agent does not acquire any further information but instead replicated $\el^{\mathsf{x}}$. At $\el^*$ we have 
\begin{align*}
    0 &< \sup_{\el > \el^*} \Ex\Big[ U(\mu_\el,\el)  - U\big(\Ex[\theta| \mathcal{F}_{\el^{\mathsf{X}}}], \el^{\mathsf{x}}\big) \Big| \mathcal{F}_{\el^*}^{\mathsf{x}} \Big] \tag{$\el^* < \el^{\mathsf{x}}$}\\
    &= \Ex\Big[ U(\mu_\el,\el)  - \underbrace{\int^{\el^{\mathsf{X}}}_{\el^*} C(I_s) ds}_{=0} - U\big(\Ex[\theta| \mathcal{F}_{\el^*}], \el^*\big) \Big| \mathcal{F}_{\el^*} \Big]  \tag{Construction of deviation} \\
    &\leq \sup_{\substack{\mathcal{F}' \in \mathbb{F}|_{\mathcal{F}_\el} \\ \el > \el^*}}  \Ex\Big[ U(\mu_\el,\el)  - \int^{\el^{\mathsf{X}}}_{\el^*} C(I_s) ds - U\big(\Ex[\theta| \mathcal{F}_{\el^*}], \el^*\big) \Big| \mathcal{F}_{\el^*}' \Big]  
    \tag{Replicating $\el^{\mathsf{x}}$ is feasible}
    \\
    &\leq 0 
\end{align*}
a contradiction, where $\mathbb{F}_{\mathcal{F}_l}$ is the set of filtrations consistent with $\mathcal{F}_l$. Hence $\el^{\mathsf{x}} = \el^*$ a.s. Moreover, by construction of $\mathcal{F}^{\mathsf{x}}_\el$ we have $\Ex[\theta| \mathcal{F}^{\mathsf{x}}_{\el^*}] = \Ex[\theta| \mathcal{F}_{\el^*}]$ so the induced joint distribution over stopping levels and beliefs are identical. We have showed that under any $\phi \in \Phi$ and any $(H,C) \in \mathcal{E}$, we can find some $\mathcal{F} \in \mathbb{F}$ so that the regulator's payoffs are identical. Then proceeding with Step 2 of the proof of \cref{thrm:robustness} completes the result.
\end{proof}

\subsection{Revenue motives} 
In the main text we supposed that our principal has no revenue motives. Now consider the case in which the principal's payoff is 
\[
V(\mu_\ell, \ell) + \lambda \cdot \phi_{\ell}
\]
for $\lambda > 0$. Suppose $U \in \mathbb{U}$ is known (so we focus on learning-robustness) and construct the sandbox $\phi^{\mathsf{R}}$ as follows: 
\[
\phi^{\mathsf{R}}_l = \begin{cases}
    \mathsf{R}(U,\mathcal{G}) \quad &\text{if $l \leq \el_{\mathsf{R}}$}\\
    +\infty &\text{otherwise.}
\end{cases}
\]
where $\el_{\mathsf{R}}$ solves
\begin{align*}
    &\sup_{\el} \Big(V(\mu_\el,\el) + \lambda \cdot U(\mu_\el,\el) \Big) \\
    &\quad \text{s.t. $\el$ is $\mathcal{G}$-adapted.}
\end{align*}
and 
\[
\mathsf{R}(U,\mathcal{G}) := \Ex\big[U(\mu_{\el_{\mathsf{R}}},\el_{\mathsf{R}}) - U(\mu_0,0)\big]. 
\]

\begin{proposition}[Robust regulation with revenue motives]\label{prop:revenue}
    $\phi^{\mathsf{R}}$ solves 
    \[
    \sup_{\phi \in \Phi} \inf_{\mathcal{F} \in \mathbb{F}} \Ex\Big[ V\big(\mu_{\el^*},\el^*(\phi,\mathcal{F},U)\big) + \phi_{\el^*}\Big].
    \]
\end{proposition}
\begin{proof}[Proof sketch] Only Step 1 of \cref{thrm:robustness} needs to be modified: it is easy to see that given mechanism $\phi^{\mathsf{R}}$ and under no additional information $(\mathcal{G} =\mathcal{F})$, the agent's participation constraint is tight since 
\begin{align*}
    \sup_{\el \leq \el_{\mathsf{R}}} \Ex\Big[ U(\mu_\el,\el) - \phi_\el \Big]     
    \geq  \Ex\Big[ U(\mu_{\el_{\mathsf{R}}},\el_{\mathsf{R}}) - \phi_{\el_{\mathsf{R}}} \Big] 
    = U(\mu_0,0)
\end{align*}
because continuing until the sandbox boundary is a feasible stopping time for the agent. Now for the principal, notice that $\phi^{\mathsf{R}}$ is the best mechanism against $\mathcal{F} = \mathcal{G}$ facing this participation constraint since 
\begin{align*}
    \sup_{\phi \in \Phi} &\Ex\Big[V\big(\mu_{\el^*}, \el^*(\phi,\mathcal{G},U)\big) + \lambda \cdot \phi_{\el^*}\Big]  \\
    &= \sup_{\phi \in \Phi} \Ex\Big[V\big(\mu_{\el^*}, \el^*(\phi,\mathcal{G},U)\big) \\
    &\qquad \qquad + \underbrace{ \lambda \cdot \phi_{\el^*} - \lambda \cdot U\big(\mu_{\el^*}, \el^*(\phi,\mathcal{G},U)\big)}_{\leq -U(\mu_0,0) \text{ from participation}} 
    + \lambda \cdot U\big(\mu_{\el^*}, \el^*(\phi,\mathcal{G},U)\big)\Big]   \\
    &\leq \sup_{\phi \in \Phi} \Ex\Big[V\big(\mu_{\el^*}, \el^*(\phi,\mathcal{G},U)\big)  + \lambda \cdot  U\big(\mu_{\el^*}, \el^*(\phi,\mathcal{G},U)\big)\Big] - \lambda \cdot U(\mu_0,0). 
\end{align*}
and it is easy to verify that $\phi_{\mathsf{R}}$ implements this upper-bound since by \cref{lem:compstat} under $\phi_{\mathsf{R}}$ and $\mathcal{G} = \mathcal{F}$, $\el^* = \el_{\mathsf{R}}$ and $\ell_{\mathsf{R}}$ is by definition the $\mathcal{G}$-stopping level that maximizes the term inside the brackets.  Now Step 2 proceeds similarly to that of \cref{thrm:robustness}, noting that for any alternate learning process $\mathcal{F} \in \mathbb{F}$, the agent's participation constraint is still fulfilled (from \cite{greenshtein1996comparison}) and a similar argument to that of \cref{thrm:robustness} yields that any deviation (weakly) increases the principal's payoff to weakly increase since revenue, being extracted upon participation, is unaffected. 
\end{proof}


\subsection{Undominated Stopping Rules}\label{app:undominated_stopping}

In the main text we assumed that the agent knew the learning process $\mathcal{F} \in \mathbb{F}$ and, facing the mechanism $\phi$, solved her optimal stopping problem \eqref{prob:OSP}.  This served as a helpful Bayesian benchmark in which the agent has (i) correct beliefs about the learning process she faces; and (ii) solves the optimal stopping problem accordingly. We now relax both of these assumptions which will allow us to capture a wider class of agent behavior e.g., misspecifcation, ambiguity aversion, myopia, etc.

We start by defining undominated stopping rules. The set of \emph{dominated} belief-level pairs under preference $f$ is: 
\[
\mathsf{DOM}[f] := \Big\{(\mu,l) \in \Delta(\Theta) \times \mathcal{L}: l \notin \text{argmax}_{s \geq l} f(\mu,s) \Big\}
\]
That is, at belief $\mu$, even in the absence of further information, the agent stops too early---she can do strictly better by pushing the technology further.

\begin{definition}
    Fix the learning process $\mathcal{F} \in \mathbb{F}$ and preference $U \in \mathbb{U}$. The $\mathcal{F}$-adapted stopping level $\el$ is \emph{undominated} if  
    \[
    \mathbb{P}\Big( (\mu_\el,\el) \in \mathsf{DOM}[U]\Big) = 0.
    \]
    and let $\mathsf{USR}[U,\mathcal{F}]$ be the set of undominated stopping rules.
\end{definition}

Undominated stopping rules include: 
\begin{enumerate}[leftmargin = 2em]
    \item \textbf{Bayes optimal} stopping levels that solve \eqref{prob:OSP} i.e., the agent has correct beliefs about the learning process. This was studied in the main text. 
    \item \textbf{Ambiguity} where the agent faces non-Bayesian uncertainty about the law of future information and solves 
    \[
    \sup_\el \inf_{\mathcal{F}' \in \mathbb{F}} \Ex^{\mathcal{F}'} \Big[U^{\phi}(\mu_\el,\el)\Big] \quad \text{s.t. $l$ is $(\mathcal{F}_l)_l$-adapted}
    \]
    where we use $\Ex^{\mathcal{F}}$ to denote the expectation under the learning process $\mathcal{F}$.\footnote{Here we suitably extend the probability space.} This is closely related to the formulation in \cite{riedel2009optimal} who studies optimal stopping under ambiguity.\footnote{The difference is that in \cite{riedel2009optimal}, the process directly specifies payoffs.} Note that here, ambiguity is about the law of the learning process i.e., how informative future experiments will be as opposed to having ambiguous priors (scenario 2 in \cite{epstein2007learning}\footnote{See also \cite*{auster2024prolonged}}) or ambiguous signals (scenario 3).
    \item \textbf{Misspecification} in which the agent has misspecified beliefs about the law of information i.e., the true law is not in the support of her prior and she solves her stopping problem under this erroneous belief.

    \item \textbf{Myopia} in which the agent does not internalize the learning value of pushing the technology, and simply chooses the best technology level under her present beliefs.
\end{enumerate}

We first establish that $\phi^{*}$ remains learning- and dually-robust whenever the agent's stopping level is undominated. We will impose an additional regularity condition that $U(\mu,\cdot)$ is quasiconcave.  

\begin{proposition}[Robustness against undominated stopping rules] 
\label{thrm:undominated_stopping_robust} Suppose that $U(\mu,\cdot)$ is quasiconcave for each $\mu \in \Delta(\Theta)$ and that the principal does not learn. Then the sandbox $\phi^*$ with hard limit $\overline{\el}$ is both learning- and dually-robust under any undominated stopping rule i.e., $\phi^*$ solves
    \begin{align*}
        \sup_{\phi \in \Phi} \inf_{\substack{\mathcal{F} \in \mathbb{F}\\
        \el \in \mathsf{USR}[U^{\phi},\mathcal{F}]}} \Ex\Big[V(\mu_\el, \el) \Big]
        \quad \text{\emph{and}} \quad 
        \sup_{\phi \in \Phi} \inf_{\substack{\mathcal{F} \in \mathbb{F}\\ U \in \mathbb{U} \\
        \el \in \mathsf{USR}[U^{\phi},\mathcal{F}]}} \Ex\Big[V(\mu_\el, \el) \Big]. 
    \end{align*}   
\end{proposition}

We next formalize the claim made in our discussion after \cref{thrm:worstcase} that the same worst-case payoff remains possible under an adversarially chosen learning process whenever the agent is using an undominated stopping rule.  

\begin{proposition}[Worst-case under undominated stopping rules] \label{prop:undominated_worstcase}
For any mechanism $\phi$ such that $U^{\phi} = g \circ V$ for some convex function $g$:
\[
        \inf_{\mathcal{F} \in \mathbb{F}}
        \sup_{\el \in \mathsf{USR}[U^{\phi},\mathcal{F}]}
        \Ex\Big[V(\mu_\el, \el) \Big] =  
        \inf_{\substack{\mathcal{F} \in \mathbb{F}}} \Ex\Big[V\Big(\mu_{\el^*}, \el^*\big(\phi, \bm{\mu}, U \big) \Big) \Big] 
\]          
i.e., 
the worst-case principal payoffs under her best undominated stopping rule is identical to that under the optimal stopping level analyzed in the main text. 
\end{proposition}

This follows quite quickly by taking the construction of weak optimism (the worst-case learning process from \cref{thrm:worstcase}) and verifying that at each interim history, the agent finds it optimal to continue under her present belief so long as conclusive bad news has not arrived. The only subtlety is that this, by itself, does not guarantee that all undominated stopping rules generate this behavior because of ties---since weak optimism is constructed to keep the agent indifferent, it is also an undominated stopping rule to break ties in favor of stopping. But we can suitably perturb the learning process to make continuation incentives strict, and verify that the principal's payoff is continuous in this perturbation to yield \cref{prop:undominated_worstcase}.

\section{Bridging Robustness and Bayes} \label{appendix:Bayes} In \cref{prop:Bayesian} we made the (informal) claim that in the Bayesian problem with a sufficiently diffuse prior $f \in \Delta(\mathbb{F} \times \mathbb{U})$, Bayes-optimal mechanisms required hard limits. We now state a more precise version of this result that requires us to clarify what it means for a prior to be `diffuse'. 

There, of course, some conceptual difficulties since the space of learning processes $\mathbb{F}$ and preferences $\mathbb{U}$ are both large and difficult to describe. Indeed, this is sometimes held as a reason for robustness. We will try nonetheless, keeping in mind that by describing a class of learning processes and preferences that we can put a prior over, we will necessarily rule out many others. Thus, our terminology `diffuse' is only with respect to the class we have chosen to describe. At the same time, the reasoning behind \cref{prop:Bayesian} is straightforward and in the interest of transparency, we will pick particularly simple parametric classes to put priors over; we can do the same exercise with fancier classes to obtain the same qualitative result. 

Continue with the setting of \cref{eg:CRRA_worstcase}: the principal has CRRA preferences over consumption with risk-aversion parameter $\gamma_P > 0$. Consumption is given by 
\[
c(\theta,l) = \begin{cases}
    \exp\big(a\cdot l\big) \quad &\text{if $\theta = 1$}\\ 
    \exp\big(-b\cdot l\big) \quad &\text{if $\theta = 0$}
\end{cases}
\]
Suppose that $\tilde{\mathbb{U}}$ consists of the set of CRRA preferences over consumption with \emph{level} $\lambda$ and risk-aversion parameter $\gamma_A \in [0,\gamma_P]$ so that for an agent of preference with parameter $( \lambda, \gamma)$ utility is 
\[
u(\theta,l) = \lambda^{1 - \gamma} \cdot \cfrac{c(\theta,l)^{1 - \gamma}}{1 - \gamma}. 
\]
we have thus written down a simple two-parameter family of agent preferences. Let $\widetilde{\mathbb{F}}$ denote the set of Levy signal processes
\[
dX_l = \theta \cdot dl + \sigma \cdot dB_l + J \cdot dN_l 
\]
where $\sigma \geq 0$ is the degree of Brownian interference and $N_l$ is a Poisson process with intensity $\lambda_\theta \geq 0$ in state $\theta$, and $\lambda_0 \neq \lambda_1$. We have thus written down the simplest three-parameter family $( \sigma, \lambda_0,\lambda_1)$ we can put a prior over.

Our prior is thus over the smaller parametric class
\[P \in  \Delta\Big(\widetilde{\mathbb{U}} \times \widetilde{\mathbb{F}}\Big) 
\]
which is equivalently over the family of parameters $\langle (\lambda,\gamma), (\sigma, \lambda_0,\lambda_1)\rangle$. Say it is \emph{diffuse} if it is fully supported on this parameter space. \cref{prop:Bayesian} follows. 

\section{Existence and selection of stopping levels} \label{appendix:existence_selection}
\label{appendix:stoppingselection} In the main text we made a specific selection of equilibrium experimentation---fixing the mechanism $\phi$, the agent's learning process $\mathcal{F}$ and preferences $U$, we focused on the largest stopping level $\ell^*(\phi,\mathcal{F},U)$ that solved the agent's problem \eqref{prob:OSP}. In so doing, we presumed (i) the existence of \emph{some} stopping level; (ii) the existence of a \emph{largest} stopping level; and, perhaps most substantively, (iii) that our selection was not driving results on instrument performance. We deal with each of these concerns in turn. 

\subsection{Existence of optimal stopping levels} The set of optimal stopping levels is: 
\[
\mathcal{O}(\phi,\mathcal{F},U) := \Big\{\ell: \text{$\ell$ $\mathcal{F}$-adapted, solves $\eqref{prob:OSP}$}\Big\}
\]
denote the set of optimal stopping levels. 

We first claim $\mathcal{O}(\phi,\mathcal{F},U) \neq \emptyset$. Recall the agent solves
\[
\sup_{\ell} \Ex\Big[\underbrace{U(\mu_{\ell},\ell) - \phi_{\ell}}_{:= X_\ell}\Big] \quad \text{s.t. $\ell$ is $\mathcal{F}$-adapted.}
\]
The process $(X_l)_l$ is (i) a.s. upper-semicontinuous since $U(\mu,\cdot)$ is continuous and $\phi_l$ is lower-semicontinuous; and (ii) class (D) since $U$ is bounded and $\phi_\ell$ is class (D) so their sum must also be class (D). This implies that the Snell envelope $(Z_t)_t$ exists where $Z_l := \sup_{\ell \geq l} \Ex[X_\ell|\mathcal{F}_l]$, and so too does an optimal stopping level (these arguments are standard; see \cite{el1979aspects}). 

We next claim that there exists a largest optimal stopping level 
\[
\ell^* := \text{ess} \sup \mathcal{O}^*(\phi,\mathcal{F},U)
\]
where here we take the essential supremum because the set of optimal stopping times is potentially uncountable. 

It will be helpful to define a helpful property for stochastic processes---\emph{upper-semicontinuous in expectation along stopping levels} (USCE)---as follows: 

\begin{definition}
    The stochastic process $(Z_l)_l$ is right-USCE (left-USCE) if for any stopping level $\ell$ and for all sequences of stopping levels $(\ell_n)$ such that $\ell_n \downarrow \ell$ ($\ell_n \uparrow \ell$): 
\[
\Ex[Z_\ell] \geq \underset{n \to \infty}{\lim \sup} \, \Ex[Z_{\ell_n}].
\]
$Z$ is USCE if it is both right- and left-USCE. 
\end{definition}

We now check that $X := V - \phi$ is USCE. Recall that we have already argued that it is of class (D), and a.s. i.e., pathwise upper-semicontinuous. We show both right- and left-USCE togeher. 

Fix the stopping level $\ell$ and take any sequence of stopping levels $(\ell_n)_{n}$ such that $\ell_n \to \ell$. This holds for either $\ell_n \uparrow \ell$ or $\ell_n \downarrow \ell$. Since paths are upper-semicontinuous 
\[
 \underset{s \uparrow l}{\lim \sup} X_s \leq X_l \quad \text{a.s. for all $l \in \mathcal{L}$.}
\]
which means that along our sequence of stopping levels $\ell_n\uparrow \ell$:
\[
 \underset{n \to \infty}{\lim \sup} X_{\ell_n} \leq X_\ell \quad \text{a.s.}
\]
and from Fatou: 
\[
\underset{n \to \infty}{\lim \sup} \Ex[X_{\ell_n}] \leq \Ex[ \underset{n \to \infty}{\lim \sup} X_{\ell_n}]  \leq \Ex[X_\ell].
\]

We have shown that $X := V - \phi$ is USCE. Now apply Theorem 2.15 of \cite{kobylanski2012optimal} to conclude that $\ell^*$ is, in fact, an $\mathcal{F}$-stopping level. 

\subsection{Adversarial selection of optimal stopping levels}
\label{appendix:adversarialstoppingselection}
Recall $\mathcal{O}(\phi, \mathcal{F},U)$ denotes the set of stopping levels that solve \eqref{prob:OSP}. 

Consider an alternative selection of optimal stopping levels in which nature also adversarially choses the agent's optimal stopping levels:
 \begin{align*}
        \sup_{\phi \in \Phi} \, \inf_{\substack{{\mathcal{F} \in \mathbb{F}} \\ {\ell^* \in \mathcal{O}(\phi,\mathcal{F},U)}}} 
        \Ex\Big[ V\big(\mu_{\el^*}, \el^*\big)
        \Big] 
    \tag{L-ADV} \label{eqn:ADV_adversarial} \\
        \sup_{\phi \in \Phi} \, \inf_{\substack{{(\mathcal{F},U) \in \mathbb{F} \times \mathbb{U}} \\ {\ell^* \in \mathcal{O}(\phi,\mathcal{F},U)}}} 
        \Ex\Big[ V\big(\mu_{\el^*}, \el^*\big)
        \Big] 
    \tag{D-ADV} \label{eqn:DR_adversarial}
    \end{align*}

Note that the proof of robustness (\cref{thrm:robustness}) does not rely on any selection of stopping levels hence $\phi^*$ remains learning- and dually-robust under this alternate selection. 

Next, we argue that the proof of \cref{prop:dualrobust_characterization} carries through for  adversarial selections of optimal stopping levels. For sufficiency of limit with option mechanisms, observe that the proof of \cref{lemma:converse_dual} only used (i) the fact that the agent was choosing \emph{a} optimal stopping level; and (ii) \cref{lem:compare_two_stoppingtimes} that applies to any pair of stopping times (optimal or not). For necessity, the argument from \cref{lemma:dual_hard_deadline,lemma:zero_marginal} on the necessity of the option and limit goes through as stated by the same construction of the agent's learning process and preferences since an adversarially chosen stopping time continues to do strictly worse than the payoff guarantee.  

We thus obtain the following versions of \cref{thrm:robustness} and \cref{prop:dualrobust_characterization}:  

\begin{proposition}\label{thrm:robustness_adversarial}
The adaptive sandbox $\phi^*$ is learning-robust and dually-robust under adversarial selections of optimal stopping time i.e., it solves \eqref{eqn:ADV_adversarial} and \eqref{eqn:DR_adversarial}. $\phi$ is dually-robust under adversarial selection of stopping levels if and only if it is a limit with option mechanism induced by $\overline{\el}$.
\end{proposition}

Next, we discuss alternative selections of stopping levels for dominance and undominated mechanisms. For two adaptive mechanism $\phi,\phi' \in \Phi$, say $\phi$ dominates $\phi'$ under adversarial selection of stopping levels (ASSL) if, for any learning process $\mathcal{F} \in \mathbb{F}$ and any agent preference $U \in \mathbb{U}$,
\begin{align*}
    \inf_{\el^* \in \mathcal{O}(\phi,\mathcal{F},U)}\Ex\Big[V(\mu_{\el^*},\el^*) \Big] \geq \inf_{\el^* \in \mathcal{O}(\phi',\mathcal{F},U)}\Ex\Big[V(\mu_{\el^*},\el^*) \Big].
\end{align*}
with strict inequality for some $(\mathcal{F},U) \in \mathbb{F} \times \mathbb{U}$. A mechanism is undominated under adversarial selection if it is not dominated under ASSL.  

The proof of \cref{thrm:dominance} holds as is under adversarial selection of the agent's stopping level which yields:
\begin{proposition}
    $\phi^*$ is undominated under ASSL. $\phi^*$ dominates under ASSL all other dually-robust and regular mechanisms. 
\end{proposition}

\section{Static vs dynamic regulation} \label{appendix:static_vs_dynamic} In the main text we analyzed a setting in which the principal faced uncertainty about the agent's learning process. This uncertainty encodes \emph{both} uncertainty over the \emph{law} of the signal process, as well as its \emph{realizations} over time. In this appendix, we consider variations of our setting. This will clarify what is distinctive---and difficult---about dynamic regulation in the face of uncertain learning, and connect our results to the work of \cite{kruse2015optimal,kruse2019inverse}. 

\textbf{Static information with unknown law.} We focus on the case in which the principal does not learn i.e., $\mathcal{G}_l = \mathcal{G}_{0-}$ for all $l \in \mathcal{L}$, and the agent's preference $U \in \mathbb{U}$ is known. Consider a special case of our model in which the agent observes information \emph{only} at level-$0$. That is, write: 
\[
\mathbb{F}^{\mathsf{s}} := \Big\{\mathcal{F} \in \mathbb{F}: \mathcal{F}_l = \mathcal{F}_0 \text{ for all $l \in \mathcal{L}$} \Big\}
\]
recalling that we allow for information to arrive at level $l = 0$ since the filtration is right-continuous i.e., we can have $\mathcal{F}_{0-} \neq \mathcal{F}_0$. Define
\[\underline{\mu} \coloneqq \sup \{\mu \in \Delta(\Theta): l^*_V(\mu) = 0\}\]
as highest belief that the principal does not want to develop to any positive level. 

We first verify that $l^*_V$ is strictly increasing over the interval $[\underline{\mu},L]$ so that the inverse $(l^*_V)^{-1}$ is well-defined when the range is restricted to the interval $[\underline{\mu},L].$

\textbf{Fact.} $l^*_V$ is strictly increasing over the interval $[\underline{\mu},L]$. 

To see this, observe  that a necessary condition for $l^*_V(\mu)$ when $l^*_V(\mu) > 0$ is that it must solve
    \begin{align*}
        \bigg|\frac{\partial V/\partial l (1,l^*_V(\mu))} {\partial V/\partial l (0,l^*_V(\mu))} \bigg| = \frac{\mu}{1-\mu}.
    \end{align*}
    But since the function $x \mapsto \frac{x}{1-x}$ is strictly increasing, then for any $\nu \neq \mu$, $l^*_V(\nu) \neq l^*(\mu)$, But since $l^*_V$ is weakly increasing in $\mu$ it must also be strictly increasing.

Observe that a necessary and sufficient condition for $\phi^{\mathsf{s}}$ to correct the externality belief-by-belief is for each $\mu \in \Delta(\Theta)$:
\begin{align*}
&U\big(\mu, l^*_V(\mu)\big) - \phi^{\mathsf{s}}\big(l^*_V(\mu) \big) = \underbrace{\max_{l} \Big\{ U(\mu,l) - \phi^{\mathsf{s}}(l)\Big\}}_{=:U^*(\mu)}.
\end{align*}
If this holds, the envelope theorem gives:  
\[
\cfrac{dU^*(\mu)}{d\mu} = U(1,l^*_V(\mu)) - U(0,l^*_V(\mu))
\]
so for each belief $\mu \geq  \underline{\mu}$ choose 
\[
 \phi^{\mathsf{s}}\big(l^*_V(\mu) \big)  = U\big(\mu, l^*_V(\mu)\big) - \int^{\mu}_{\underline{\mu}} \Big[ U(1,l^*_V(\nu)) - U(0,l^*_V(\nu)) \Big] d\nu.
\]
We can write $\phi^{\mathsf{s}}$ entirely in terms of $l$ by changing variables:
\[
\phi^s(l) = U\big((l^*_V)^{-1}(l), l\big) - \int_0^l  \Big[ U(1,s) - U(0,s) \Big] d \big((l^*_V)^{-1}\big)(s)
\]
for every $l \geq 0.$
It remains to verify that choosing $l^*_V(\mu)$ under belief $\mu$ is optimal for the agent. 

Consider that, for every $\mu \geq \overline{\mu}.$
\begin{align*}
    \cfrac{d \phi^{\mathsf{s}} \circ l^*_V}{d\mu} &= \mu \cdot \cfrac{\partial U}{\partial l}\Big(1,l^*_V(\mu)\Big) \cdot \cfrac{d l^*_V(\mu)}{d\mu} + (1-\mu) \cdot \cfrac{\partial U}{\partial l}\Big(0,l^*_V(\mu)\Big) \cdot \cfrac{d l^*_V(\mu)}{d\mu} \\
    &=\cfrac{d l^*_V(\mu)}{d\mu}\cdot \cfrac{\partial U}{\partial l} \Big(\mu,l^*_V(\mu)\Big),
\end{align*}
where the first equality is from noting that the terms $ U(1,l^*_V(\mu)) -  U(0,l^*_V(\mu)) $ cancel out with itself; the second is from the definition of $U(\mu,\cdot)$. However,
\begin{align*}
    &\cfrac{d \phi^{\mathsf{s}} \circ l^*_V}{d\mu} = \cfrac{d l^*_V(\mu)}{d\mu} \cdot (\phi^s)'\big(l^*_V(\mu) \big) \\
    &\implies (\phi^s)'\big(l^*_V(\mu) \big) = \frac{\partial U}{\partial l} \Big(\mu,l^*_V(\mu)\Big) \\ 
    &\implies (\phi^s)'(l) = \frac{\partial U}{\partial l} \Big((l^*_V)^{-1}(l),l\Big).
\end{align*}
The first-order condition of the agent's payoff net of the mechanism $\phi^{\mathsf{s}}$ is thus: 
\begin{align*}
    \frac{\partial}{\partial l} \Big( U(\mu,l) - \phi^s(l) \Big) &= \frac{\partial U}{\partial l}(\mu,l) - \frac{\partial U}{\partial l} \Big((l^*_V)^{-1}(l),l\Big) \\
    &= \underbrace{\Big(\frac{\partial U}{\partial l}(1,l) - \frac{\partial U}{\partial l}(0,l) \Big)}_{>0} \cdot \big( \mu - (l^*_V)^{-1}(l) \big).
\end{align*}

If $\mu < \overline{\mu},$ then $(l^*_V)^{-1}(l) \geq \overline{\mu} > \mu$ for every $l.$ Thus, the agent's payoff net of the mechanism is decreasing in $l$, so it is maximized at $l = 0$, as required.

Next, consider the case that $\mu \geq \overline{\mu}.$

Since $l^*_V$ is strictly increasing over $[\underline{\mu},L]$ so is $(l^*_V)^{-1}$. Hence, the agent's first-order condition is weakly greater than $0$ if and only if $\mu \geq (l^*_V)^{-1}(l)$. This is equivalent to $l \leq l^*_V(\mu)$ since the principal's optimal technology level $l^*_V(\mu)$ is increasing in $\mu$. Hence, the agent's payoff net of the mechanism is quasiconcave in $l$ and maximized at $l = l^*_V(\mu)$. We have thus found a nonlinear tax schedule that is ex post optimal---it corrects the externality belief-by-belief. The following proposition formalizes this: 


\begin{proposition}\label{prop:static_firstbest}
    For the mechanism $\phi^{\mathsf{s}}$ constructed above: 
    \[
    \Ex\big[V(\mu_{\ell^*},\ell^*(\phi^{\mathsf{s}}, \mathcal{G}, \mathcal{F})\big]  = 
\begin{array}[t]{l}
     \sup_{\ell} \Ex\big[V(\mu_\ell,\ell) \big]  
     \\
    \text{s.t. $\ell$ is $\mathcal{F}$-adapted}
    \end{array}  \quad \text{for all $\mathcal{F} \in \mathbb{F}^{\mathsf{s}}$}
    \]
    i.e., the principal fully internalizes the externality belief-by-belief.
\end{proposition}
\cref{prop:static_firstbest} states that the principal can pick a nonlinear tax carefully to fully correct the externality for any realization of the agent's beliefs. What is more, this mechanism does not depend on the information structure---$\phi^{\mathsf{s}}$ depends only on $U$ and $V$. This means that (static) informational robustness has no bite in our static environment. 

\textbf{Dynamic information with known law.} Now suppose that $\mathcal{F}$ is Markov:  
\[
\mathbb{F}^{\mathsf{m}} := \Big\{ \mathcal{F} \in \mathbb{F}: \big(\Ex[\theta | \mathcal{F}_l]\big)_{l \in \mathcal{L}} \text{ has the strong Markov property} \Big\}.
\]
If $\mathcal{F} \in \mathbb{F}^{\mathsf{m}}$ is \emph{known} to the principal, we can once again achieve the first-best outcome---this is the insight of \cite{kruse2015optimal} who establish this in discrete time, and \cite{kruse2019inverse} in continuous time when the belief process is a diffusion. We expect the same is true of the whole set $\mathbb{F}^{\mathsf{m}}$ (including those without continuous sample belief paths e.g., jump processes) but we will not attempt a proof since the present paper is already far too long. 

\begin{table}[H]
\vspace{-0.5em}
\centering
\begin{center}
\begin{tabular}{l|c|c|}
  \multicolumn{1}{c}{} & \multicolumn{1}{c}{$\mathcal{F}$ known} & \multicolumn{1}{c}{$\mathcal{F}$ unknown} \\
  \cline{2-3}
  Static & \multicolumn{2}{c|}{\parbox[t]{4cm}{
   \centering First-best \\(\cref{prop:static_firstbest})   
  }} \\
  \cline{2-3}
  Dynamic & \parbox[t]{4cm}{First-best if Markov  \citep{kruse2015optimal,kruse2019inverse}} & \textbf{Our focus} \\
  \cline{2-3}
\end{tabular}
\end{center}
\vspace{-1em}
\caption{A taxonomy of regulation problems} \label{table:taxonomy}
\end{table}

\vspace{-1em}
\cref{table:taxonomy} puts these observations together. We further note that if $\mathcal{F}$ is known but non-Markov, the learning process can contain information about the informativeness of \emph{future} signals so beliefs are no longer a sufficient statistic for the principal's optimal stopping decisions---this prevents the designer from carefully choosing the mechanism to internalize the externality. Likewise if $\mathcal{F}$ is unknown then simply correcting externalities belief-by-belief does not mean that we have closed the wedge between the agent's and principal's stopping levels.

\end{document}